\documentclass[iop]{emulateapj}
\usepackage{amsmath}
\usepackage{color}
\shorttitle{The Young Solar Analogs Project}
\shortauthors{Gray et al.}

\begin{document}

\title{The Young Solar Analogs Project: I. Spectroscopic and Photometric 
Methods and Multi-year Timescale Spectroscopic Results}

\author{R. O. Gray}
\affil{Department of Physics and Astronomy, Appalachian State University,
Boone, NC 26808}

\author{J. M. Saken}
\affil{Department of Physics and Physical Science, Marshall University, Huntington, WV 25755}

\author{C.J. Corbally}
\affil{Vatican Observatory Research Group, Steward Observatory, Tucson, AZ
85721-0065}

\author{M. M. Briley}
\affil{Department of Physics and Astronomy, Appalachian State University,
Boone, NC 26808}

\author{R. A. Lambert, V.A. Fuller, I. M. Newsome, M. F. Seeds}
\affil{Department of Physics and Astronomy, Appalachian State University,
Boone, NC 26808}

\and

\author{Y. Kahvaz}
\affil{Department of Physics and Physical Science, Marshall University, Huntington, WV 25755}

\begin{abstract}
This is the first in a series of papers presenting methods and results from
the Young Solar Analogs Project, which began in 2007.  This project monitors
both spectroscopically and photometrically a set of 31 young (300 - 1500~Myr)
solar-type stars with the goal of gaining insight into the space environment
of the Earth during the period when life first appeared. From our 
spectroscopic observations we derive the Mount Wilson $S$ chromospheric 
activity index ($S_{\rm MW}$), and describe the method we use to transform our 
instrumental indices to $S_{\rm MW}$ without the need for a color term.  We 
introduce three {\it photospheric} indices based on strong absorption
features in the blue-violet spectrum -- the G-band, the \ion{Ca}{1} resonance
line, and the Hydrogen-$\gamma$ line -- with the expectation that 
these indices might prove to be useful in detecting variations in the surface 
temperatures of active solar-type stars. We also describe our photometric 
program, and in particular our ``Superstar technique'' for differential 
photometry which,
instead of relying on a handful of comparison stars, uses the photon flux in
the entire star field in the CCD image to derive the program star magnitude.
This enables photometric errors on the order of 0.005 -- 0.007 magnitude.
We present time series plots of our spectroscopic data for all four 
indices, and carry out extensive statistical tests on those time series
demonstrating the reality of variations on timescales of years in all four 
indices.  We  also statistically test for and discover correlations and 
anti-correlations between the four indices.  We discuss the physical basis of 
those correlations. As it turns out, the ``photospheric'' indices
appear to be most strongly affected by emission in the Paschen continuum.  We
thus anticipate that these indices may prove to be useful proxies for 
monitoring 
emission in the ultraviolet {\it Balmer} continuum.  Future papers in this 
series will discuss variability of the program stars on medium 
(days -- months) and short (minutes to hours) timescales.  
\end{abstract}

\keywords{stars: activity stars: chromospheres stars: fundamental parameters
stars: individual(\objectname{HD 166}, \objectname{HD 5996}, \objectname{HD 9472},
\objectname{HD 13531}, \objectname{HD 16673}, \objectname{HD 27685},
\objectname{HD 27808}, \objectname{HD 27836}, \objectname{HD 27859},
\objectname{HD 28394}, \objectname{HD 42807}, \objectname{HD 76218},
\objectname{HD 82885}, \objectname{HD 96064}, \objectname{HD 101501},
\objectname{HD 113319}, \objectname{HD 117378}, \objectname{HD 124694},
\objectname{HD 130322}, \objectname{HD 131511}, \objectname{HD 138763},
\objectname{HD 149661}, \objectname{HD 152391}, \objectname{HD 154417},
\objectname{HD 170778}, \objectname{HD 189733}, \objectname{HD 190771},
\objectname{HD 206860}, \objectname{HD 209393}, \objectname{HD 217813},
\objectname{HD 222143}) stars: late-type stars: rotation}

\section{Introduction}

The Young Solar Analogs Project is a long-term spectroscopic and
photometric effort to monitor a sample of Young Solar Analogs (YSAs) in
order to gain a deeper understanding of their magnetically related stellar
activity.  YSAs give us a window into the conditions in the early solar
system when life was establishing a foothold on the Earth.  That early life
had to contend with a hostile space environment, including strong ultraviolet
fluxes from a young active sun (without the benefit of an ozone layer), 
an enhanced solar wind, strong and frequent flares, as well
as significant variability in the solar irradiance.  By studying solar-type 
stars with ages corresponding to this period ($\sim$0.3 -- 1.5 Gyr) in the
history of the solar system, we can gain insight not only into the conditions 
on the early Earth, but a better understanding of the space environment 
experienced by Earth analogs, and the implications that might have for the 
development of life on those worlds. 

Stellar activity is closely related to the dynamics of the magnetic field of 
the star.  The existence of the chromosphere and corona and the associated
far-ultraviolet (FUV), extreme-ultraviolet (EUV) and X-ray emissions 
of a solar-type star are the result of magnetic heating, and solar and 
stellar active regions are associated with strong local enhancements in 
the stellar magnetic field.  The direct detection of the magnetic fields of 
solar-type stars is difficult and direct measurement of FUV (both emission-line
and Balmer continuum), EUV, and X-ray fluxes requires space-based 
observations, so the monitoring of 
magnetic activity and FUV, EUV, and X-ray fluxes in those stars depends upon 
more easily measured proxies such as, traditionally, the chromospheric 
flux in the cores of the \ion{Ca}{2} H \& K lines.  Recent studies have shown 
that \ion{Ca}{2} H \& K fluxes are correlated in solar-type stars with 
both X-ray luminosities \citep[61 Cyg A \& B]{hempelmann03}; 
\citep[HD~81809]{favata04} and FUV excesses \citep{smith10,gray11}.  Thus 
ground-based monitoring of \ion{Ca}{2} H \& K fluxes has played and 
continues to 
play a vital role in the study of stellar magnetic activity, and serves as 
a valuable proxy for the direct measurement of ultraviolet and X-ray fluxes.  

Long-term monitoring of the \ion{Ca}{2} H \& K fluxes in a sample of F-, G-, 
and K-type dwarfs began at Mount Wilson in 1966 \citep{wilson78,baliunas95}, and
continued until 2003.  That program monitored about 100 stars on a
continuous basis.  The stars in the Mount Wilson program range from young 
stars with very active chromospheres to old stars with minimal activity.  
The program discovered stellar activity cycles similar to that of the 
Sun in about 60\% of the sample, with a further 25\% varying with no 
well-defined cycle, and the remainder showing little variation at all.

The Lowell Observatory SSS (solar-stellar
spectrograph) program started in 1988 and continues today 
\citep{hall09}. It employs a fiber-fed spectrograph that enables \ion{Ca}{2} 
H \& K
measurements to be carried out on both the Sun and stars with the same
instrument.  That program, unlike the Mount Wilson project, focuses closely
on 28 stars that are most like the Sun in terms of spectral type 
(F8 -- G8, with most in the range G0 -- G2).  Many in this sample are ``solar
twins'', and thus have ages and metallicities similar to that of the Sun,
but a few of the program stars may be described as ``young solar analogs'' with
activity levels much higher than the Sun.  Lowell Observatory, unlike the
Mount Wilson project, carries out near-contemporaneous precision photometric
observations, in the Str\"omgren $b$ and $y$ bands, of a number of the 
SSS-program solar-type stars as well as
others \citep{lockwood07}.  They have found, as might be expected in
analogy with the Sun, that many of the SSS stars are brightest when at the
highest activity levels, but, surprisingly, others are faintest when most 
active.  It is the most active stars in their sample that show an inverse 
correlation between brightness and activity, suggesting that stars, as they 
age and decline in activity, flip from inverse- to direct-correlation behaviors.

The ``Sun in Time'' project \citep{guinan09} carried out, over the course 20 years,
 multi-wavelength studies of a small sample of solar analogs (G0 - G5) with ages 
ranging from $\sim 50$ Myr to 9 Gyr.  That project 
found that the early Sun was most likely rotating 10 times faster 
than at present and 
that its coronal X-ray and transition-region/chromospheric EUV and FUV 
fluxes were several hundred times higher than the present.  This project 
as well confirmed that \ion{Ca}{2} H \& K observations are useful proxies for 
estimating X-ray, EUV, and FUV fluxes and variability.

Spectroscopic features in the optical other than the \ion{Ca}{2} H \& K lines 
may yield useful stellar activity data.  The core of the H$\alpha$ line samples
the chromosphere, but other strong features in the spectrum may be sensitive to
photospheric manifestations of stellar activity.  Prime among these in the
blue-violet region of the spectrum are the 4305\AA\ G-band (a molecular
feature arising from the CH molecule), the 4227\AA\ 
\ion{Ca}{1} resonance line, and the 4340\AA\ H$\gamma$ line.  These three
features are temperature sensitive in late-F, G, and early K-type stars,
with the G-band increasing in strength through the F and G-type stars, coming 
to a broad maximum in the late G-type through early K-type stars and then 
declining toward later types.   
The \ion{Ca}{1} resonance line is negatively correlated with the effective 
temperature, and the H$\gamma$ line positively correlated.
Thus these spectral features may be useful in tracking the presence and 
areal coverage of sunspots and faculae on the photospheric disk.  In addition, 
high-resolution images of the solar surface taken in the G-band show bright 
points (GBPs) that are strongly correlated with magnetic structures such as 
intergranular lanes and extended facular regions \citep{berger01,schussler03}.
We will discuss in Sections \ref{sec:Gband}, \ref{sec:CaI}, and \ref{sec:HG} 
our definition of spectroscopic indices for the measurement of the G-band, 
the \ion{Ca}{1} resonance line, and the H$\gamma$ line.  In
\S 5.1 we test the sensitivity of these photospheric indices to temperature
variations, and in \S 5.4 examine correlations between these indices and with
the Mount Wilson chromospheric activity index.  These tests enable us to
evaluate the usefulness of these indices as temperature indicators.   

For the purpose of this project, we define a YSA as an
F8 -- K2 dwarf with an age between 0.3 and 1.5 Gyr.  A sample of 40 candidate 
YSAs north of $-10$\degr\ were chosen from the NStars project \citep{gray03} on
the basis of the following criteria: 1) Their spectral types should lie
between F8 -- K2, as we are interested in solar-type stars, and not late-K 
and M-type dwarfs.  In addition, within that spectral-type range, the
``photospheric'' features we have identified (G-band, \ion{Ca}{1} resonance
line, and the H$\gamma$ line) may be measured with sufficient accuracy. 2) The
stars should be north of $-10^\circ$ declination, and sufficiently bright
($V < 8.0$) that they may be observed at high signal-to-noise (S/N $\ge 100$)
on a routine basis in a reasonable length of time with our equipment (see 
\S\ref{sec:obs}) and 3) they should have
ages approximately between 0.3 and 1.5 Gyr, for the reasons explained above.
Initial ages were estimated on the basis of the ``snapshot'' \ion{Ca}{2} H \& K
activity measures provided by the Nearby Stars project, and the calibration of 
\citet{soderblom91} and, later, when it became available, and we had derived
better average activity measures of our program stars, that of 
\citet{mamajek08}.  Some ages were also refined via the determination of 
rotational periods \citep{barnes07}.  The list was thus culled to 31 YSAs 
(see Table \ref{tbl:bod}).  Many of these stars have been monitored 
spectroscopically since 2007.  We note that this list includes the star
HD~189733, even though that star apparently has an age $> 4$Gyr.  The
activity age of HD~189733 is approximately 600 Myr \citep{melo06}, but
this young age is inconsistent with the low X-ray flux of its M-dwarf
companion \citep{pillitteri11}.   Its rapid rotation and high activity 
presumably derives from the transfer of angular momentum from a close-orbiting 
hot jupiter \citep{pillitteri11,santapaga11}.   We have
retained this star in our program not only because of its intrinsic interest,
but because insights may come from comparing its activity behavior to young
stars with similar rotation periods and activity levels.

\begin{deluxetable*}{llllll}
\tablecolumns{6}
\tablewidth{0pc}
\tablecaption{Young Solar Analog Stars\\Basic Observational Data\label{tbl:bod}}
\tablehead{
\colhead{Name} & \colhead{SpT\tablenotemark{a}} & \colhead{V} & \colhead{B-V} & \colhead{Duplicity\tablenotemark{b}} & \colhead{Program\tablenotemark{c}}}
\startdata
HD 166    & G8 V       & 6.10   & 0.75   &  s,a 	  & \\
HD 5996   & G9 V (k)   & 7.67   & 0.75   &  s		  & \\
HD 9472   & G2+ V      & 7.63   & 0.68   &  s		  & \\
HD 13531  & G7 V       & 7.36   & 0.70   &  s		  & \\
HD 16673  & F8 V       & 5.78   & 0.52   &  s		  & \\
HD 27685  & G4 V       & 7.84   & 0.67   &  s,c 	  & \\
HD 27808  & F8 V       & 7.13   & 0.52   &  s,c 	  & \\
HD 27836  & G0 V (k)   & 7.61   & 0.60   &  s,c 	  & \\
HD 27859  & G0 V (k)   & 7.80   & 0.60   &  s,c 	  & \\
HD 28394  & F8 V       & 7.02   & 0.50   &  SB,c	  & \\
HD 42807  & G5 V       & 6.44   & 0.66   &  s		  & SSS \\
HD 76218  & G9- V (k)  & 7.69   & 0.77   &  s		  & \\
HD 82885  & G8+ V      & 5.41   & 0.77   &  V(B\tablenotemark{d}) & MtW,SSS \\
HD 96064  & G8+ V      & 7.64   & 0.77   &  V(B: M0+ Ve)  & \\
HD 101501 & G8 V       & 5.32   & 0.72   &  s		  & MtW,SSS \\
HD 113319 & G4 V       & 7.55   & 0.65   &  s		  & \\
HD 117378 & F9.5 V     & 7.64   & 0.56   &  s		  & \\
HD 124694 & F8 V       & 7.19   & 0.52   &  cpm 	  & \\
HD 130322 & G8.5 V     & 8.04   & 0.78   &  Ex;hj	  & \\
HD 131511 & K0 V       & 6.01   & 0.83   &  SB  	  & \\
HD 138763 & F9 V       & 6.51   & 0.58   &  s		  & \\
HD 149661 & K0 V       & 5.76   & 0.83   &  V ? 	  & MtW \\
HD 152391 & G8.5 V (k) & 6.64   & 0.76   &  s		  & MtW \\
HD 154417 & F9 V       & 6.01   & 0.58   &  s		  & MtW \\
HD 170778 & G0- V (k)  & 7.52   & 0.59   &  s		  & \\
HD 189733 & K2 V (k)   & 7.65   & 0.93   &  V,Ex;hj	  & \\
HD 190771 & G2 V       & 6.17   & 0.64   &  V		  & \\
HD 206860 & G0 V       & 5.94   & 0.59   &  V (T2.5\tablenotemark{e}),Ex;j & MtW \\
HD 209393 & G5 V (k)   & 7.97   & 0.68   &  s		  & \\
HD 217813 & G1 V       & 6.64   & 0.60   &  s		  & \\
HD 222143 & G3 V (k)   & 6.58   & 0.65   &  s		  & \\ 
\enddata
\tablenotetext{a}{Spectral types from \citet{gray03} and \citet{gray06} unless
otherwise indicated.}
\tablenotetext{b}{Key to duplicity notes: s = single, a = member of association,
c = member of cluster, SB = spectroscopic binary, V = visual binary (along
with spectral types of companions, if known), cpm = common proper motion
companion, Ex = exoplanet host: hj = hot jupiter; j = jupiter-mass planet.}
\tablenotetext{c}{The stars indicated are in common with other spectroscopic
activity programs, in particular MtW = Mount Wilson project 
\citep{baliunas95} and the Solar/Stellar spectrograph project \citep{hall09}.}
\tablenotetext{d}{Simbad lists a spectral type of M5 V for HD~82885B, but gives
no source.}
\tablenotetext{e}{Brown dwarf companion \citep{luhman07}.}
\end{deluxetable*}

The Lowell SSS project has shown the importance and value of contemporaneous
photometry, and so we added a photometric component to our project in 2011.
We monitor our program stars in 5 photometric bands, the Str\"{o}mgren-$v$
($\lambda_{\rm eff} = 4100$\AA), Johnson-Cousins $B$ 
(4450\AA), $V$ (5510\AA), and $R$
(6530\AA) bands, and a 3~nm-wide passband centered on the
H$\alpha$ line (6563\AA).  This photometric system is optimized to detect
stellar-activity variations.  For instance, it is well-known that late-type
active stars show greater variability at shorter wavelengths; this is
related to a greater contrast between the photosphere and the spots, and
a similar increase in the contrast between the photospheric faculae and
the photosphere at those wavelengths.  During flare events, emission in 
the Paschen continuum rises sharply with decreasing wavelength.  For both 
these reasons, it is expected that photometric variability will be more 
apparent in the Str\"{o}mgren-$v$ filter than in the Str\"{o}mgren-$b$ 
($\lambda_{\rm eff} = 4670$\AA) filter employed by the Lowell SSS project.  
Variation in stellar activity,
especially during flare events, should also be apparent in the H$\alpha$ 
line.  We will examine the relationship between these photometric data and
the spectroscopic indices we present in this paper in Paper II of this
series.

\section{Observations}
\label{sec:obs}
\subsection{Spectroscopy}
\label{sec:spectra}

Spectroscopic observations for this project have been carried out primarily 
with the G/M  spectrograph on the Dark Sky Observatory (Appalachian State 
University) 0.8-m reflector.  Except for early in the endeavor, observations 
for this
project on that instrument have been obtained with the 1200~g\,mm$^{-1}$
grating in the first order.  That grating gives a spectral range of
3800 -- 4600\AA, with a resolution of 1.8\AA/2 pixels ($R \sim 2300$).  This
spectral range includes the \ion{Ca}{2} H \& K lines as well as the
\ion{Ca}{1} resonance line, the G-band, and the H$\gamma$ line.  Exposures
have been calculated to give a S/N of at least 100 in the continuum near
the \ion{Ca}{2} H \& K lines, which means that the S/N near the G-band
is consistently better than 150.  A few early observations were made with
the 600~gmm$^{-1}$ grating (used in the first order), yielding a resolution
of 3.6\AA/2 pixels and the 1000~g\,mm$^{-1}$ grating (used in the second order)
giving a resolution of $\sim 1$\AA/2 pixels.  Before April 2009, our spectra 
were recorded on a thinned, back-illuminated $1024 \times 1024$ pixel 
Tektronics CCD operated in the multipinned-phase mode.  Since April 2009, 
we have been using an Apogee
camera with a $1024 \times 256$ pixel e2v technologies CCD30-11 chip with
enhanced ultraviolet sensitivity.  These two chips have very similar pixel
sizes and spectral sensitivities, and we have detected only minor changes 
in the instrumental systems (detailed below) in the transition between the two
CCDs.  

An Fe-Ar hollow-cathode comparison lamp was observed for wavelength 
calibrations, and the spectroscopic data were reduced with 
IRAF\footnote{IRAF is distributed by the National Optical Astronomy 
Observatory, which is operated by the Association of Universities for 
Research in Astronomy, Inc. under cooperative agreement with the National 
Science Foundation.} using standard techniques.

Since January 2013 the VATTspec spectrograph on the Vatican Advanced 
Technology Telescope (VATT; 1.8-m, located on Mount Graham, Arizona) has 
also been used for this project, primarily for
high-cadence, high-S/N observations designed to detect flares and other 
short-term events on these stars.  Those observations will be discussed in a
later paper in this series.  For these observations,
the VATTspec is used with a 1200 g mm$^{-1}$ grating which gives a resolution
of 0.75\AA/2 pixels in the vicinity of the \ion{Ca}{2} H \& K lines, with 
a spectral
range of 3640 -- 4630\AA. The spectra are recorded on a low-noise STA0520A
CCD with 2688$\times$512 pixels (University of Arizona Imaging Technology
serial number 8228).  Two hollow cathode lamps, Hg and Ar, were observed
simultaneously for wavelength calibrations, and the spectroscopic data were
again reduced with IRAF using standard techniques.

We have also obtained high-resolution echelle data for six of our stars with
the FIES spectrograph on the Nordic Optical Telescope \citep{telting14}.  These
data, which were obtained under the Nordic Optical Telescope Service Observing
Program employed the FIES spectrograph with the high-resolution fiber,
yielding a resolution of 65,000, and a spectral range from 3640 -- 7360\AA.
Spectra from the FIES spectrograph were reduced with FIEStool.

\subsection{Photometry}

An important component of the Young Solar Analogs project is concurrent 
multiband 
photometry of our program stars.  The analysis of this photometry and how it
relates to our spectroscopic observations will be the subject of Paper II in
this series. 
In March 2011 we began obtaining photometric observations in the
Str\"{o}mgren-$v$, Johnson-Cousins $B$, $V$, $R$ and narrowband H$\alpha$
filter system, described in the previous section, by employing a CCD camera on
a  0.15-m 1300mm focal-length astrograph attached to the 0.8-m Dark Sky
Observatory reflector.  The detector is a KAF-8300 monochrome CCD, operated 
with on-chip $2 \times 2$ binning
to give an effective pixel size of $10.8 \times 10.8\mu$m.  The CCD utilizes
an SBIG ``even illumination shutter'' which ensures uniform exposures over
the entire field even for very short exposures.  Flat fields are obtained
every night with a ``Flipflat'' luminescent panel which offers more consistent
flats than sky flats.  This instrument, which has
a 48\arcmin\ $\times$ 36\arcmin\ field of view, is known as the ``Piggy-back'' 
telescope.  It enables us to obtain photometry simultaneously with the 
spectroscopy.  

In April 2012 we installed a small robotic dome at the Dark Sky Observatory
containing a clone of the Piggy-back telescope mounted on a German equatorial
mount.  This robotic telescope employs the CCDAutopilot5 and Pinpoint 
software which, when combined, allow fully automated operation with precise
and consistent centering of the object to within a few arcseconds.  This
telescope enables us to obtain photometry on every clear night, as the
YSA project has access to the 0.8-m and Piggy-back telescopes only 
$\sim 11 - 12$ nights a month.

Both the Robotic and the Piggy-back telescopes are operated very slightly
out of focus so that the star image is spread over a number of pixels.  This
enables more precise photometry.  Multiple exposures are obtained for each
target, which are reduced and then combined using the IRAF {\it xregister} 
function. 

Since August 2014 we have also obtained photometry with a wide-field
imager mounted on the Robotic telescope.  This wide-field imager consists of
an ST-8300 SBIG CCD, a filter wheel with Johnson $B$, $V$ and $R$ filters, and
a Pentax 150mm f/3.5 camera lens.  This setup yields a 
$6.9^\circ \times 5.3^\circ$
field of view, and supplements the Robotic telescope data for program stars 
which do not have sufficient comparison stars in the 48\arcmin\ $\times$ 
36\arcmin\ field of view of the main telescope.

\subsubsection{Photometric Reduction Technique}

Reducing the photometric data from the Piggy-back and Robotic telescopes is
challenging in a number of ways.  First, despite the small aperture (0.15-m),
some of our program stars are bright ($V < 6$), which requires short exposures.
To mitigate these difficulties, the telescopes are slightly defocused, and we
obtain multiple exposures which are stacked using IRAF routines which preserve
the stellar flux.  None of our fields are crowded, and so photometry is 
carried out on the stacked images using the IRAF APPHOT package.  

We utilize differential photometry to determine the magnitudes of
our program stars.  In most cases, the program star is the brightest in the
field.  Suitable comparison and check stars are typically one or two magnitudes
fainter than the program star, so the standard differential photometry 
technique leads to unacceptably large photometric errors.  To achieve 
better photometric
accuracy we have devised an improved method, which we call the ``Superstar 
technique'' (SST).  The SST, instead of utilizing a handful of comparison 
stars, considers
the photon flux in the entire star field in the image.  Thus the SST adds up the
flux from many different sources, both 
bright and faint, and constructs from that summed flux a ``super'' comparison 
star that often has comparable flux to the program star.  The technique 
compares each individual source against the summed flux, thus enabling, in
an interactive fashion, the elimination of variable stars from the final summed
flux.  In this way a reference file of comparison stars, often 20 -- 50 objects,
(the ``reference stars'') is constructed.  The individual fluxes in that 
reference file are based on
averages over a large number of nights, so the relative fluxes are known to
high precision.  To determine the magnitude of the program star
for a given night, the SST identifies as many of the reference stars as
possible on the stacked frame for that night (it is not necessary to identify
all of the reference stars) and uses those identified to construct the
``super'' comparison star.  The summed flux for that super comparison is
compared to the summed flux of the identified stars in the reference file,
and that ratio enables the calculation of a $\Delta m$ for that 
particular observation.  That $\Delta m$ is added to the instrumental 
magnitude of the program star to give the magnitude for that observation.  
The magnitudes so determined are not yet on the standard system, but are 
offset by a constant zeropoint shift.  If a number of the reference stars 
have measured magnitudes on a standard system, they can be used to calculate 
that zeropoint shift.  However, most of our work can
be carried out in the instrumental system.

The Superstar technique gives best results when the program star is situated
in a rich stellar field, enabling the summation of scores of stellar fluxes
into the single super comparison star.  For those stars in our program for
which 20 or more reference stars are available, the typical photometric 
errors in the individual Johnson-Cousins $B$, $V$, and $R$ 
magnitudes are on the order of 0.005 -- 0.007 mag.  The errors in the 
Str\"{o}mgren-$v$ and H$\alpha$ bands tend to be somewhat higher: 
0.007 -- 0.010 mag.  For the brightest stars in our program and stars with 
sparse fields ($ < 20$ reference stars) 
the errors are higher, and typically range, on good nights, from 0.010 - 0.015
magnitude, with slightly higher errors in Str\"{o}mgren-$v$ and H$\alpha$.
These are the stars that will benefit from the photometry obtained with the
wide-field imager that is mounted on the Robotic telescope (see above).

We defer a deeper discussion of the photometric errors until Paper II which
will be devoted to an analysis of the photometric data as well as its 
relationship to the spectroscopic data discussed in this paper.
  
\section{Basic Physical Parameters}

\begin{deluxetable*}{llllcll}
\tablecolumns{7}
\tablewidth{0pt}
\tablecaption{Young Solar Analog Stars\\Basic Physical Data\label{tbl:bpp}}
\tablehead{
\colhead{Name} & \colhead{$T_{\rm eff}$(K)} & \colhead{$\log g$} & \colhead{[M/H]} & \colhead{$\xi_t$} & \colhead{$v\sin i$ {\scriptsize km s$^{-1}$}} &  
\colhead{Source\tablenotemark{a}}\\
   &    &    &    & \colhead{{\scriptsize km s$^{-1}$}}   
& \colhead{(error)} & }
\startdata
HD 166    & 5454 & 4.52 & $+0.05$ & 1.3  & \phn4.5 (0.2) & Keck            \\
HD 5996   & 5463 & 4.60 & $+0.01$ & 0.7  & \phn0.0 (1.5) & Elodie          \\
HD 9472   & 5705 & 4.46 & $-0.03$ & 1.1  & \phn3.1 (0.2) & Keck           \\
HD 13531  & 5595 & 4.54 & $-0.02$ & 1.1  & \phn6.1 (0.1) & Keck            \\
HD 16673  & 6241 & 4.38 & $-0.05$ & 1.3  & \phn7.3 (0.2) & Elodie          \\
HD 27685  & 5681 & 4.43 & $+0.13$ & 1.0  & \phn1.6 (1.0) & Elodie          \\
HD 27808  & 6217 & 4.31 & $+0.11$ & 1.2  & 12.7 (0.2) & Elodie             \\
HD 27836  & 5843 & 4.35 & \nodata & \nodata  & \nodata    & \nodata      \\
HD 27859  & 5887 & 4.36 & $+0.06$ & 1.2  & \phn7.3 (0.2) & Keck            \\
HD 28394  & 6243 & 4.31 & $+0.09$ & 1.2  & 22.0 (1.0) & Keck               \\
HD 42807  & 5722 & 4.55 & $-0.03$ & 1.2  & \phn5.0 (0.2) & Keck            \\
HD 76218  & 5380 & 4.56 & $+0.07$ & 1.0  & \phn3.4 (0.2) & Keck            \\
HD 82885  & 5487 & 4.43 & $+0.29$ & 1.3  & \phn3.2 (0.2) & Keck             \\
HD 96064  & 5402 & 4.54 & $+0.13$ & 0.6  & \phn2.8 (0.5) & Elodie          \\
HD 101501 & 5535 & 4.55 & $-0.04$ & 1.0  & \phn2.8 (0.4) & Keck            \\
HD 113319 & 5736 & 4.53 & $-0.05$ & 1.1  & \phn3.6 (0.2) & Keck            \\
HD 117378 & 6000 & 4.51 & $-0.07$ & 1.3  & 10.2 (0.2)    & NOT    \\
HD 124694 & 6195 & 4.44 & $+0.05$ & 1.2  & 17.6 (0.5)    & NOT             \\
HD 130322 & 5385 & 4.53 & $+0.05$ & 1.0  & \phn0.0 (1.5) & Elodie          \\
HD 131511 & 5215 & 4.51 & $+0.07$ & 1.2  & \phn4.7 (0.2) & NOT             \\
HD 138763 & 6040 & 4.43 & \nodata & \nodata  & \nodata    & \nodata      \\
HD 149661 & 5255 & 4.57 & $-0.01$ & 1.0  & \phn1.5 (0.2) & Paranal         \\
HD 152391 & 5443 & 4.53 & $+0.02$ & 1.2  & \phn4.3 (0.2) & NOT             \\
HD 154417 & 6022 & 4.42 & $-0.02$ & 1.4  & \phn6.8 (0.2) & Keck            \\
HD 170778 & 5925 & 4.48 & $+0.01$ & 1.3  & \phn7.9 (0.2) & NOT      \\
HD 189733 & 5049 & 4.59 & $+0.04$ & 1.1  & \phn2.9 (0.2) & Keck            \\
HD 190771 & 5789 & 4.45 & $+0.12$ & 1.5  & \phn5.4 (0.2) & NOT          \\
HD 206860 & 5986 & 4.49 & $-0.07$ & 1.5  & 10.0 (0.2) & Keck    \\
HD 209393 & 5670 & 4.58 & $-0.10$ & 1.0  & \phn4.0 (0.2) & Keck    \\
HD 217813 & 5876 & 4.45 & $+0.00$ & 1.5  & \phn4.4 (0.2) & Keck    \\
HD 222143 & 5787 & 4.43 & $+0.06$ & 1.3  & \phn3.2 (0.2) & Keck    \\
Sun       & 5774 & 4.44 & $+0.00$ & 1.0  & \phn1.8 (0.2) & NSO      \\
\enddata
\tablenotetext{a}{Keck: The Keck Observatory Archive 
https://koa.ipac.caltech.edu/cgi-bin/KOA/nph-KOAlogin; Elodie: The Elodie
Archive http://atlas.obs-hp.fr/elodie/, \citet{moultaka04}; NOT: Nordic 
Optical Telescope Service Observing Proposal P50-410; Paranal: The UVES 
Paranal Observatory Project (POP), \citet{bagnulo03},
http://www.eso.org/sci/observing/tools/uvespop.html; NSO: \citet{kurucz84}.}
\end{deluxetable*}

Table \ref{tbl:bpp} presents basic physical data, namely effective 
temperatures, 
surface gravities ($\log g$), metallicities ([M/H]), microturbulent velocities
($\xi_t$), and projected rotational velocities ($v\sin i$) for the program 
stars.  The effective temperatures were determined using the
infrared flux method formulae of \citet{cassagrande10}, specifically, those
for $b-y$, $B-V$, and $V-K_s$, where $K_s$ is the 2MASS K-magnitude 
\citep{skrutskie06}.  The effective temperatures presented are straight means
of the values based on those three indices, except for some of the brighter
stars for which $K_s$ is saturated and thus unreliable.  The statistical error
associated with these temperatures is on the order of $\pm 70$K, with an
additional systematic error in the zeropoint of the system of about $15 - 20$K
\citep{cassagrande10}.  The gravities were calculated via the absolute 
bolometric magnitudes, based on Hipparcos parallaxes as recalculated by
\citet{vanleeuwen07} and bolometric corrections from \citet{flower96} along
with the mass-luminosity relationship from \citet{andersen91}, and have
errors on the order of $\pm 0.10$ in the $\log$.  Metallicities, 
microturbulent velocities, and projected rotational velocities were calculated
from measurements of high-resolution archival spectra from the HIRES 
spectrograph on the Keck 10-m telescope, the Elodie spectrograph on the 193-cm
telescope at the Observatoire de Haute-Provence, the UVES spectrograph on the
ESO VLT provided by the UVES Paranal Observatory Project, as well as new 
observations with the FIES spectrograph on the Nordic Optical Telescope.  

Projected rotational velocities were calculated with the cross-correlation
method.  To do this, we first estimated the line-spread function (LSF) for each
spectrum by measuring the FWHM in \aa ngstroms of a number of telluric lines 
in the atmospheric $\alpha$-band of oxygen, centered $\sim 6300$\AA\ or, in
some cases the $\alpha^\prime$ band centered near 5800\AA, and 
then transformed that FWHM to the echelle orders containing the spectral range
6050 -- 6200\AA\ where most of the measurements for calculating $v\sin i$ and
[M/H] were made.  Once the LSF was characterized, we computed synthetic
spectra in the 6050 -- 6200\AA\ range with the 
SPECTRUM\footnote{http://www.appstate.edu/$\mathtt{\sim}$grayro/spectrum/spectrum.html} 
code of \citet{gray94} and solar-metallicity 
ATLAS12 models \citep{castelli03} calculated with the effective 
temperatures and gravities in Table \ref{tbl:bpp}.  Those synthetic 
spectra were then 
convolved with the LSF. Cross correlations were obtained between the
synthetic spectrum and the observed spectrum, and the synthetic spectrum and
rotationally broadened versions of itself for a range of rotational velocities.
These cross correlations were normalized at a common point and compared
to derive the rotational velocity of the program star.  Our results are in
very good agreement with those of \citet{mishenina12} who used the 
cross-correlation method of \citet{queloz98}.

Once the LSF and the $v\sin i$ were known, we used a $\chi^2$ minimization
method comparing the observed and synthetic spectra to determine both the 
metallicity and the microturbulent velocity for
each program star.  For the synthetic spectra, we used a spectral line list
in the region 6050 -- 6200\AA\ with updated $\log(gf)$ values from the 
NIST Atomic Spectra Database, version 5.2 \citep{nist}. Broadening parameters
and $\log(gf)$ values were adjusted, when necessary, by 
reference to the Solar Flux Atlas \citep{kurucz84}. The metallicities and
microturbulent velocities are recorded in Table \ref{tbl:bpp}.  We 
estimate errors in
that Table to be $\pm 0.05$ dex for the metallicity, and about 
$\pm 0.3$ km s$^{-1}$ for the microturbulent velocity.

The projected rotational velocities will be used in a later paper in 
this series to
interpret periodicities observed in our activity and photometric data.

\section{Spectroscopic Indices for Stellar Activity}
\label{sec:indices}

Our project measures four spectroscopic indices from the spectra obtained on
the G/M spectrograph.  These are the \ion{Ca}{2} H \& K chromospheric 
activity index, based
on the Mount Wilson ``$S$'' index (hereinafter $S_{\rm MW}$), and indices for 
the \ion{Ca}{1} 4227\AA\
resonance line, the 4305\AA\ G-band, and the 4340\AA\ H$\gamma$ line.  

\subsection{\ion{Ca}{2} H \& K chromospheric activity indices}

\subsubsection{Definition and Measurement of the Instrumental Indices}

\citet{wilson68,wilson78} and \citet{vaughan78} introduced the Mount Wilson 
chromospheric activity index, $S_{\rm MW}$, which recorded the chromospheric flux
in the cores of the \ion{Ca}{2} H \& K lines in ratio with flux in the
``continuum'' on either side of those lines.  Their instrument employed
effective triangular bands with full width at half peak of 1.09\AA\ centered
on the cores of the H \& K lines, and continuum bands of 20\AA\ width to the
violet side (3891.067 -- 3911.067\AA) and the red (3991.067 -- 4011.067\AA).
The fluxes measured through these bands are ratioed to give the $S_{\rm MW}$ 
index.
We measure two instrumental indices from the DSO spectra, the $S_2$ index
which measures the flux in the cores of the H \& K lines with 2\AA-wide
rectangular bands and the $S_4$ index which employs 4\AA-wide rectangular
bands in the H \& K cores.  Both indices utilize the same continuum
bands as the Mount Wilson Project.  The indices are calculated (in analogy
with the Mount Wilson index)  with the equations
\begin{displaymath}
S_2 = 5\frac{f_{K2} + f_{H2}}{f_v + f_r}
\end{displaymath}
\begin{displaymath}
S_4 = 5\frac{f_{K4} + f_{H4}}{f_v + f_r}
\end{displaymath}
where the $f$'s are the {\it monochromatic} fluxes (i.e. the integrated flux
divided by the bandwidth) through the various bands described above.  In
particular, $f_{K2}$ and $f_{K4}$ are the fluxes measured in the core of the
\ion{Ca}{2} K-line using 2 and 4\AA\ bandpasses respectively; $f_{H2}$ and 
$f_{H4}$ are the same for the \ion{Ca}{2} H-line, and $f_v$ and $f_r$ are the
fluxes in the two continuum bands. The
DSO spectra do not have sufficient resolution to directly measure 1\AA\ 
fluxes in
the cores of the \ion{Ca}{2} H \& K lines.

However, the 0.75\AA/2 pixel resolution of the VATTspec spectra does 
allow direct
measurement of an $S_1$ index, which employs rectangular 1\AA\ passbands in 
the cores of the H \& K lines.  The advantage of the $S_1$ index is that it 
is closer to the original instrumental system of the Mount Wilson project 
(although that project utilized a triangular passband) and the transformation
from $S_1$ to $S_{\rm MW}$ is linear and does not involve a color ($B-V$) term, 
whereas 
the $S_2 \rightarrow S_{\rm MW}$ and $S_4 \rightarrow S_{\rm MW}$ transformations
are both nonlinear and require a color term (see below).

Steps in the measurement of the $S_1$, $S_2$, and $S_4$ indices include
transforming the stellar spectrum in question to the rest frame of the star,
the rebinning of the spectrum to a uniform spacing of 0.1\AA, followed by 
the numerical integration of the spectrum in the various passbands.  We 
employ the raw (non-flux-calibrated) spectrum for these calculations.  The 
division by the sum of the continuum fluxes ($f_v + f_r$) in the above 
equations accounts for changes in the slope of the continuum due to 
differing amounts of atmospheric extinction, although for routine observations
we attempt to observe the star as close to the meridian as possible.  For 
moderately high S/N spectra (S/N $> 100$), all
three indices may be measured to a precision of $\sim 0.001$ in the 
index.  
 
\subsubsection{Calibration of the Instrumental Indices: Transformation to the
Mount Wilson index}

\begin{figure} 
\includegraphics[width=3.25in]{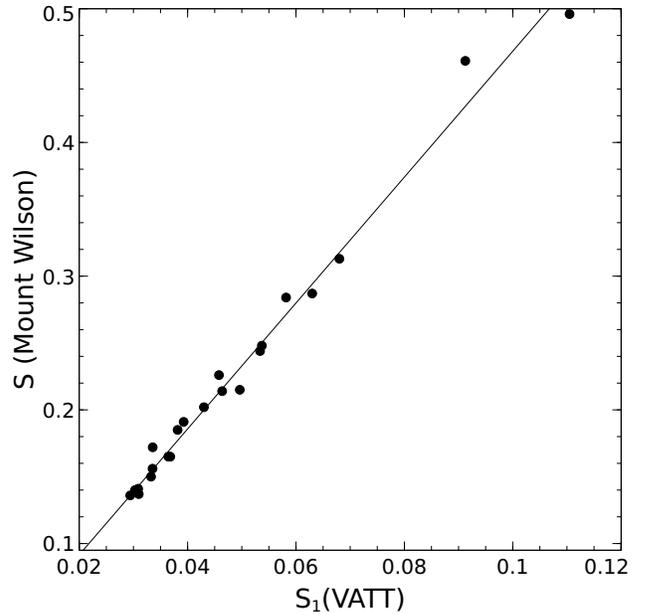} 
\caption{The $S_1 \rightarrow S_{\rm MW}$ (Mount Wilson) transformation for 
VATTspec
spectra.  The calibration is linear, and has no significant color term.}
\label{fig:VATTS1Smw}
\end{figure}

The transformation of $S_4$ to $S_{\rm MW}$, as described in \citet{gray03} is
problematical, as the relationship is highly nonlinear.  In addition, it
was not appreciated at the time that there is a small but significant color
term in the transformation.  The transformation for $S_2$ is better behaved,
but is still non-linear, and a color term is still required.  As stated
above, the $S_1$ indices measured in the VATTspec spectra are linearly
correlated with the Mount Wilson $S_{\rm MW}$, and that transformation does not
involve a color term.  To derive that transformation, we have observed
with the VATTspec a number of the chromospheric activity calibration stars 
used by \citet{gray03} in their original calibration of $S_{18}$, which is the
same as the $S_4$ index of the present paper.  
The relationship between the VATTspec $S_1$ index and the mean $S_{\rm MW}$ 
indices
recorded for those calibration stars in \citet{baliunas95} is given by:
\begin{displaymath}
S_{\rm MW} = -0.0011 + 4.6920S_1   \quad  \sigma = 0.0119
\end{displaymath}
and illustrated in Figure \ref{fig:VATTS1Smw}.  The goodness of fit is not
improved with a quadratic term, and the residuals show no correlation with
$B-V$.  Most of the scatter in that relationship may be traced to the 
variability of the calibration stars, especially the more active
calibration stars.

\begin{figure} 
\includegraphics[width=3.25in]{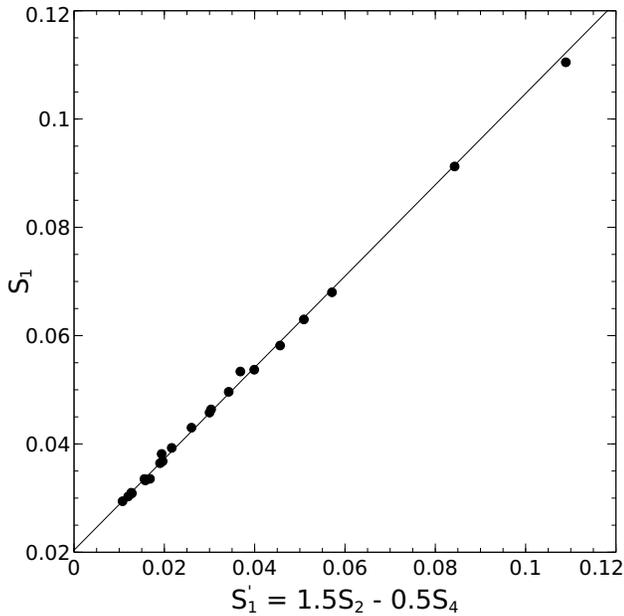} 
\caption{The relationship between the directly measured $S_1$(VATT) activity 
index and the {\it extrapolated} $S_1^\prime$ index, based on the $S_2$ and
$S_4$ indices.}
\label{fig:VATTS1}
\end{figure}

As mentioned above the $S_2 \rightarrow S_{\rm MW}$ and the $S_4 \rightarrow 
S_{\rm MW}$
transformations are both non-linear and require a color term.  The non-linear
nature of these transformations is problematical when attempting an 
extrapolation of the transformation to very active stars.  Because the 
resolution of the DSO spectra
is $\sim 1.8$\AA/2 pixels, we cannot directly measure a DSO $S_1$ index.  
However, 
experimentation with the VATTspec spectra suggests a solution.  The actual 
\ion{Ca}{2} H \& K chromospheric emission in main-sequence stars is 
intrinsically narrow 
(FWHM $\sim 0.5$\AA), narrower than even the 1\AA\ passband employed by 
the Mount Wilson project.
That flux is entirely contained in the H \& K passbands employed in the 
$S_1$, $S_2$, and $S_4$ indices, but those passbands involve successively 
larger 
amounts of {\it photospheric} flux.  This suggests that it should be possible 
to use the $S_2$ and the $S_4$ indices to {\it extrapolate} linearly to 
an $S_1$ index: $S_1 = 1.5S_2 - 0.5S_4$.  That this is feasible can be 
demonstrated with the VATTspec spectra.  Figure \ref{fig:VATTS1} shows the 
correlation between the directly measured
VATTspec $S_1$ index, and $S_1^\prime$ extrapolated from $S_2$ and $S_4$. The
two are linearly related, and $S_1^\prime$ can predict the directly measured
$S_1$ index to better than $\pm$ 1\%.  

\begin{figure} 
\includegraphics[width=3.25in]{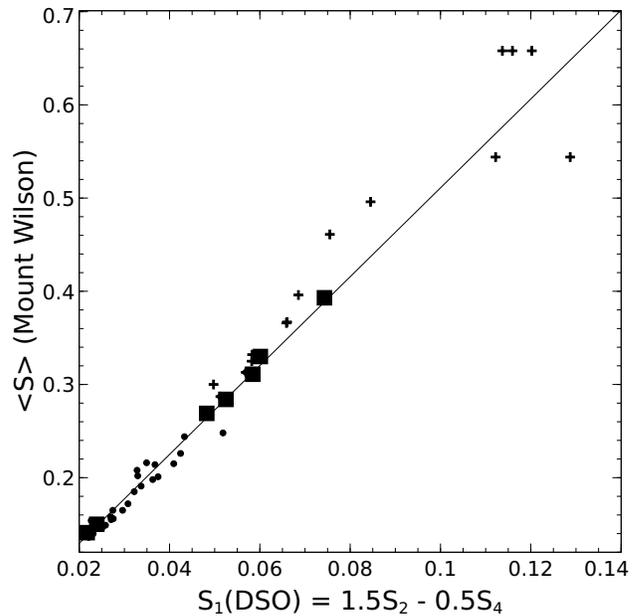} 
\caption{The DSO $S_1 \rightarrow S_{\rm MW}$(Mount Wilson) calibration.  
The ordinate 
is the mean Mount Wilson $\langle S_{\rm MW} \rangle$ index \citep{baliunas95}. 
The 
small circles represent snapshot (single to a few) observations of the Mount 
Wilson calibration stars \citep{gray03}.  The large squares represent Mount 
Wilson stars that have been regularly observed at DSO since 2007.  For 
these stars the 
8-year mean $S_1$ index (in some cases derived from over a hundred 
observations) is used.  These stars are given five times the weight of the
snapshot stars in deriving the calibration.  Finally, the crosses represent
individual snapshot observations of very active Mount Wilson stars.  These 
stars were
not used in the derivation of the calibration, but indicate that extrapolation
of the calibration is adequate even for very active stars. }
\label{fig:DSOS1Smw}
\end{figure}

This provides a way to derive a linear transformation with no color term
between the instrumental DSO system and the Mount Wilson system.  An $S_1$
extrapolated index is formed from the $S_2$ and $S_4$ instrumental indices,
and that $S_1$ index is calibrated to the Mount Wilson $S_{\rm MW}$ index via 
observations of the chromospheric activity calibration stars of \citet{gray03}.
For most of those calibration stars we have only a few ($< 5$) observations
scattered over the past 15 years.  These we refer to as ``snapshot'' 
observations.  However, as part of the YSA project we have intensively 
observed eight Mount Wilson stars -- HD~45067, HD~143761, HD~207978, HD~82885,
HD~101501, HD~152391, HD~154417, and HD~206860.  The first three of these
stars are regularly observed ``chromospherically stable stars'' used to monitor
the stability of our instrumental system (see below), and the
latter five are active G-type stars.  For these stars, we can form multi-year 
means for the instrumental indices that are much better correlated with
the Mount Wilson means than the snapshot observations of the other calibration
stars. In deriving the calibration, we give the snapshot observations a
weight of 1 and the multi-year means a weight of 5.  This yields the 
calibration (see Figure \ref{fig:DSOS1Smw}):
\begin{displaymath}
S_{\rm MW} = 0.0323 + 4.8335S_1 \quad \sigma = 0.0077
\end{displaymath}
The residuals from the calibration show no evidence for a color term.  In
addition, as the figure illustrates, extrapolation of this linear relationship 
seems to hold for very active stars.

\begin{figure} 
\includegraphics[width=3.25in]{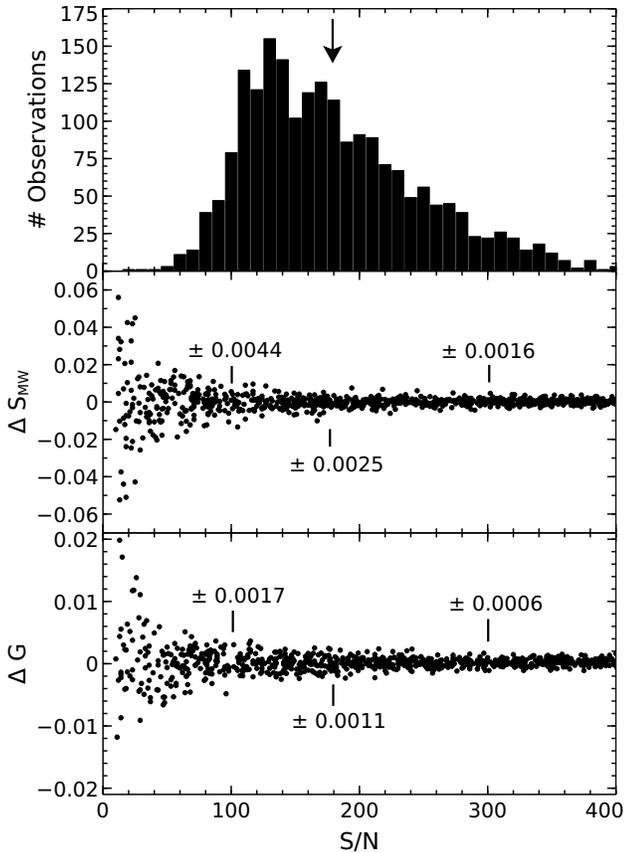} 
\caption{The top panel shows the histogram of the S/N values of our 
observations.  The S/N values are estimated in the continuum just longwards of
the \ion{Ca}{2} H line. The arrow indicates the average S/N, about 180.  The
central panel shows the results of a Monte Carlo simulation of the measurement
error of $S_{\rm MW}$ as a function of S/N.  At S/N = 180, the measurement
error is about $\pm 0.0025$.  The bottom panel shows a similar simulation
for the G-band index.  While the S/N in the continuum at the G-band is 
$\sim 1.3$ -- $1.4$ times that just longwards of the \ion{Ca}{2} H line,
we have plotted, for simplicity, the G-band errors against the H-line S/N.
At S/N = 180, the measurement error in the G-band index is approximately
$\pm 0.0005$.}
\label{fig:SN}
\end{figure}

The precision of our determinations of $S_{\rm MW}$ depend on the S/N of the
observations.  We have attempted to estimate those precisions via a Monte-Carlo
method that begins with a synthetic spectrum of the \ion{Ca}{2} H \& K region
smoothed to a resolution of 1.8\AA/2 pixels (the resolution of the 
DSO spectra).  
The Monte-Carlo technique simulates exposing on the spectrum until a certain
S/N is achieved in the continuum just longwards of \ion{Ca}{2} H.  That
exposure is processed through our measuring programs in exactly the same
way as the real spectra, including the velocity correction (the synthetic
spectra are given random radial velocity shifts between $-30$ and $+30$ km/s), 
measurements of $S_2$, $S_4$, the calculation of $S_1$, and the transformation 
to the Mount Wilson system) 
enabling a calculation of the error $\Delta S_{\rm MW}$
for a given simulation.  Those errors are plotted against S/N in the middle 
panel of Figure \ref{fig:SN}.  In the top panel of that same Figure is a
histogram of the S/N values of our observations.  The average S/N $\sim 180$,
for which a measurement precision of $\pm 0.003$ in the $S_{\rm MW}$ index
is estimated.  Indeed this error estimate (which does not include any possible
systematic errors in the transformation of our instrumental system to the
Mount Wilson system) is consistent with our measurements of $S_1$ in
the set of ``chromospherically stable'' stars (see below).  The bottom panel
of the figure shows a similar calculation for the G-band index (see below).

\subsubsection{Stability of the Dark Sky Observatory Instrumental System}

\begin{figure} 
\includegraphics[width=3.25in]{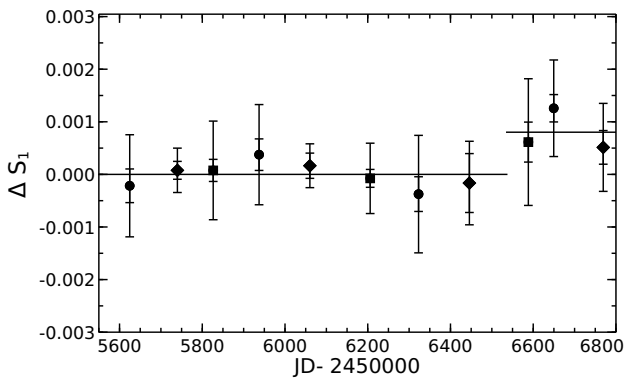} 
\caption{The seasonal mean residuals in the instrumental $S_1$ index 
observed for the three chromospherically ``stable'' stars, HD~45067 
(filled circles), HD~143761 (diamonds), HD~207978 (squares). The outer ``error''
bars indicate the standard deviation in the measured index for a given
season.  The inner error bars indicate the standard error of the mean.
This diagram and similar ones for the other instrumental indices for the
G-band, \ion{Ca}{1}, and H$\gamma$ can be used to assess the stability of 
the instrumental system and to derive corrections to apply to the observed 
indices.}
\label{fig:residuals}
\end{figure}

To monitor the stability of the Dark Sky Observatory instrumental system, we 
have regularly observed for the past 5 years, every clear night, at least 
one  chromospherically ``stable'' star, chosen from a set of stars showing 
flat activity on the Mt. Wilson project \citep{baliunas95}.  The stable 
stars that we observe are 
HD 45067, HD 143761, and HD 207978. During the course of an observing season, 
the standard deviations for night-to-night variations of those stars range 
from 0.0004 -- 0.0012 in $S_1$.  The lower figure in that range translates to
a standard deviation in $S_{\rm MW} \sim 0.0019$, in line with our Monte Carlo
estimates for the observational error in that index.  To monitor any changes 
in the instrumental system, we have adopted the period July 1, 2011 
(MJD = JD - 2450000 = 5743) to June 30, 2013 (MJD = 6445) as the
reference zeropoint baseline for the instrumental system.  Residuals in the
seasonal means of the instrumental indices relative to that baseline
will then reveal changes in the instrumental system.  This is illustrated
in Figure \ref{fig:residuals} for the $S_1$ index.  

That figure shows that 
the instrumental system has remained very stable from the time that we 
began regular monitoring of the chromospherically stable stars.  However, 
beginning September 1, 2013 (MJD = JD - 2450000 = 6536), there was a very 
small but abrupt shift in the
instrumental system.  That shift can be traced to the return of the CCD to the
manufacturers for repairs because of the failure of the vacuum seal.  During
that visit, not only was the vacuum seal repaired, but a new driver was
installed that fixed a very low-level but variable bias pattern.  In addition,
improved optical baffling was installed in the interior of the CCD housing
which may have slightly reduced the already very low level of scattered light.
To correct for this shift in the instrumental system, we subtract 0.0007 from
the $S_1$ indices obtained since September 1, 2013.  That correction may be
propagated, if required, to the $S_2$ and $S_4$ indices using the relationships
between those indices.  We have derived similar very small corrections to 
the other observed indices.  

Before April 2009, the spectroscopic data for this project were obtained with 
a Tektronics CCD (see \S \ref{sec:spectra}) on the same spectrograph.  We 
have investigated the difference in the  instrumental system between the 
two CCDs using spectra of inactive F-, G- and K-type stars taken with both 
CCDs and find a small systematic difference between the two systems of 0.0019
in the measurement of the $S_1$ index.  This correction has been applied to 
the earlier data.

\subsection{The G-band Index}
\label{sec:Gband}

\begin{deluxetable}{lll}
\tablecolumns{3}
\tablewidth{0pc}
\tablecaption{Band definitions for the Photospheric Indices\label{tbl:bands}}
\tablehead{
\colhead{Band Name} & \colhead{Violet Edge} & \colhead{Red Edge}}
\startdata
Continuum ($c_1$) & 4208.0\AA & 4214.0\AA \\
\ion{Ca}{1} 4226.7\AA & 4225.7\AA & 4227.7\AA \\
Continuum ($c_2$) & 4239.4\AA & 4245.4\AA \\
Continuum ($c_3$) & 4263.0\AA & 4266.0\AA \\
G-band            & 4298.0\AA & 4312.0\AA \\
Continuum ($c_4$) & 4316.0\AA & 4320.5\AA \\
Continuum ($c_5$) & 4329.0\AA & 4334.0\AA \\
H$\gamma$         & 4339.5\AA & 4341.5\AA \\
Continuum ($c_6$) & 4345.0\AA & 4349.5\AA \\
\enddata
\end{deluxetable}

\begin{figure}
\includegraphics[width=3.25in]{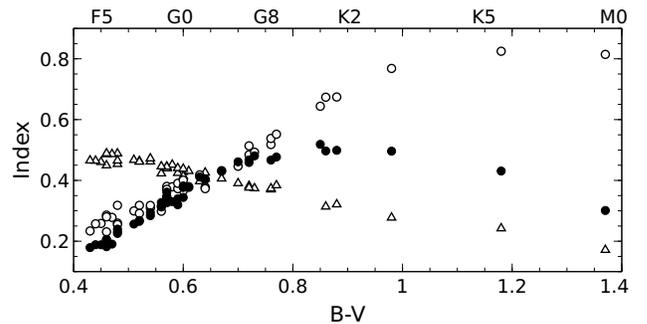}
\caption{The variation of the three {\it photospheric} indices defined in this
paper as a function of $B-V$ color and spectral type.  The G-band (solid 
circles) comes to a maximum in the early K-type stars, and then declines.
The \ion{Ca}{1} index (open circles) grows with increasing $B-V$  
linearly until the mid K-type stars, after which it appears to saturate.  The
H$\gamma$ index (open triangles) decreases linearly with increasing $B-V$.
The stars used for this diagram are the Mount Wilson calibration
stars of \citet{gray03}, and the $B-V$ data are from \citet{mermilliod97}.}
\label{fig:BV}
\end{figure}

At the suggestion of \citet{hall08}, an index has been designed to measure 
the G-band molecular feature in the
blue-violet region of the spectrum.  This wide, deep feature 
arises from the blended Q-branches of the 0-0 and
1-1 vibrational bands of the diatomic CH molecule.  The G-band appears 
first in the
early F-type stars, strengthens through the F- and G-type stars, comes to a
broad maximum in the early K-type stars on the main sequence, and then 
weakens toward later types \citep[see][]{gray09}.  The G-band index is 
measured by numerically  integrating
the stellar monochromatic flux in a 14\AA\ rectangular band centered at
4305\AA\ (corresponding closely to the visible extent of the G-band in
low-resolution spectra, and similar to the passband of G-band interference 
filters used in observations of the sun) and 
ratioing that with ``continuum'' fluxes measured in two bands on either side
of the G-band (see Table \ref{tbl:bands}).  The G-band index is defined as:
\begin{displaymath}
1 - \frac{\frac{1}{14\textrm{\AA}}\int_{4298\textrm{\AA}}^{4312\textrm{\AA}} I(\lambda)d\lambda}
{0.247c_3 + 0.753c_4}
\end{displaymath}
where $c_3$ and $c_4$ represent the monochromatic fluxes in the two
continuum bands, respectively.  Because the continuum bands are not situated
symmetrically relative to the G-band passband, the weightings in the
denominator are designed to give the ``continuum'' value at the wavelength
of the center of the G-band passband.  The ratio is subtracted from 
unity to give
an index that varies between 0 and 1: 0 when the G-band is absent, 
1 when the G-band is perfectly black.  As expected, the G-band index is a
strong function of $B-V$ (see Figure \ref{fig:BV}) and the spectral type.
The G-band index will also be a function of metallicity and $\log g$ 
\citep[see][]{gray09}.  We investigate in \S\ref{sec:var} the relationship of the 
G-band index to stellar activity.

\begin{figure}
\includegraphics[width=3.25in]{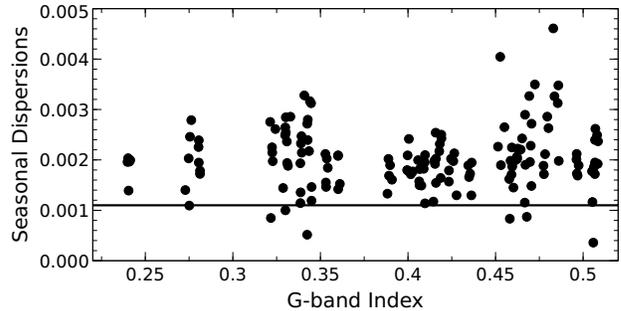}
\caption{Verification of the Monte Carlo error analysis for the G-band index.
This figure plots the seasonal dispersions ($\sigma$) of the G-band indices
for all of the program stars plus the chromospherically stable reference stars
against the G-band index.  The horizontal line, which corresponds well with
the lower envelope of the distribution of points, represents the Monte Carlo
error calculation for S/N $=180$, the average S/N of our spectra.}
\label{fig:G7}
\end{figure}

A Monte Carlo error analysis similar to that described for the $S_{\rm MW}$ index
was carried out for the G-band index.  This is illustrated in the lower panel of
Figure \ref{fig:SN}. The typical measurement error for the G-band index 
at S/N $= 100$ is $\pm 0.0017$ and at S/N $= 180$ is $\pm 0.0011$.  The
Monte Carlo analysis appears to have captured the important sources of measurement
error for the G-band, as may be deduced from Figure \ref{fig:G7}, where the 
standard deviations of the seasonal G-band data for all the program stars
and the ``chromospherically stable'' reference stars are plotted against the
G-band index.  The horizontal line in that figure, which corresponds well with
the lower envelope of the points, is the Monte Carlo G-band error for S/N $=180$.
The dispersions that lie above that line presumably arise from actual stellar
variability, a point that will be considered in \S\ref{sec:var} below.

\subsection{The \ion{Ca}{1} Index}
\label{sec:CaI}

Another prominent absorption feature in the blue-violet spectrum of G- and
K-type stars is the resonance line of \ion{Ca}{1} at 4226.7\AA. This absorption
line grows steadily in strength toward later types, at least up to mid K-type
stars.  It is also sensitive to surface gravity, especially in the K-type
stars \citep[see][]{gray09}.  We have devised an index similar to that of
the previously defined G-band index.  The \ion{Ca}{1} index is measured by
integrating over a 2\AA-wide band centered on the \ion{Ca}{1} line and ratioing
that with fluxes in two symmetrically placed continuum bands. The formula
used is:
\begin{displaymath}
1 - \frac{\frac{1}{2\textrm{\AA}}\int_{4225.7\textrm{\AA}}^{4227.7\textrm{\AA}} I(\lambda)d\lambda}
{0.5(c_1 + c_2)}
\end{displaymath}
where $c_1$ and $c_2$ are the continuum bands defined in Table \ref{tbl:bands}.
As can be seen in Figure \ref{fig:BV}, the \ion{Ca}{1} index behaves as 
designed; it grows linearly from the F-type stars into the K-type stars, 
only saturating after a spectral type of K3. 

A Monte Carlo error analysis similar to that illustrated in Figure 
\ref{fig:SN} was carried out for the \ion{Ca}{1} index, giving a
measurement error of $\pm 0.0027$ at S/N $= 180$.  This value again corresponds
well with the lower envelope of \ion{Ca}{1} index seasonal dispersions (see 
discussion in \S\ref {sec:Gband} above).

\subsection{The H$\gamma$ Index}
\label{sec:HG}

Both the G-band index and the \ion{Ca}{1} index grow with decreasing 
temperature (at least up to the early K-type stars), and so it is useful to 
define
another index that decreases with the temperature.  The hydrogen lines behave
in exactly this way in the F-, G-, and K-type stars.  The best hydrogen line
to use in the spectral range provided by our spectra from the Dark Sky
Observatory is H$\gamma$.  An index based on the H$\beta$ line would probably
be preferable, because of the less crowded surroundings, but that line is
outside our spectral range.  The H$\gamma$ index is defined similarly to the
\ion{Ca}{1} index, with a 2\AA-wide band centered on the H$\gamma$ line and
flanking ``continuum'' bands (specified in Table \ref{tbl:bands}).  The formula
used is

\begin{displaymath}
1 - \frac{\frac{1}{2\textrm{\AA}}\int_{4339.5\textrm{\AA}}^{4341.5\textrm{\AA}} I(\lambda)d\lambda}
{0.4286c_5 + 0.5714c_6}
\end{displaymath}

The H$\gamma$ index behaves as designed, declining in strength with declining
temperature (Figure \ref{fig:BV}).  However, it appears to have only about 
half the temperature sensitivity of the \ion{Ca}{1} index.

A Monte Carlo error analysis similar to that illustrated in Figure 
\ref{fig:SN} was carried out for the H$\gamma$ index, giving a
measurement error of $\pm 0.0020$ at S/N $= 180$.  The larger errors for the
\ion{Ca}{1} and H$\gamma$ indices relative to the G-band index arise 
primarily from the narrower ``science'' bands.

\begin{figure*}
\begin{tabular}{ccc}
\centering
\includegraphics[width=2.0in]{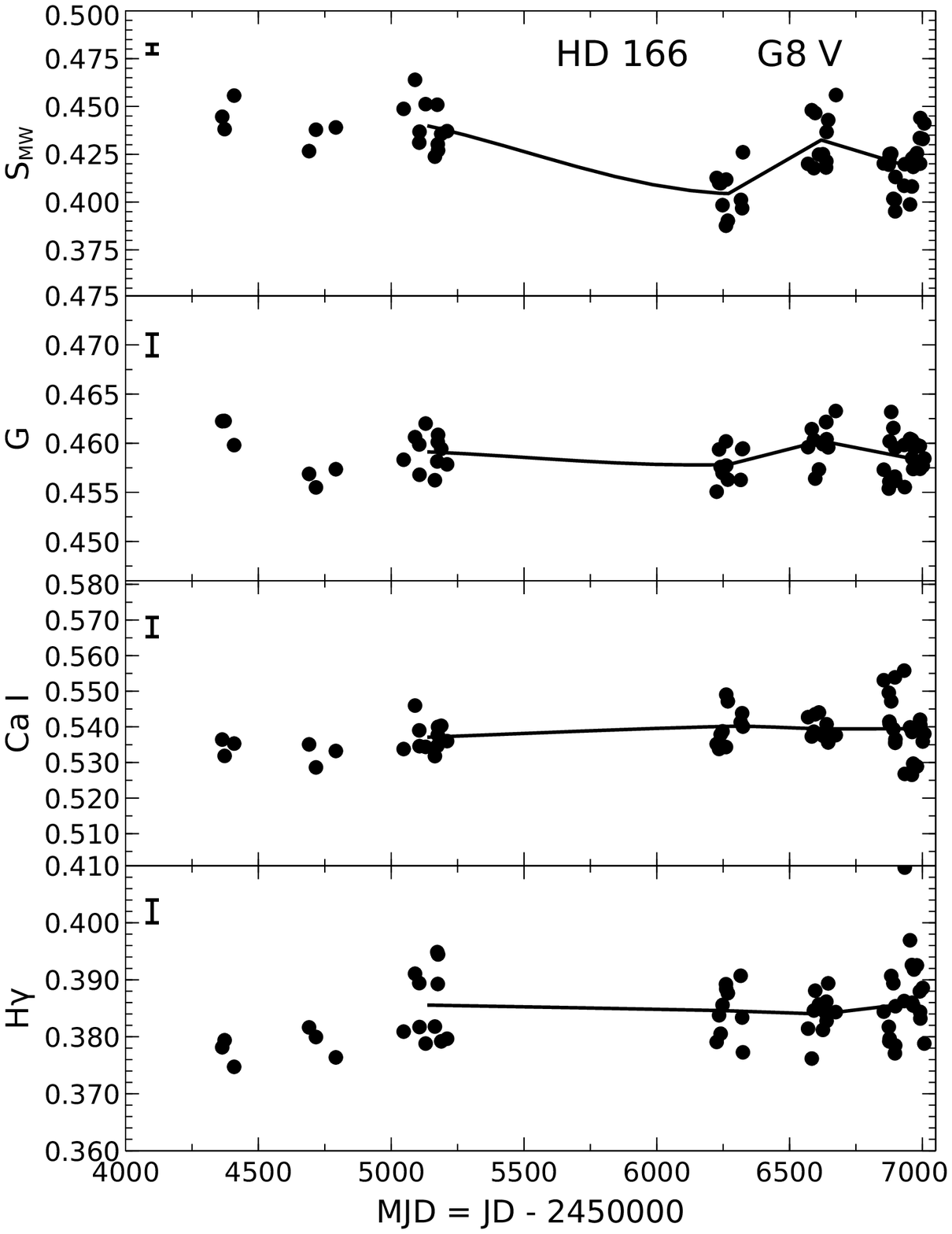} & \includegraphics[width=2.0in]{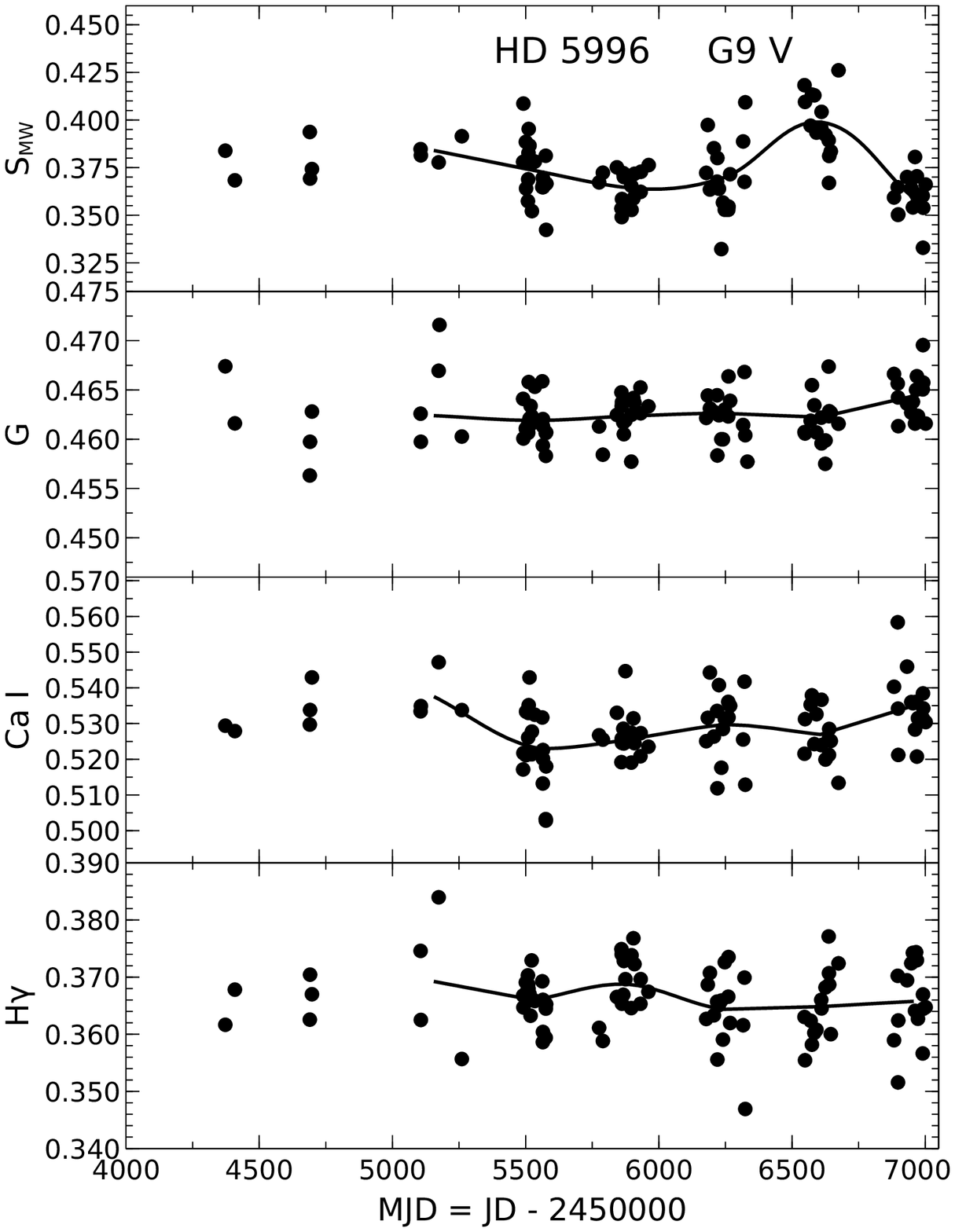} & \includegraphics[width=2.0in]{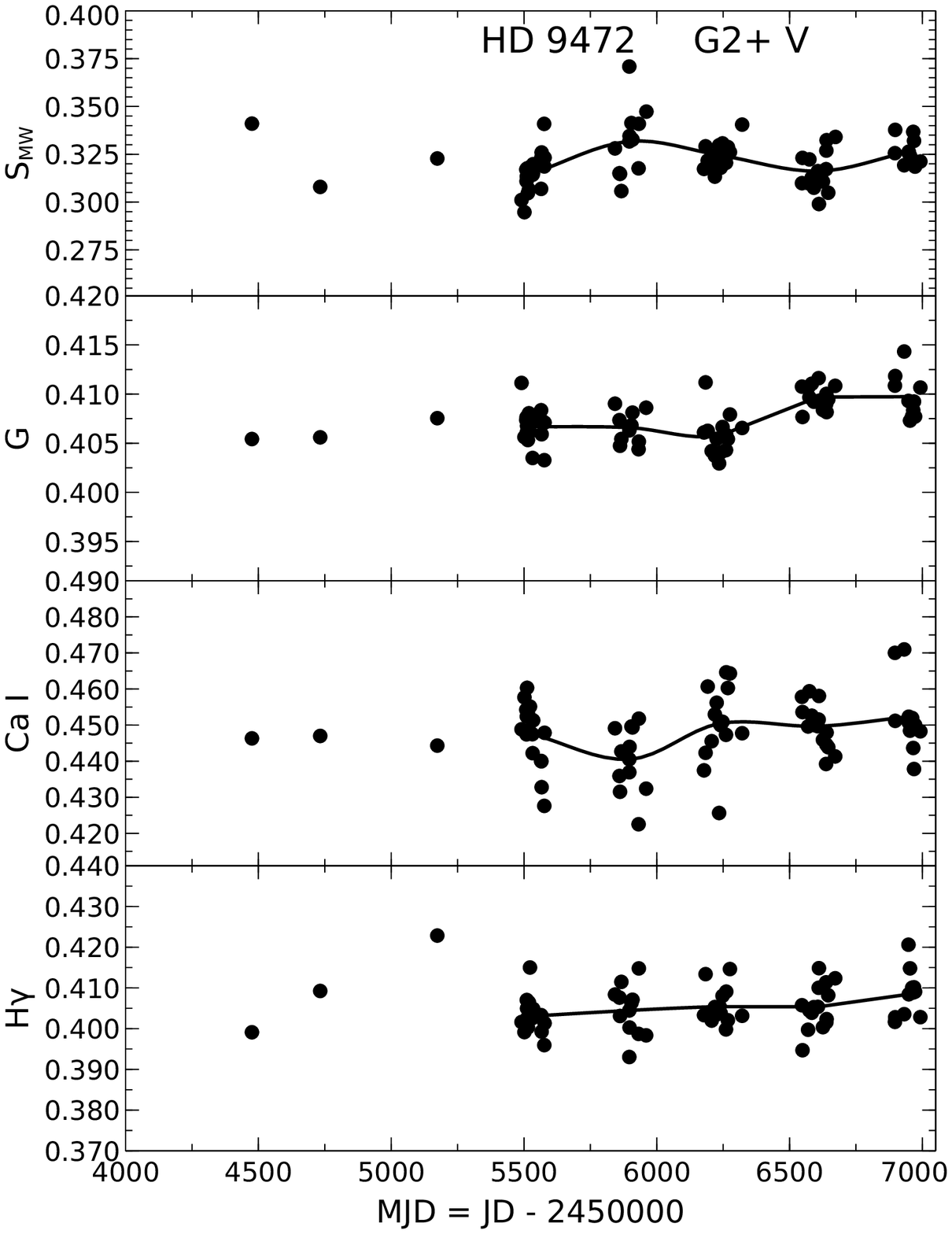} \\
\includegraphics[width=2.0in]{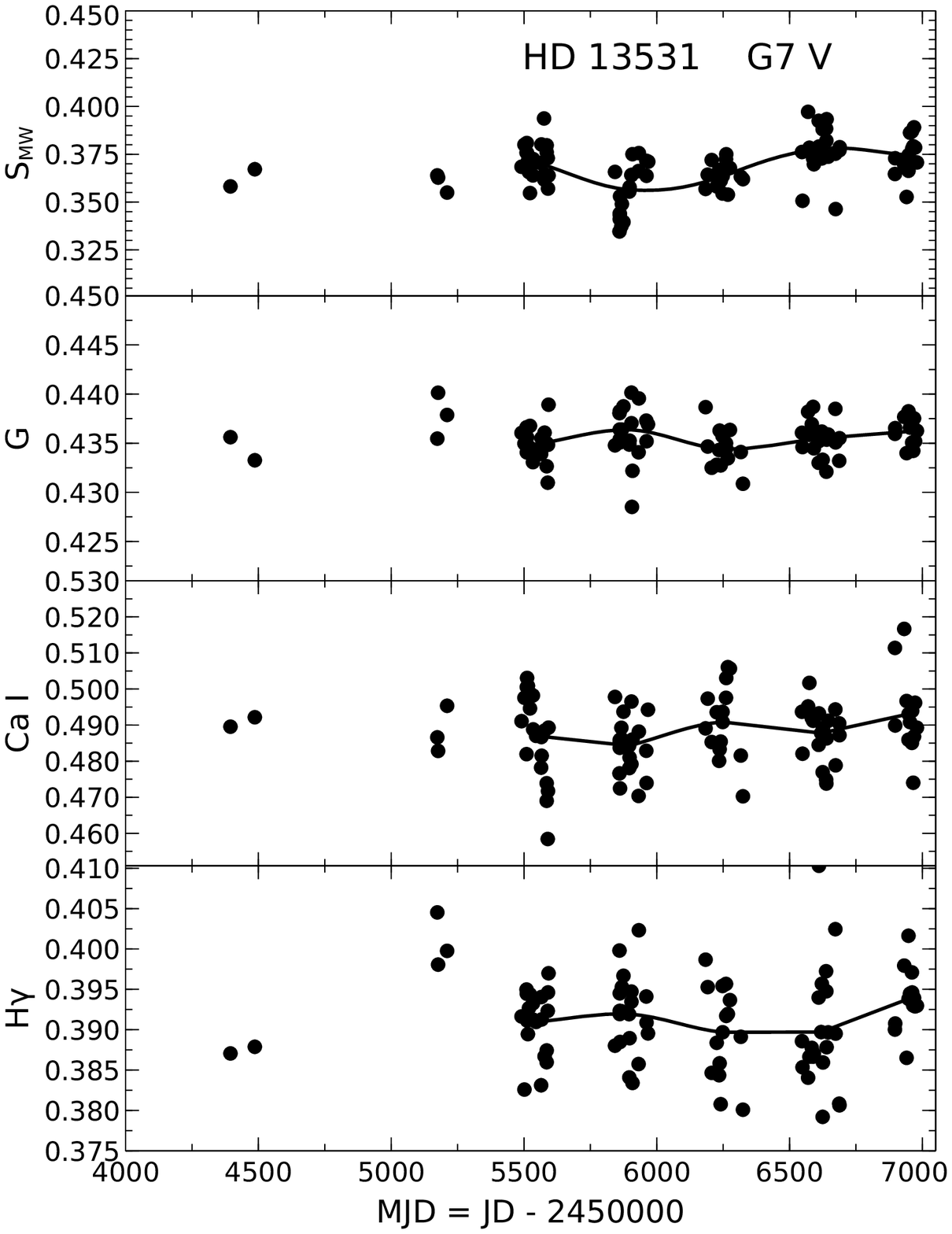} & \includegraphics[width=2.0in]{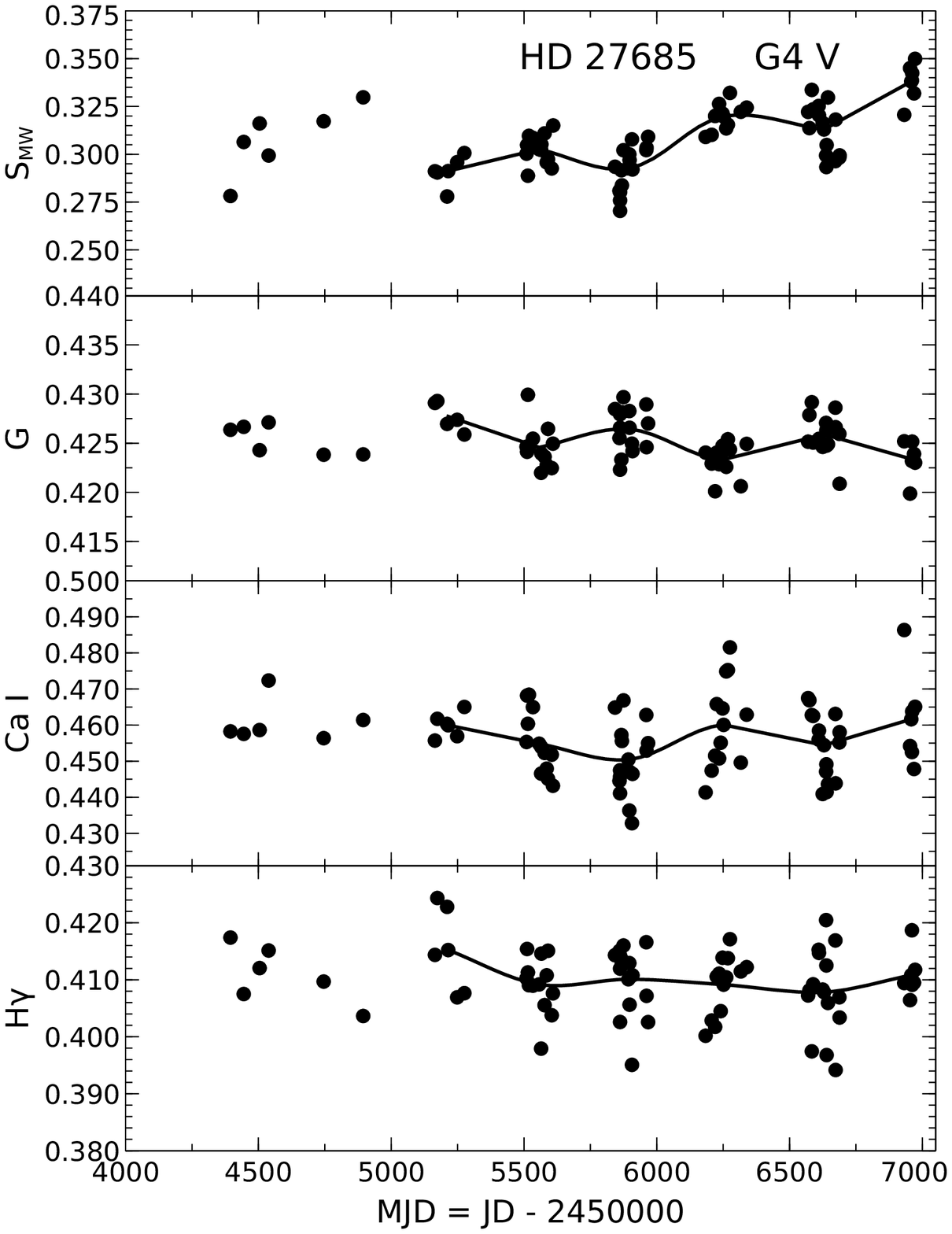} & \includegraphics[width=2.0in]{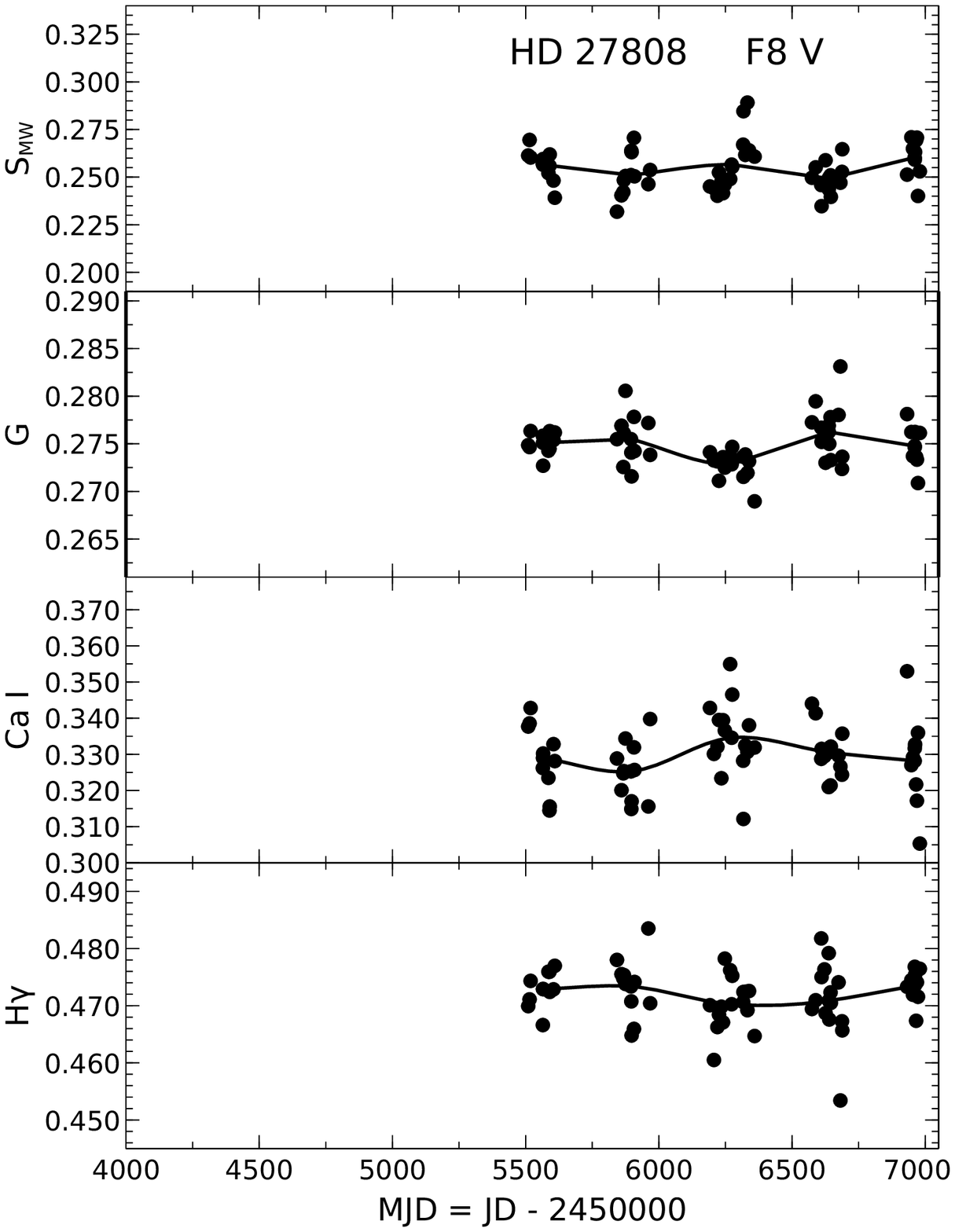} \\
\includegraphics[width=2.0in]{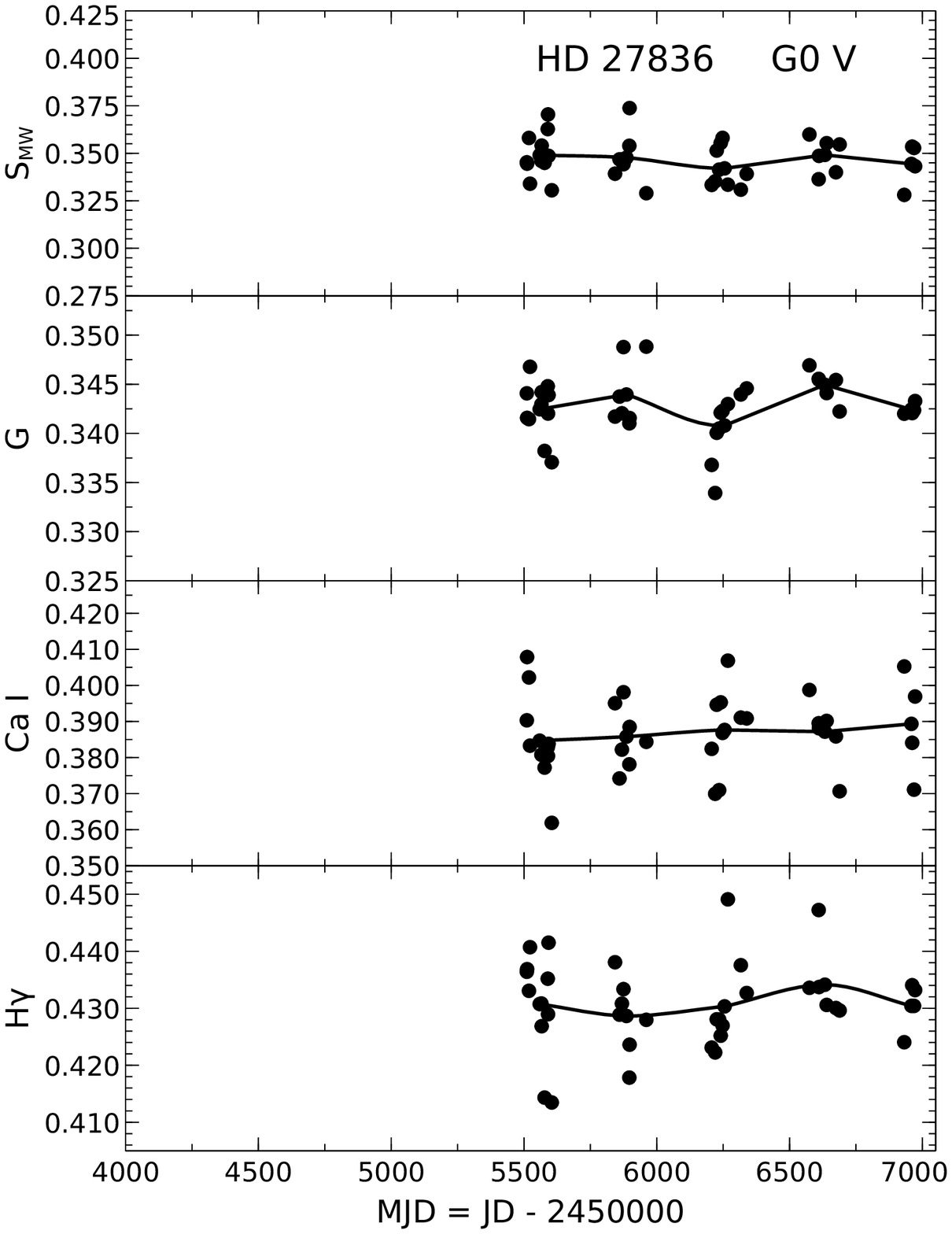} & \includegraphics[width=2.0in]{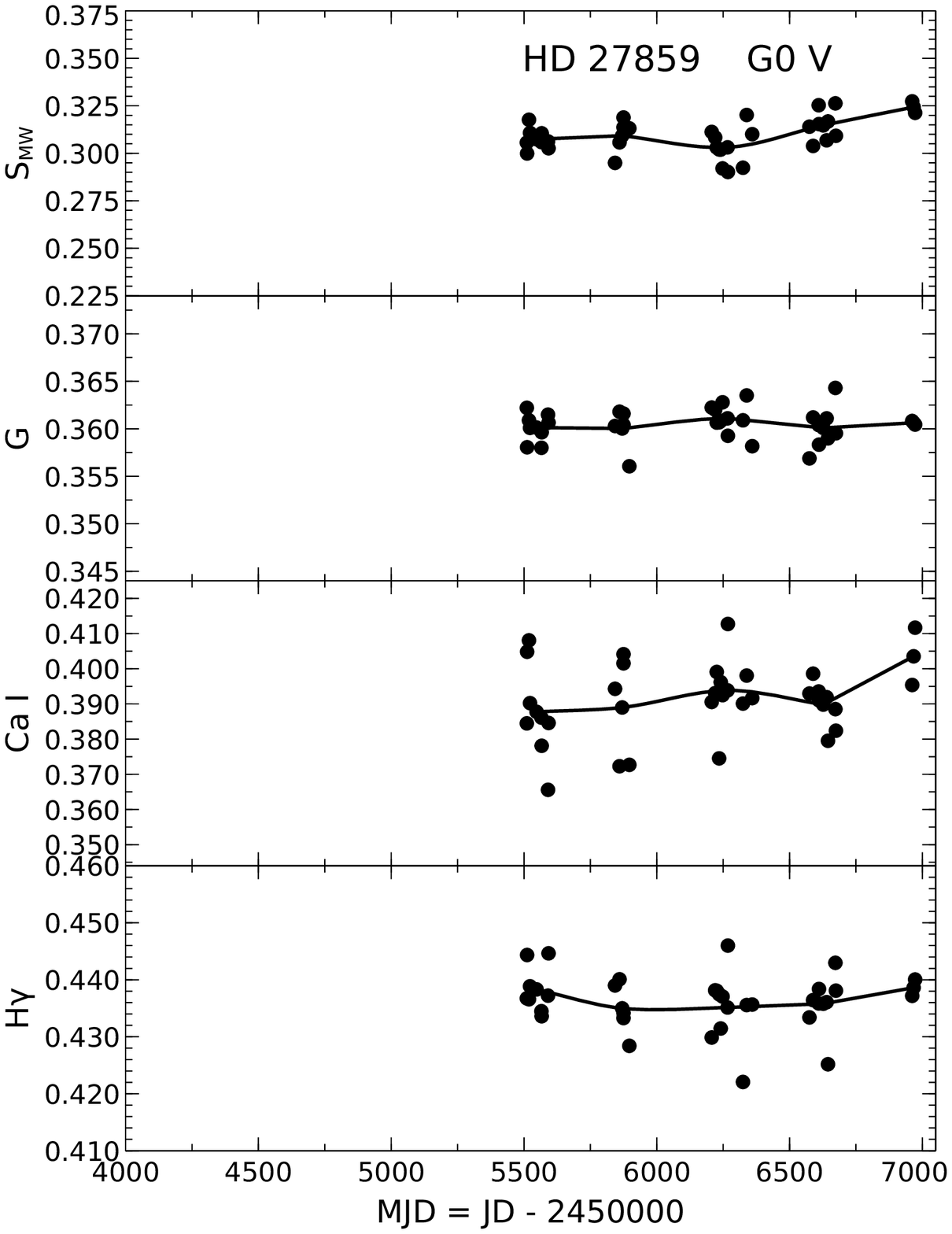} & \includegraphics[width=2.0in]{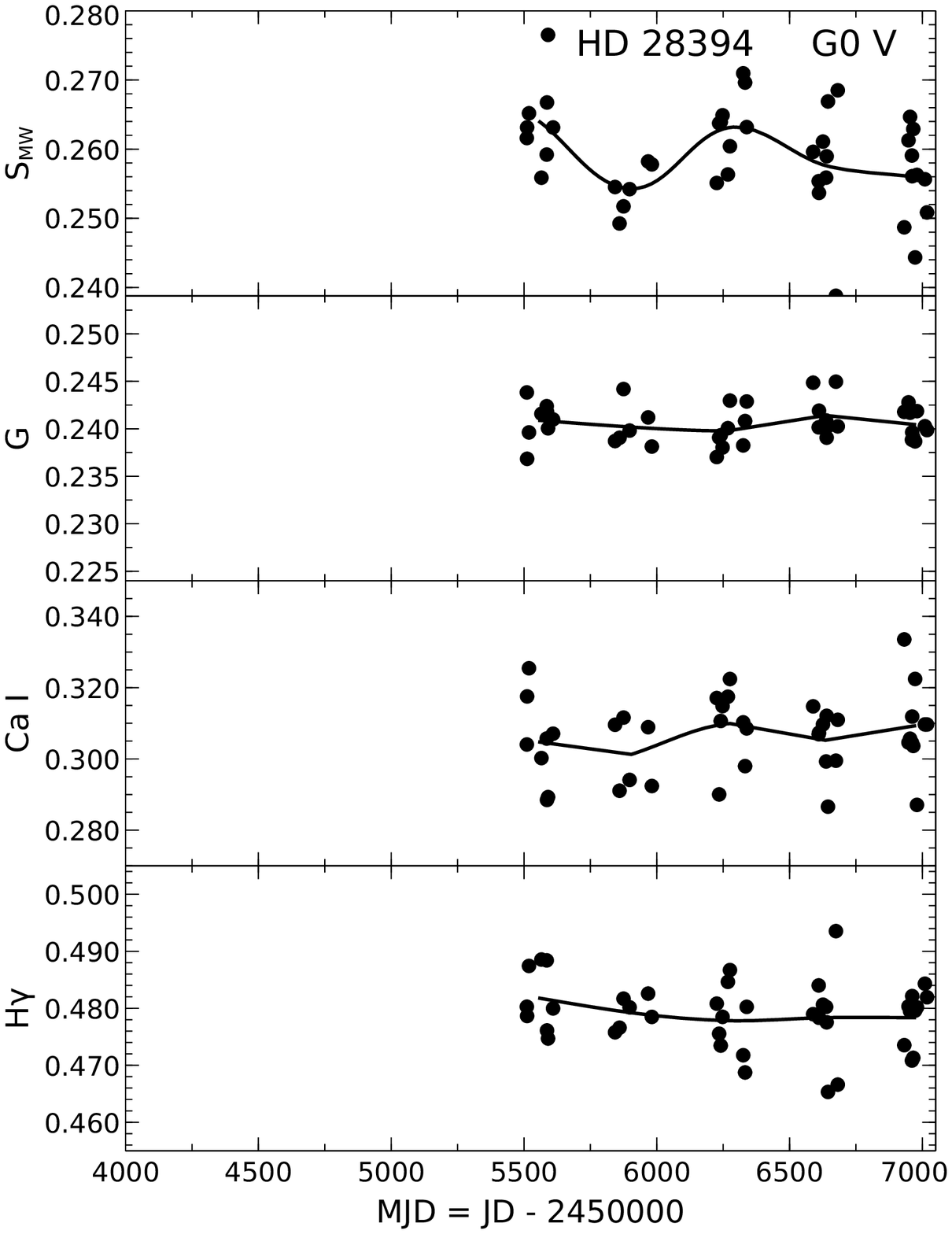} \\
\end{tabular}
\caption{A montage of \ion{Ca}{2} H \& K activity index ($S_{\rm MW}$) time series (upper 
panel), and G-band index, \ion{Ca}{1}, and H$\gamma$ times series (lower panels) 
for our program stars (montage continued in Figures \ref{fig:KHG2}, \ref{fig:KHG3}, and \ref{fig:KHG4}).  All the graphs are 
scaled identically, with a range of 0.15 in $S_{\rm MW}$, 0.03 in the G-band 
index, 0.08 in the \ion{Ca}{1} index, and 0.05 in the H$\gamma$ index so 
that amplitudes of variations and seasonal dispersions can be intercompared 
directly.  The solid lines are Bezier curves drawn through the seasonal means.  Typical error bars
for S/N $= 180$ spectra are shown in the upper left-hand corner of the panels for the first star. }
\label{fig:KHG1}
\end{figure*}

\begin{figure*}
\begin{tabular}{ccc}
\centering
\includegraphics[width=2.0in]{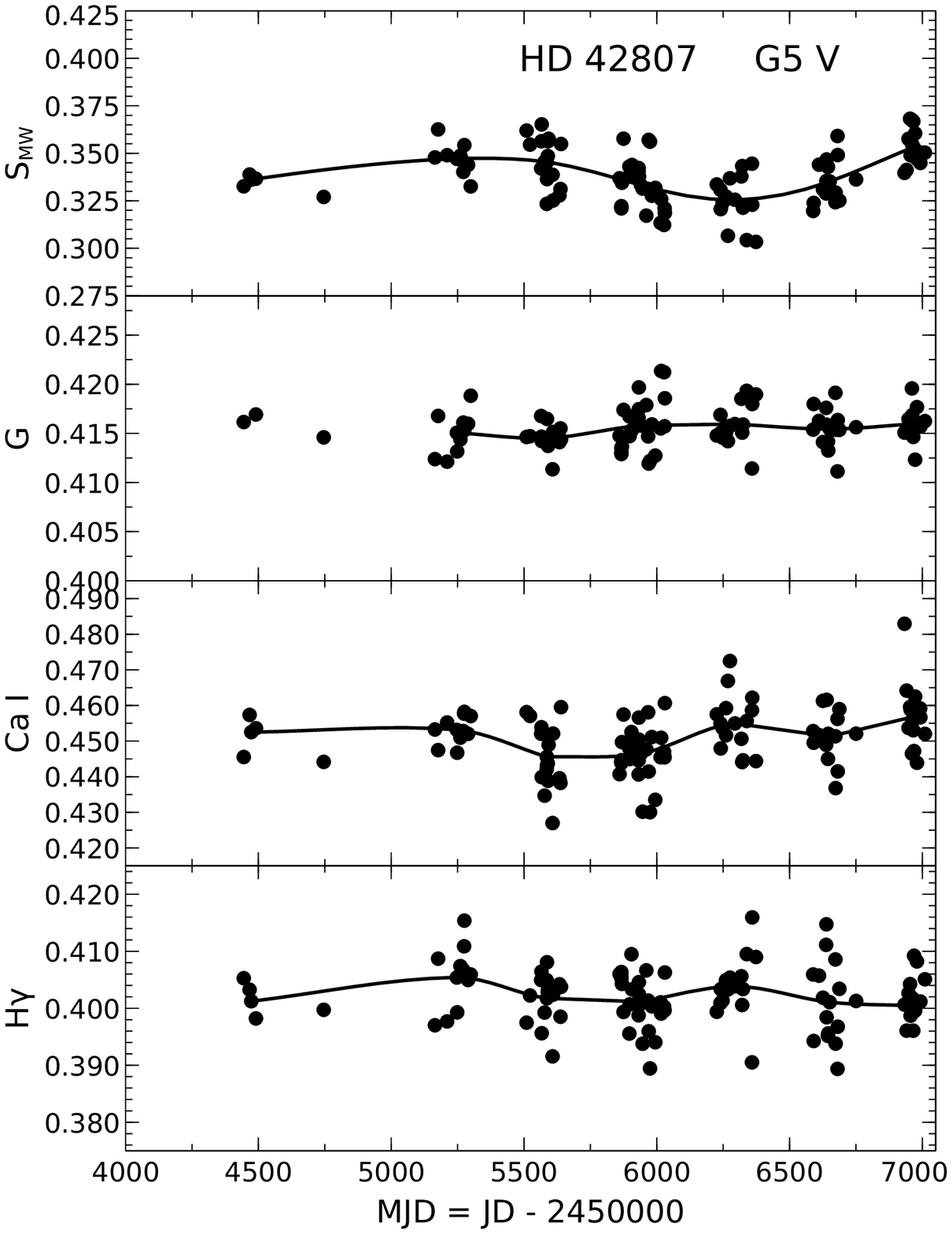} & \includegraphics[width=2.0in]{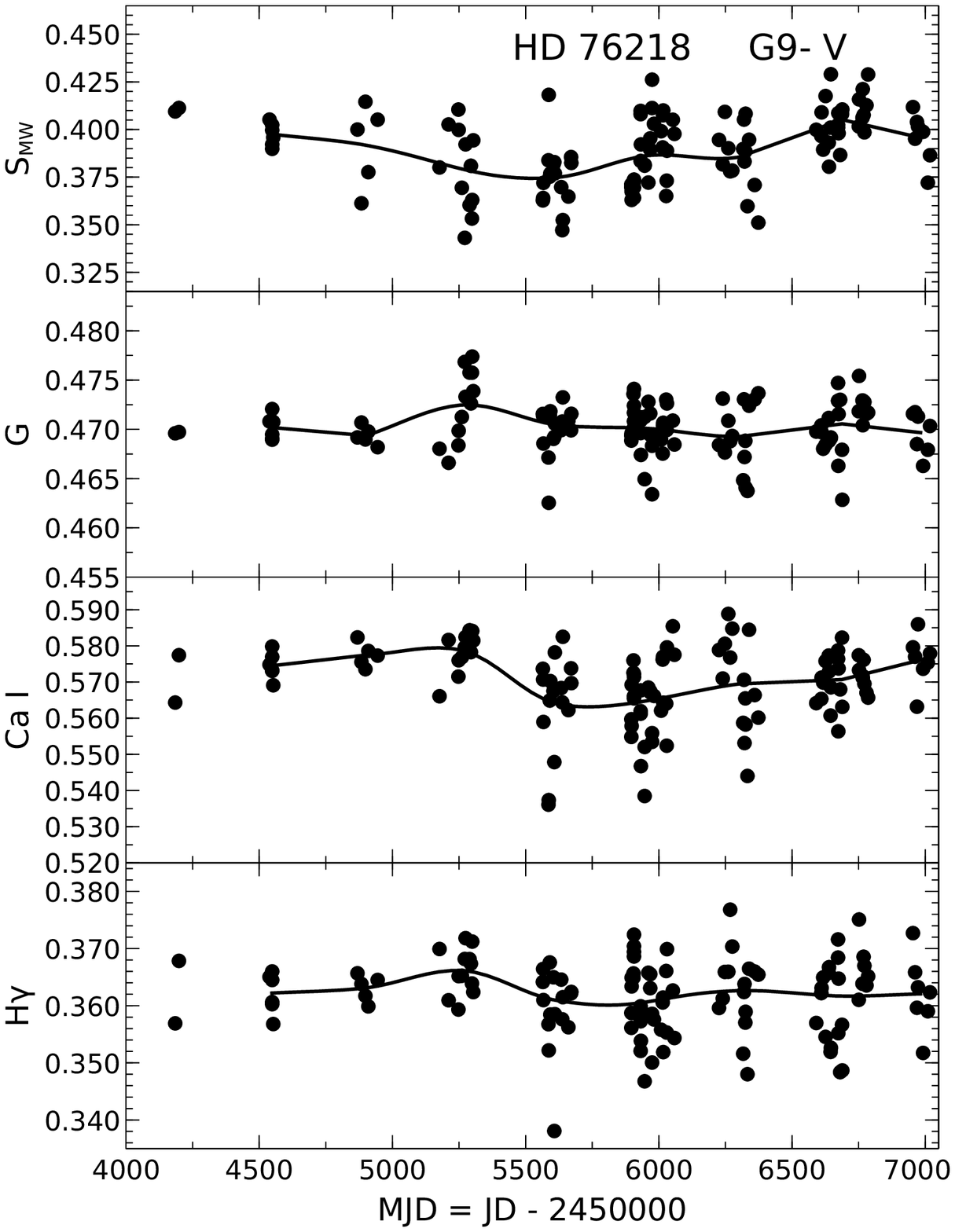} & \includegraphics[width=2.0in]{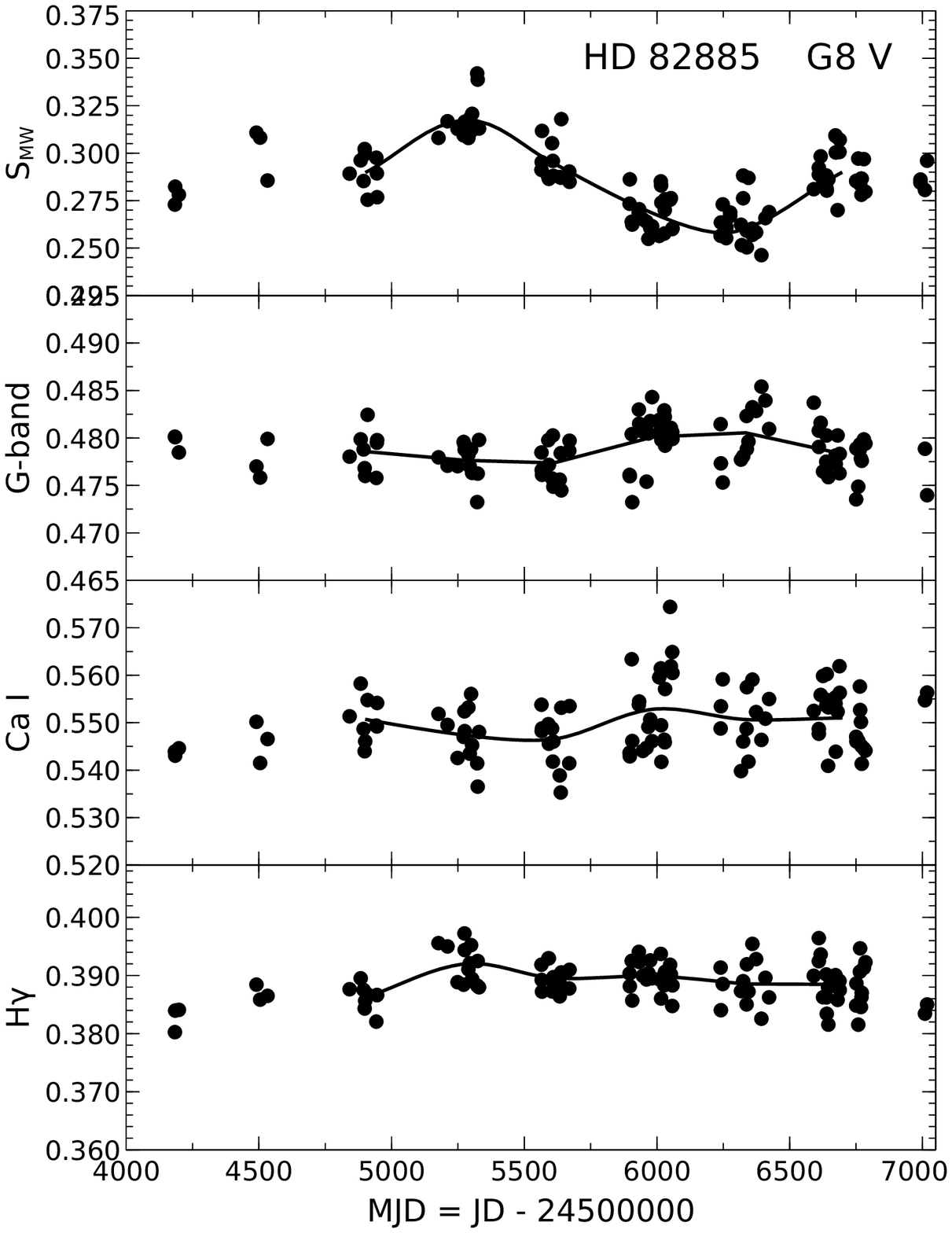} \\
\includegraphics[width=2.0in]{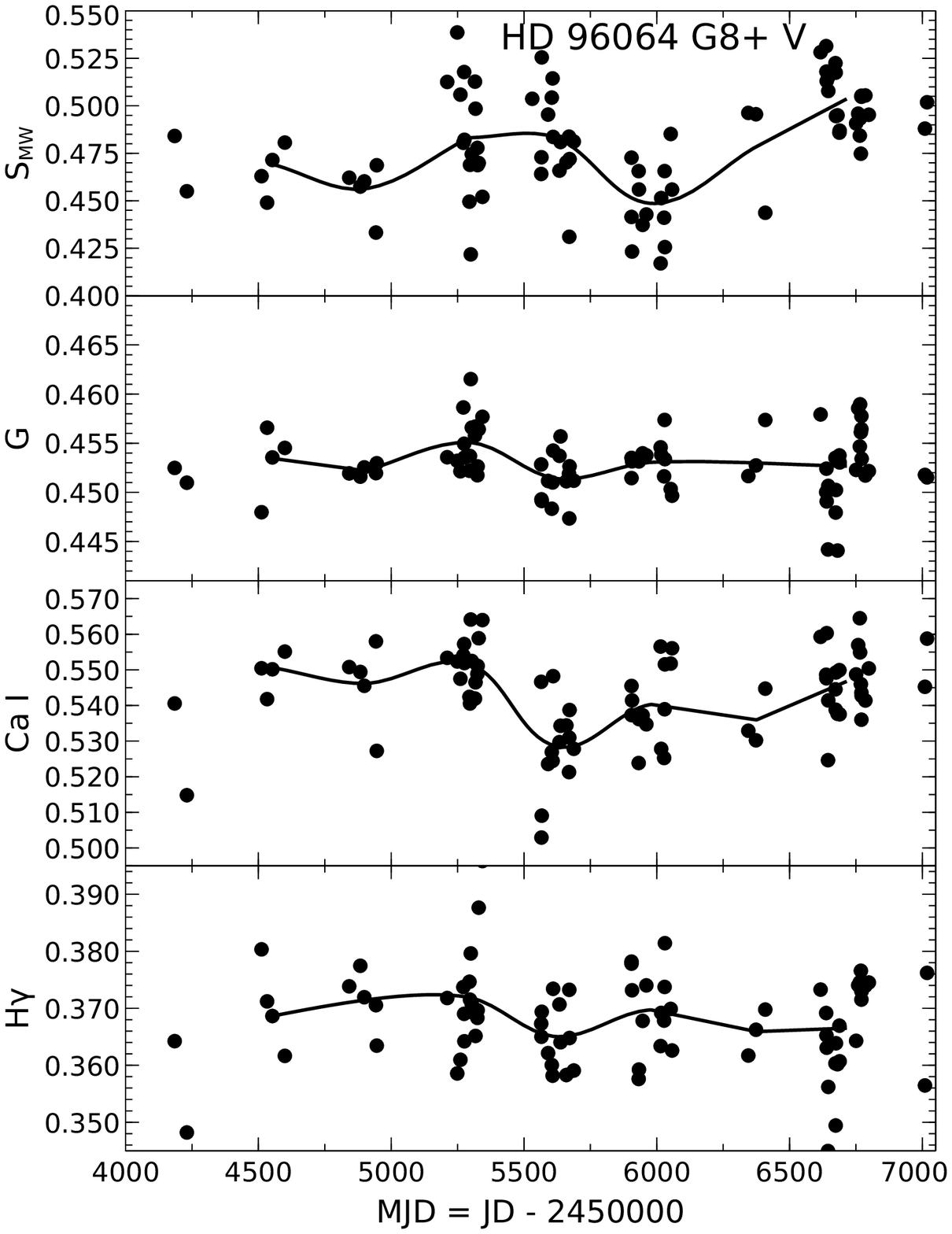} & \includegraphics[width=2.0in]{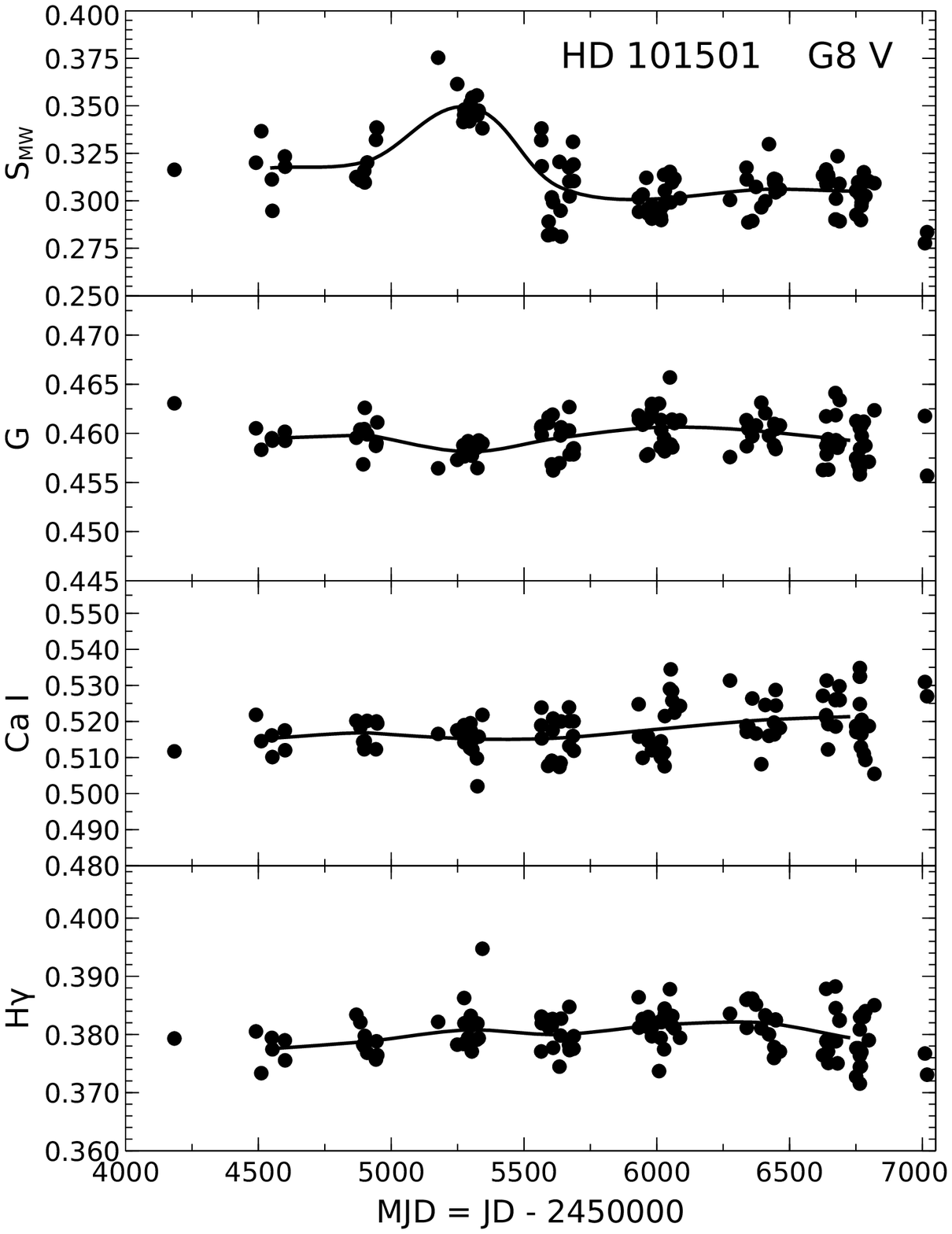} & \includegraphics[width=2.0in]{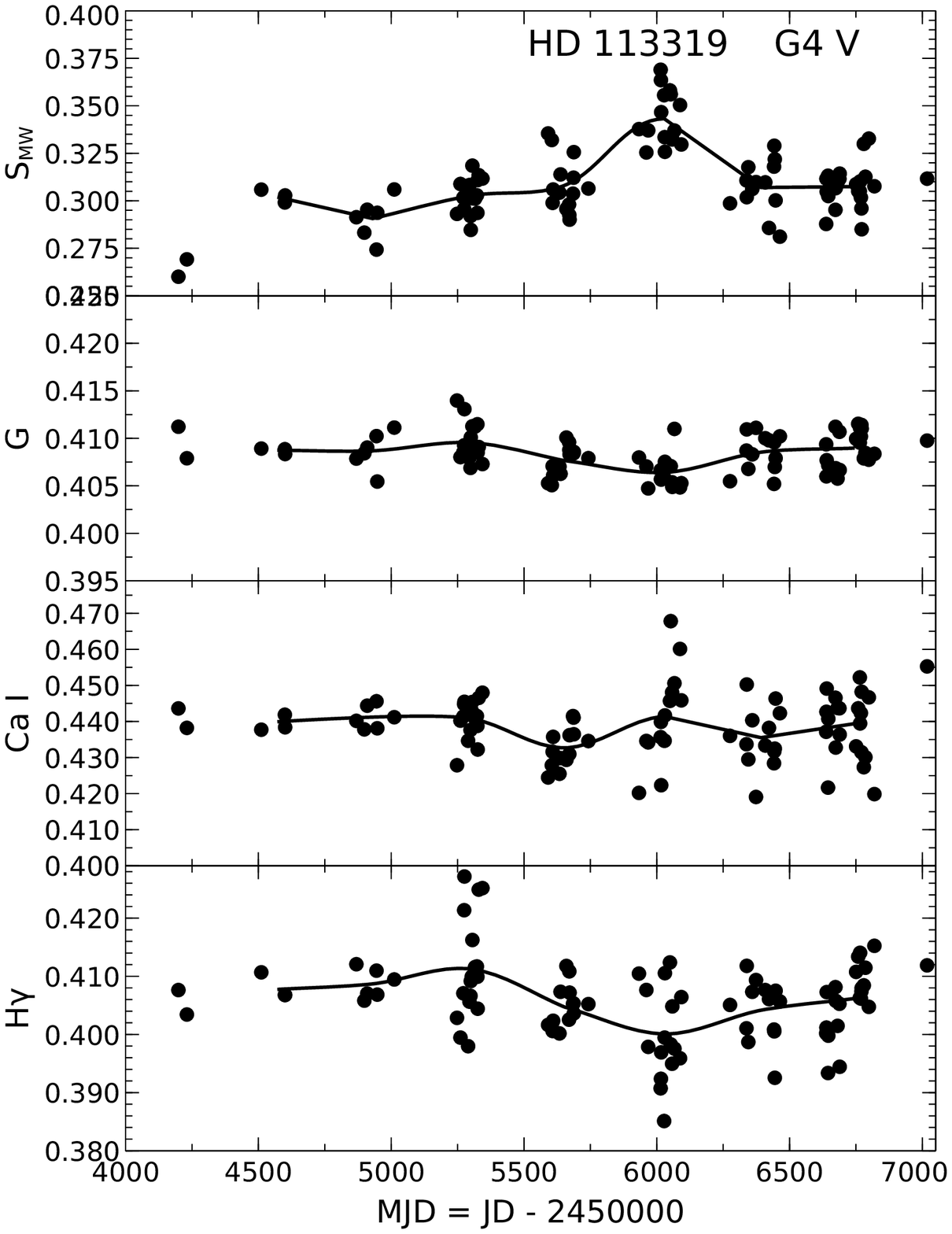} \\
\includegraphics[width=2.0in]{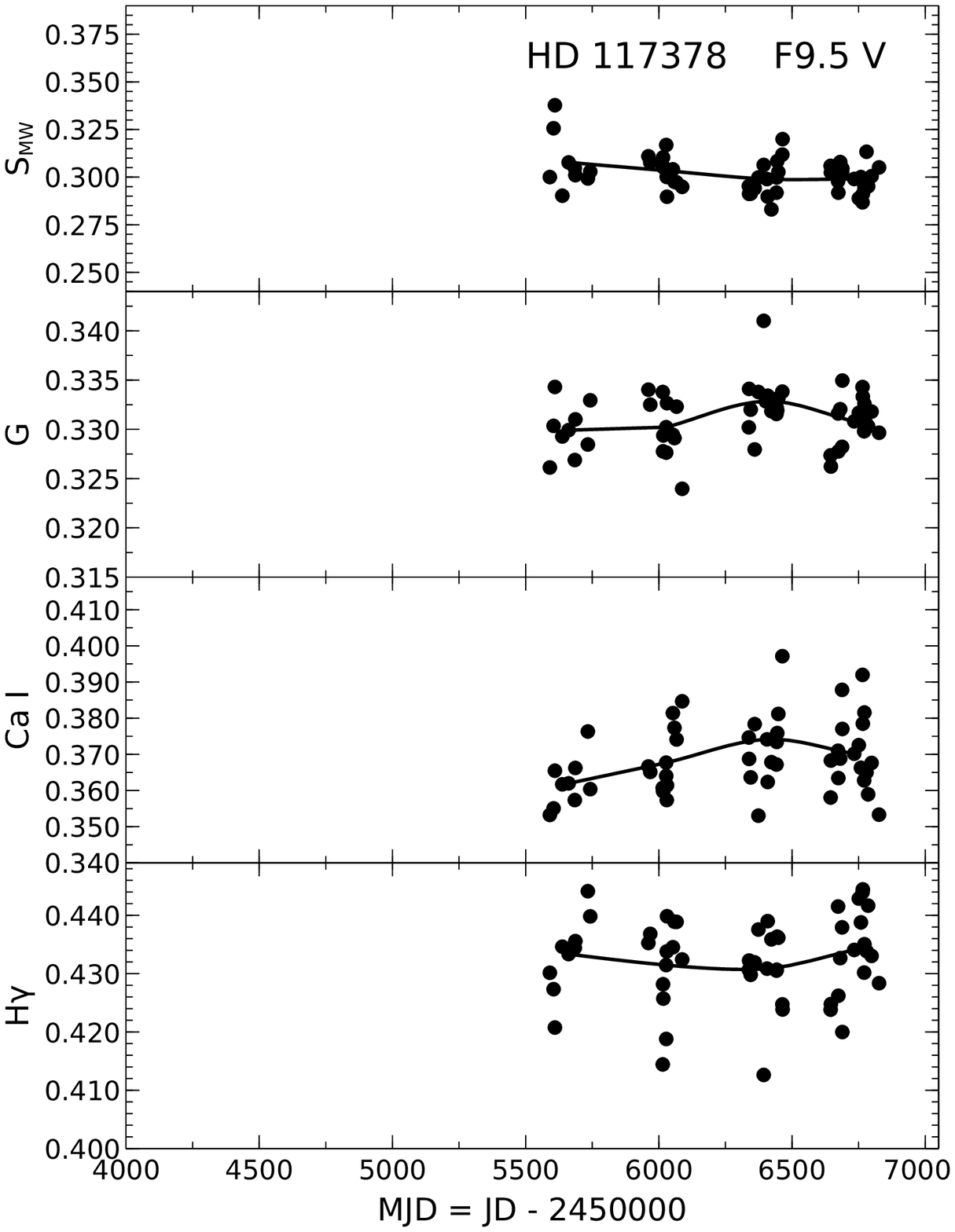} & \includegraphics[width=2.0in]{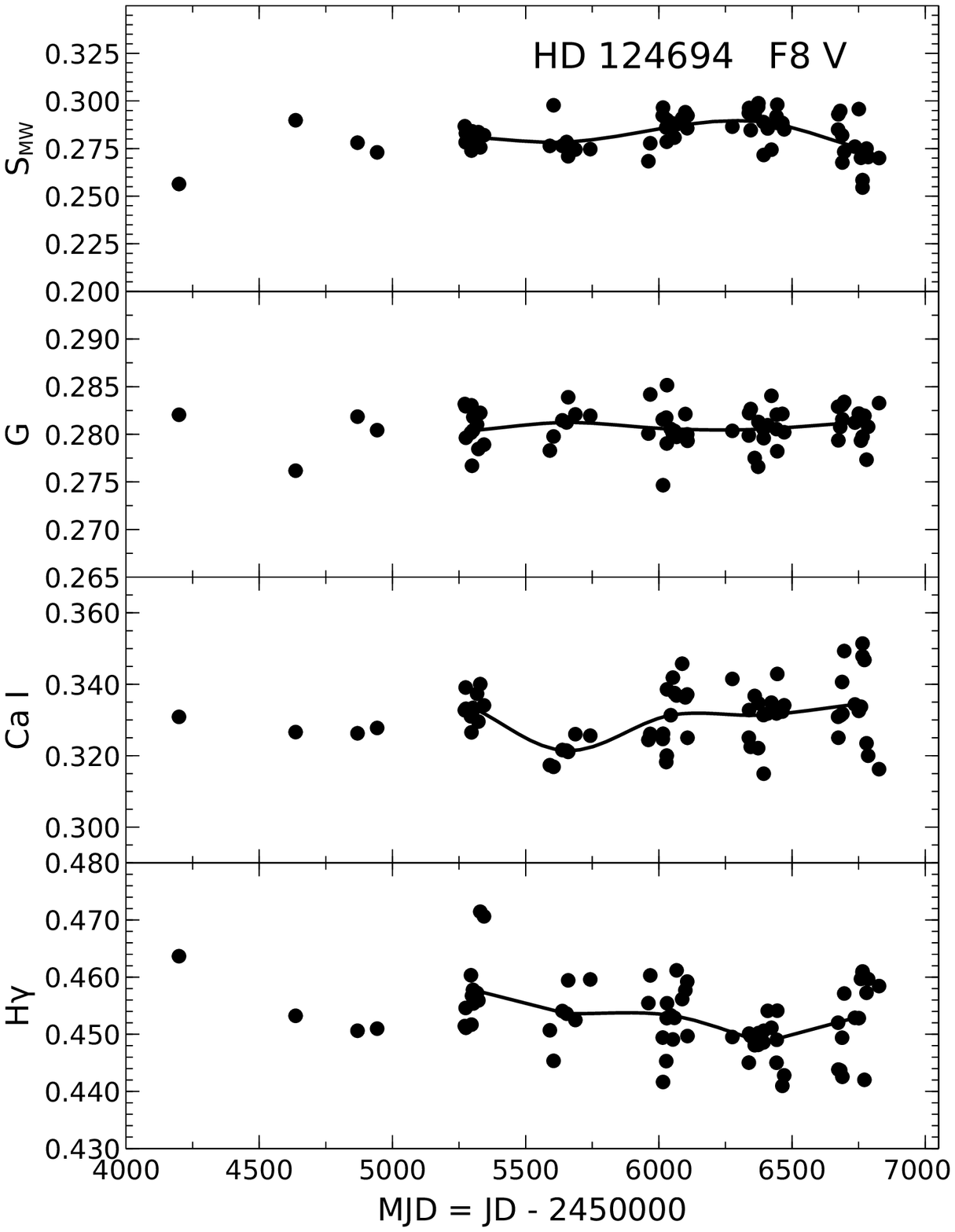} & \includegraphics[width=2.0in]{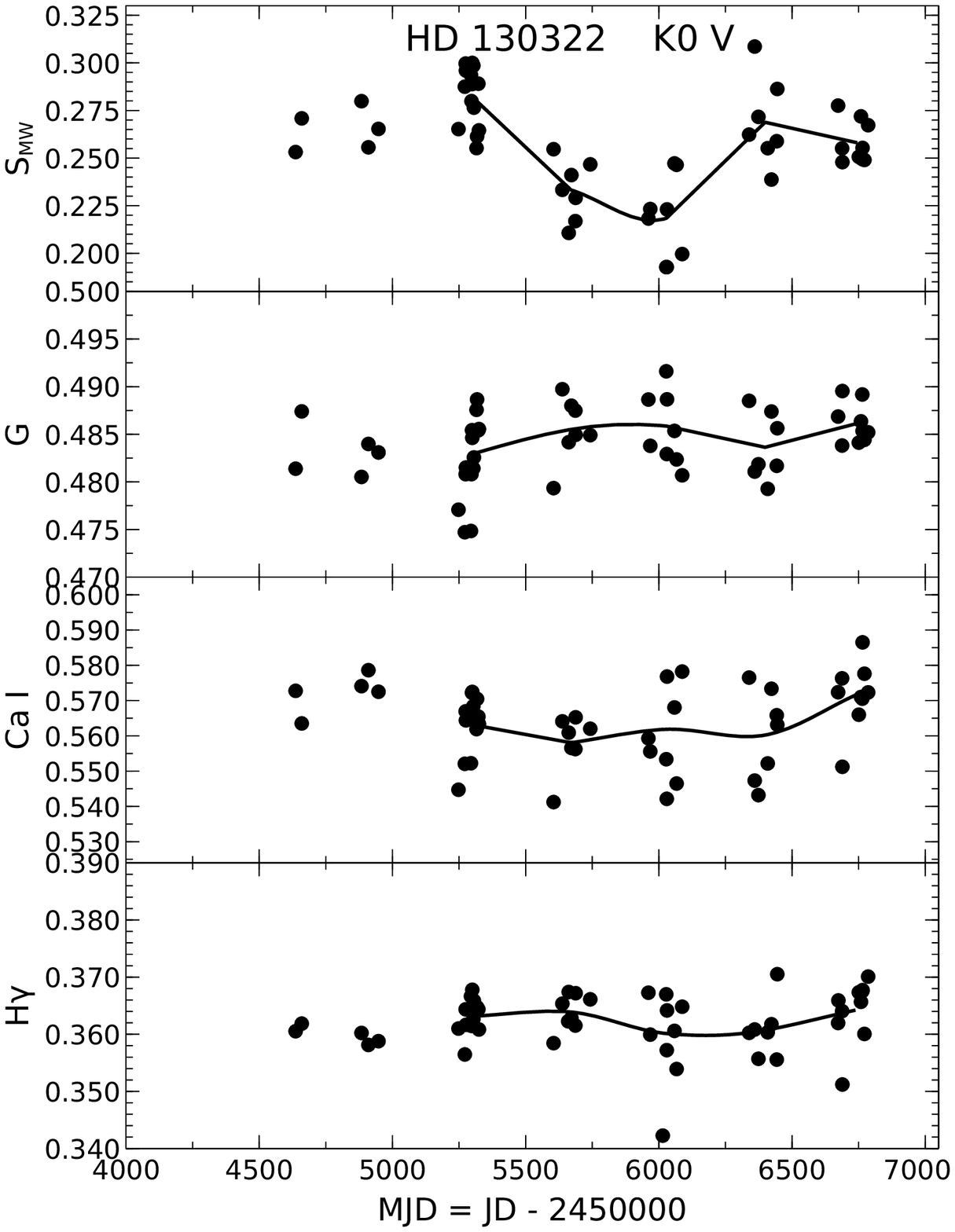} \\
\end{tabular}
\caption{Continuation of the montage in Figure \ref{fig:KHG1}.}
\label{fig:KHG2}
\end{figure*}

\begin{figure*}
\begin{tabular}{ccc}
\centering
\includegraphics[width=2.0in]{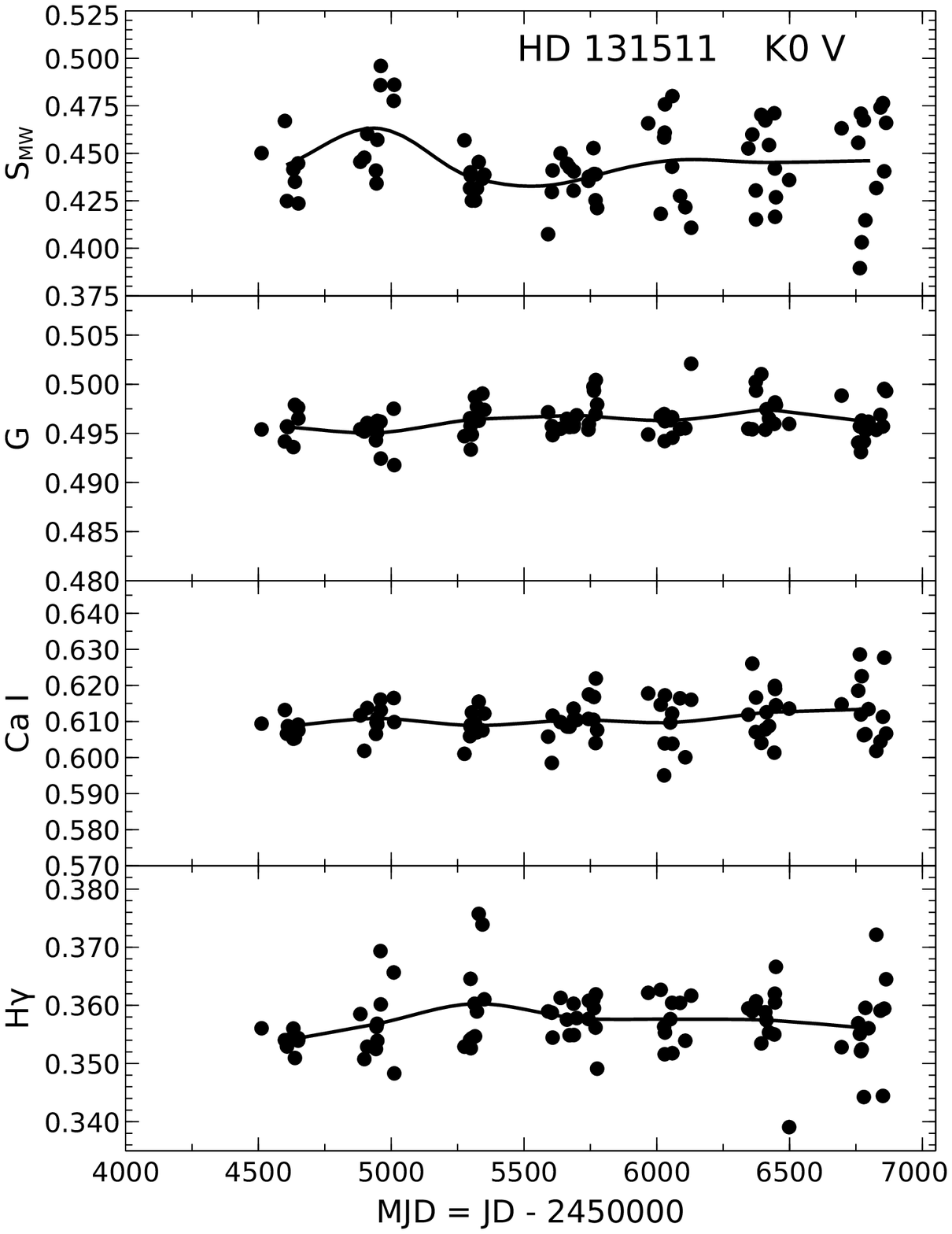} & \includegraphics[width=2.0in]{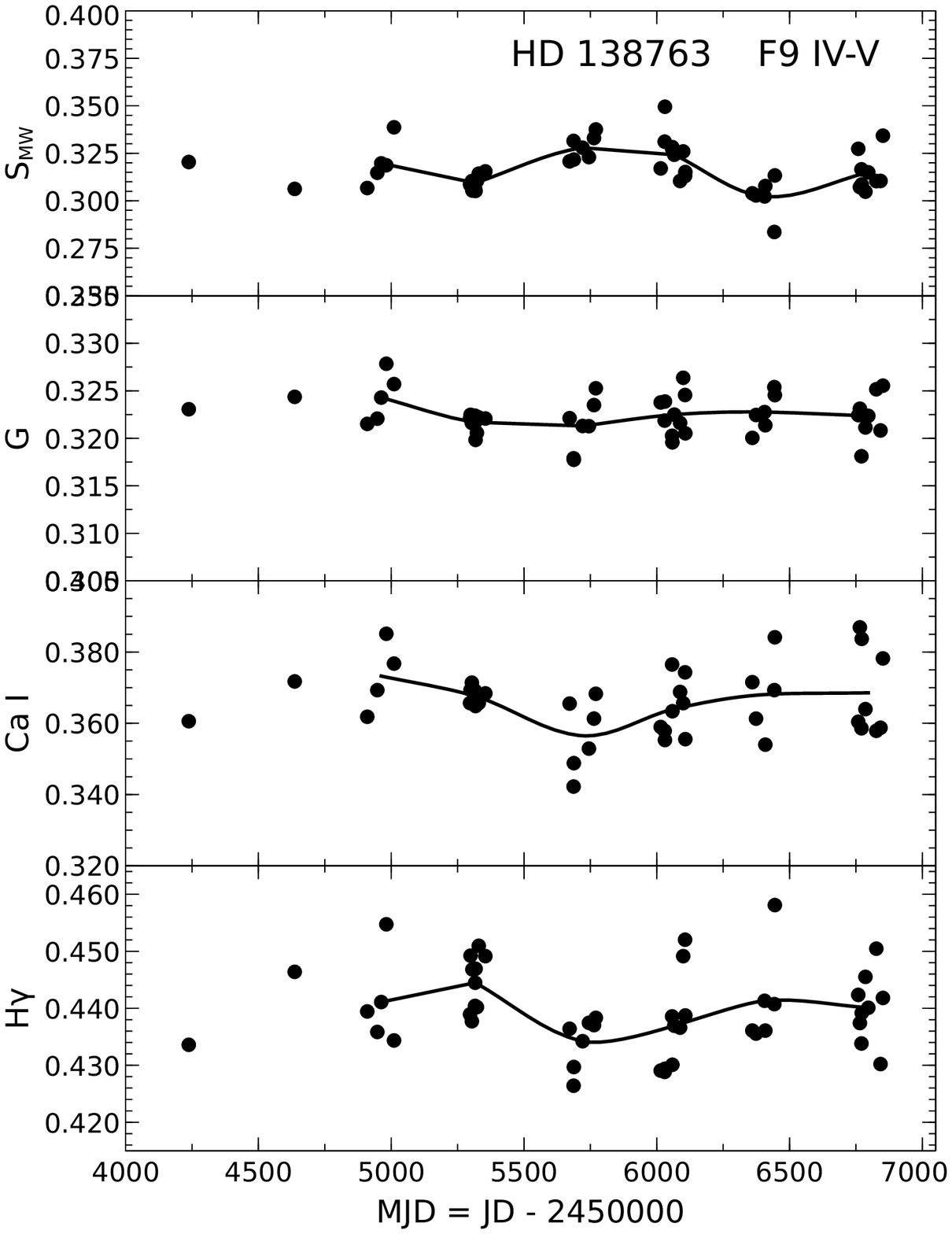} & \includegraphics[width=2.0in]{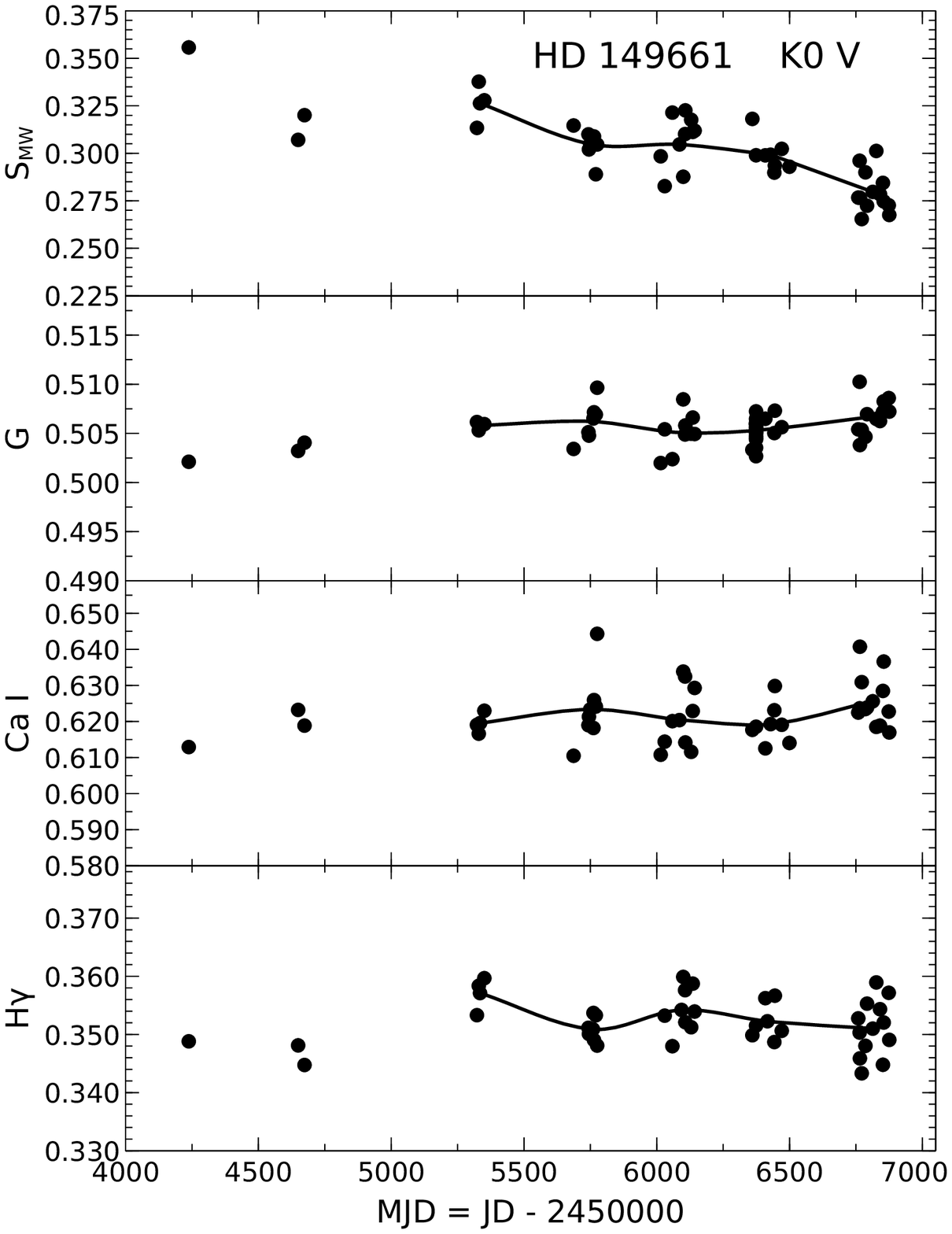} \\
\includegraphics[width=2.0in]{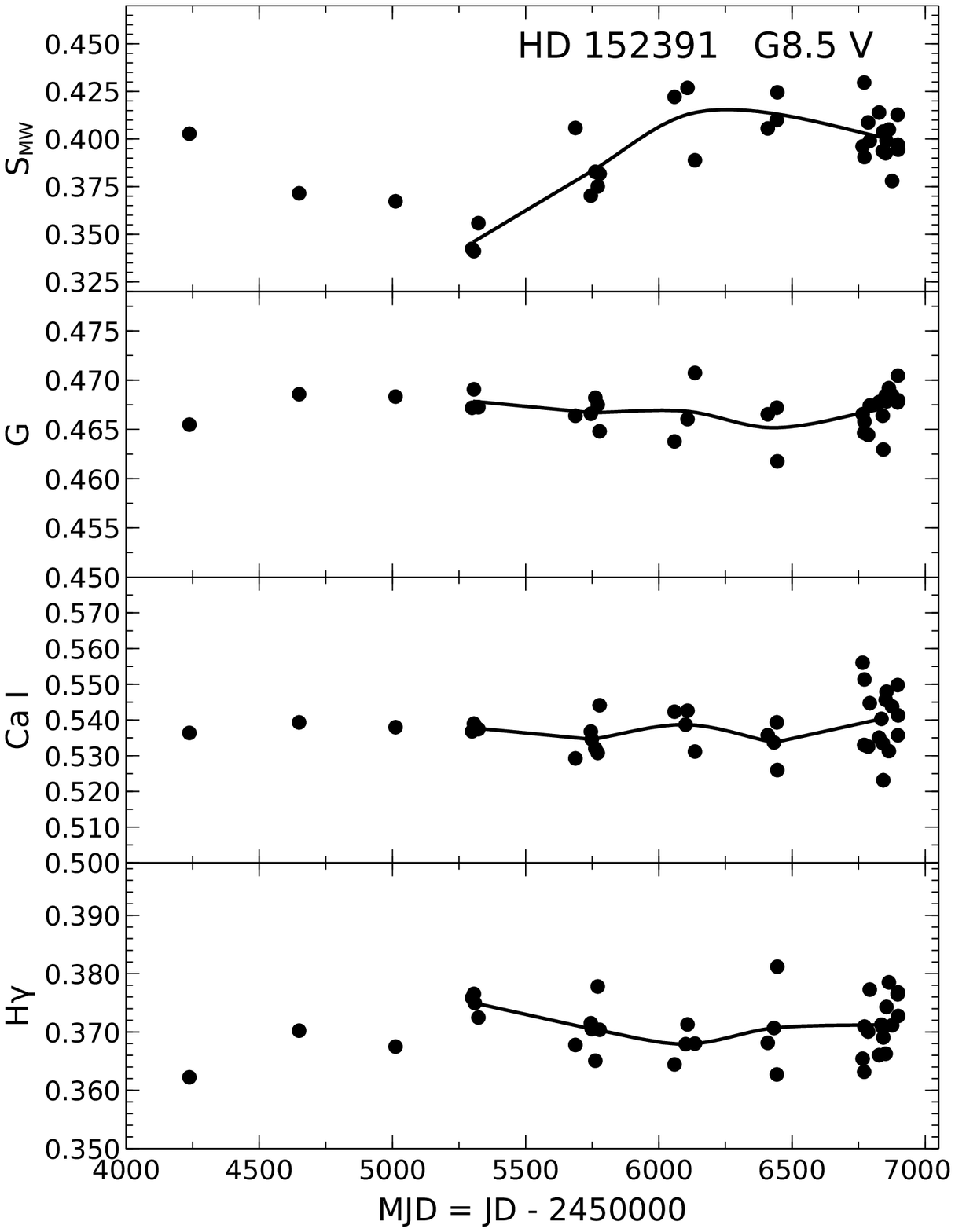} & \includegraphics[width=2.0in]{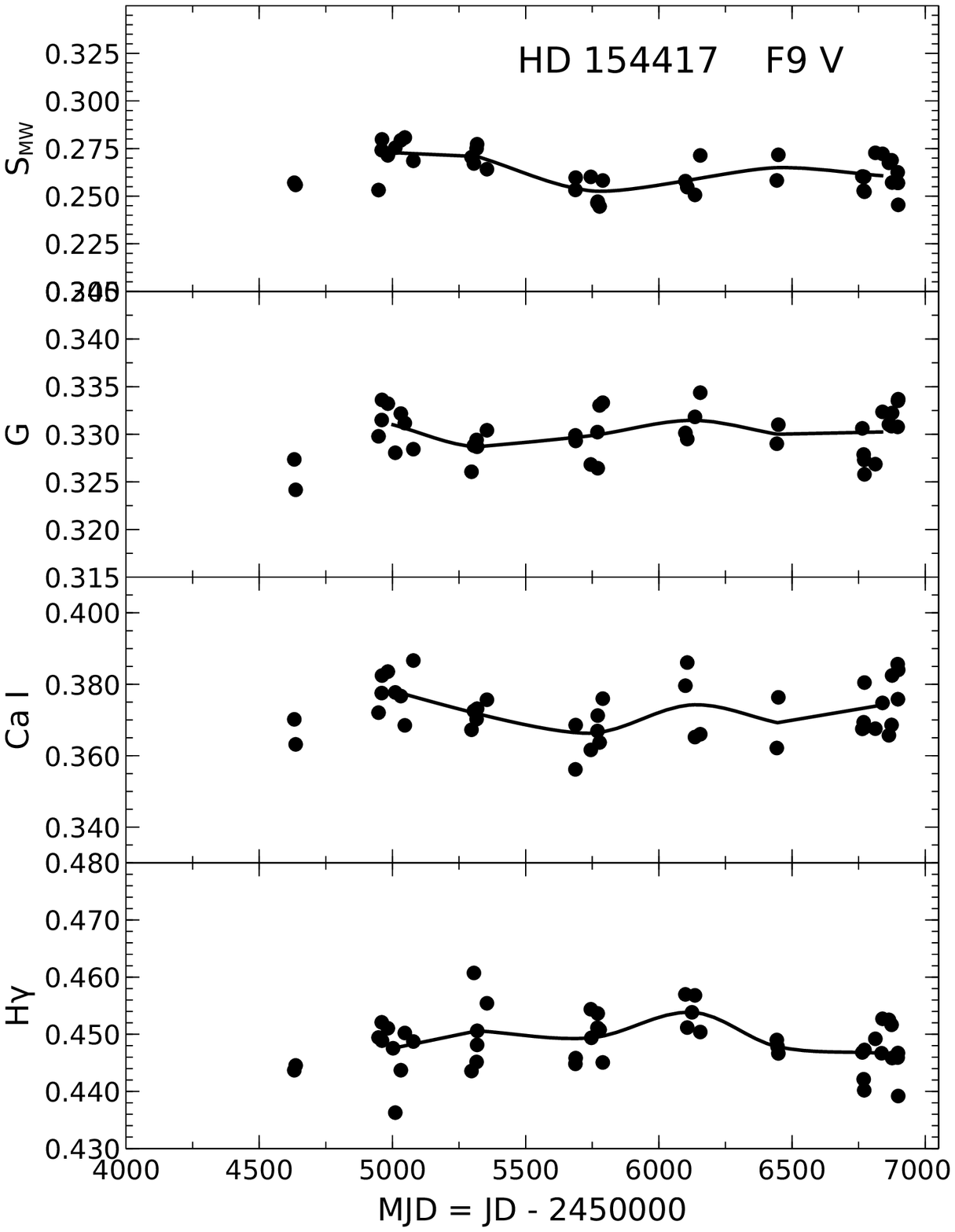} & \includegraphics[width=2.0in]{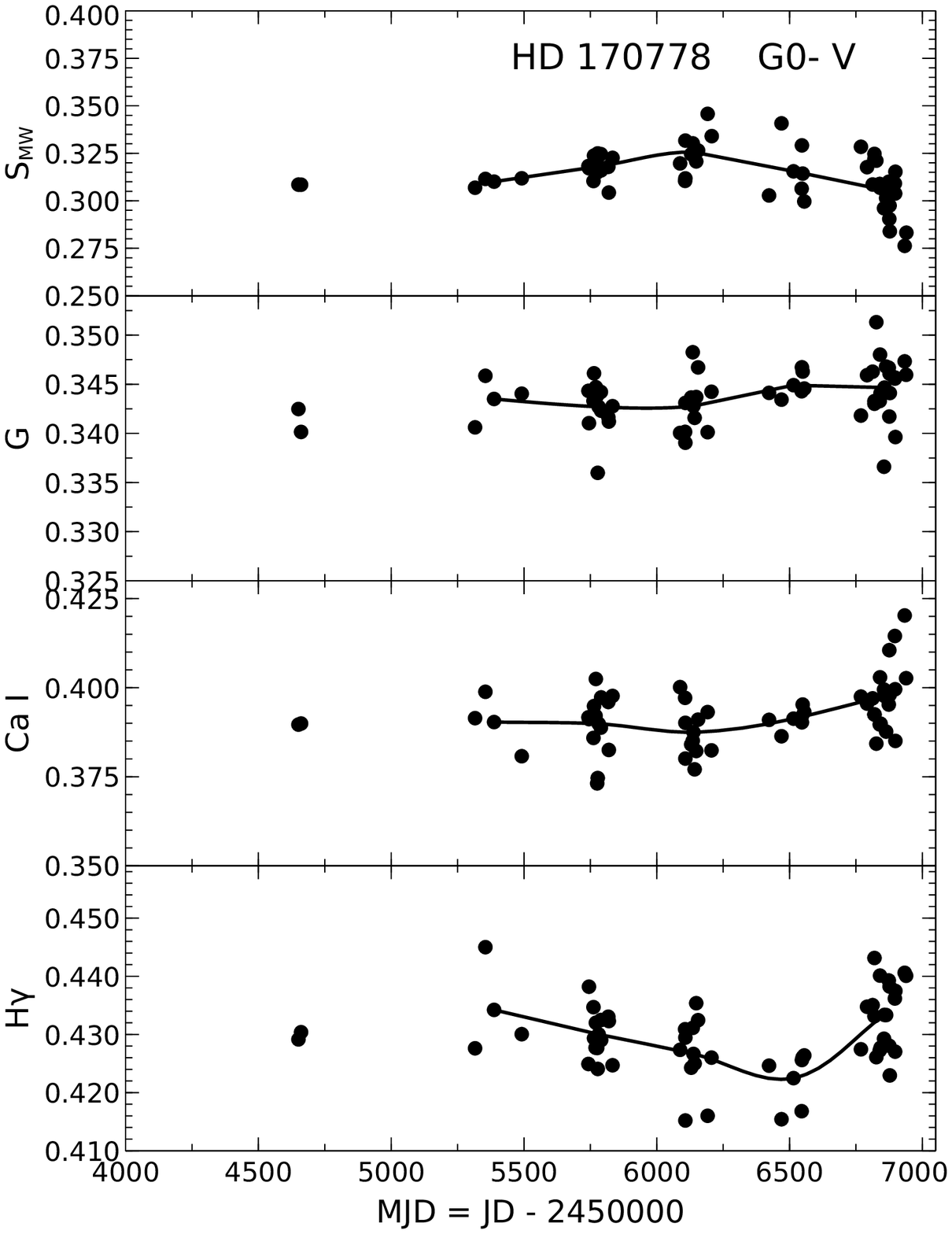} \\
\includegraphics[width=2.0in]{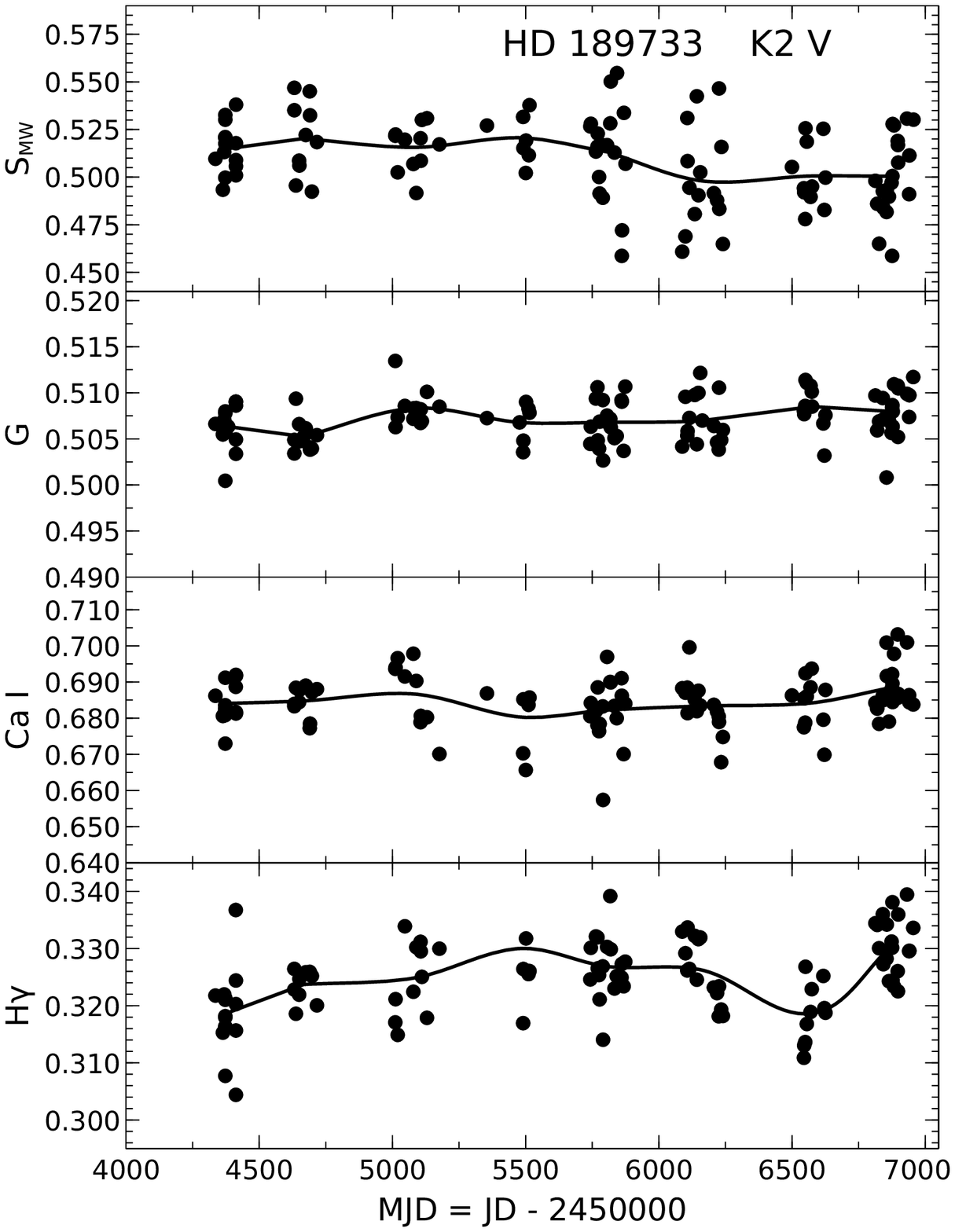} & \includegraphics[width=2.0in]{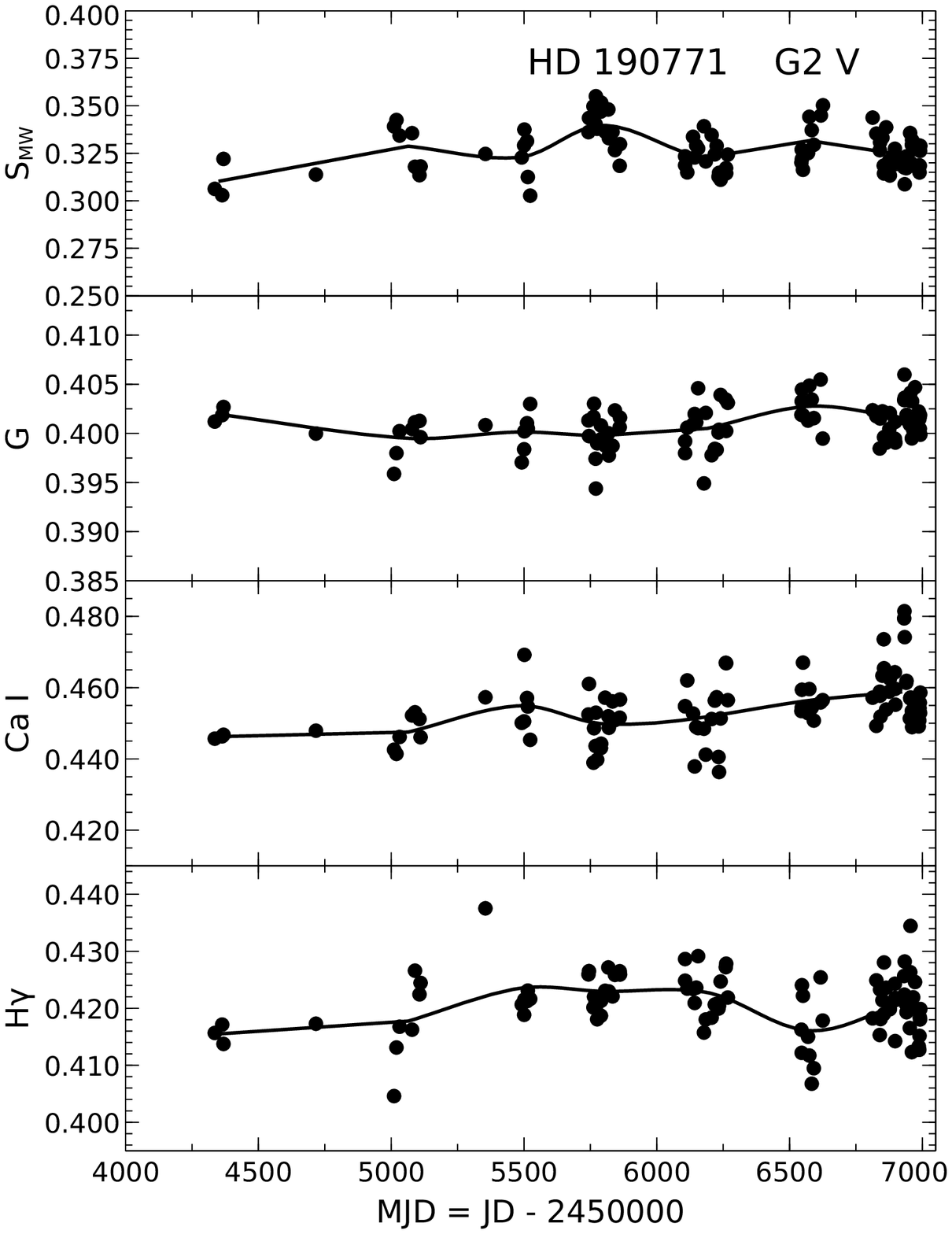} & \includegraphics[width=2.0in]{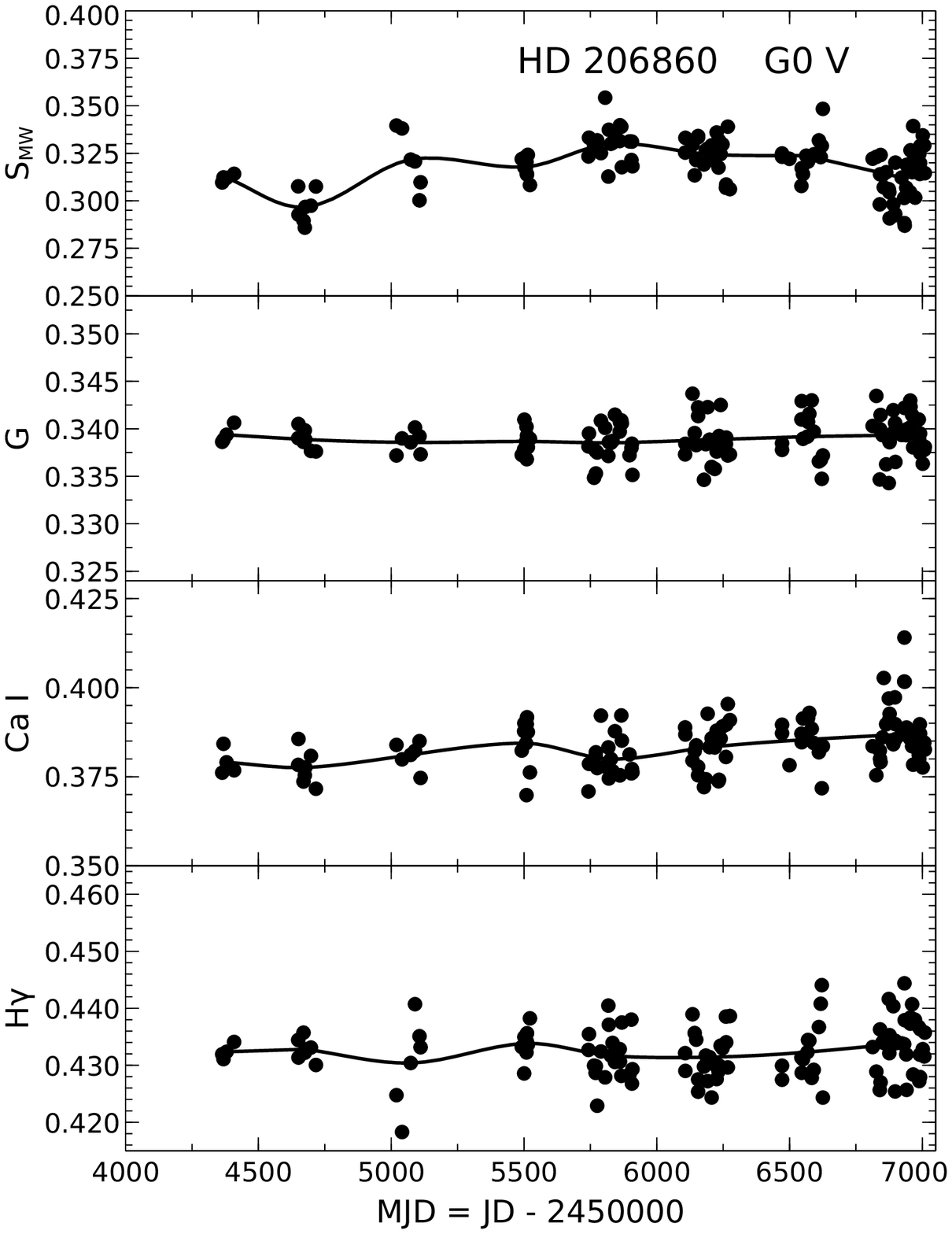} \\
\end{tabular}
\caption{Continuation of the montage in Figures \ref{fig:KHG1} and \ref{fig:KHG2}.}
\label{fig:KHG3}
\end{figure*}

\begin{figure*}
\begin{tabular}{ccc}
\centering
\includegraphics[width=2.0in]{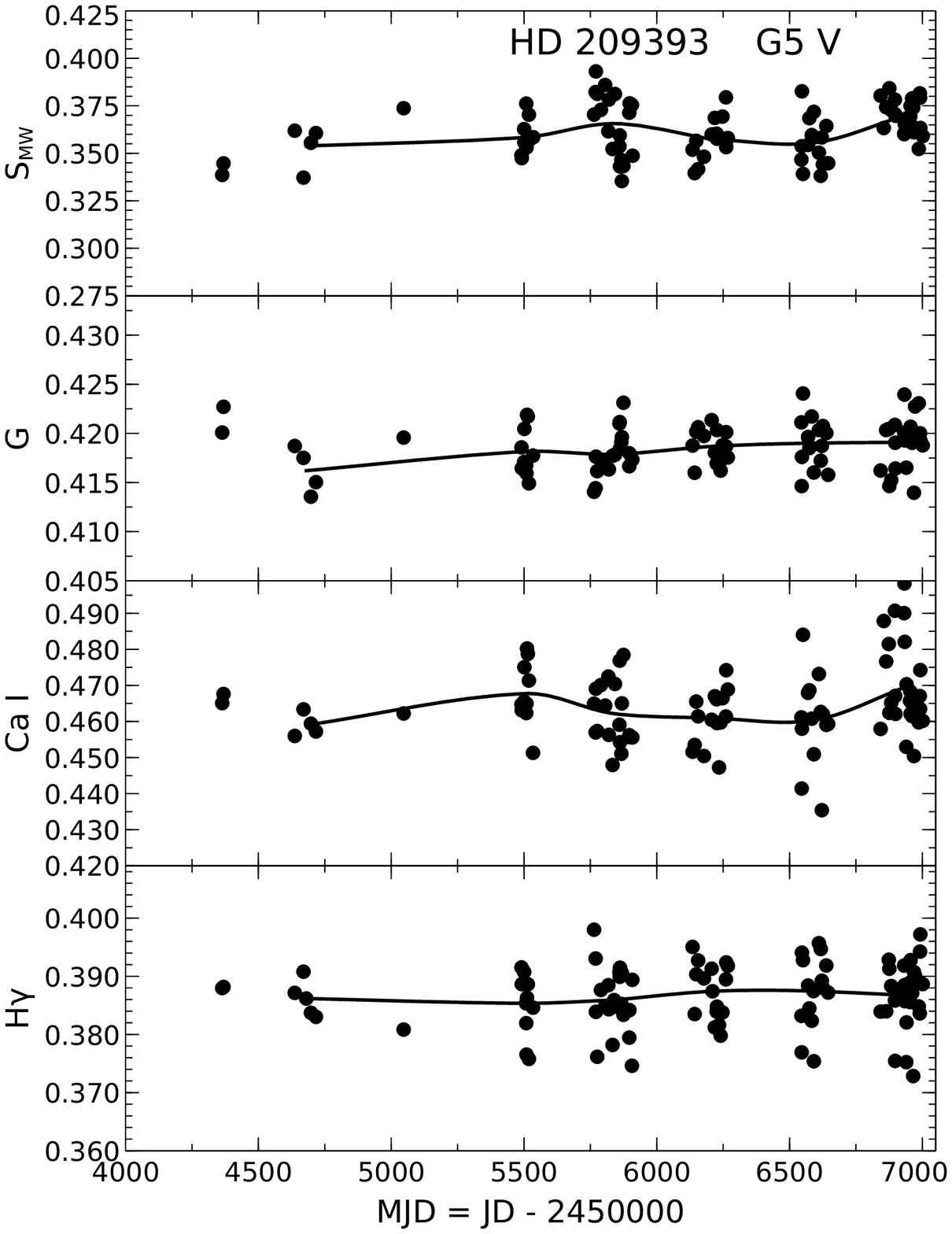} & \includegraphics[width=2.0in]{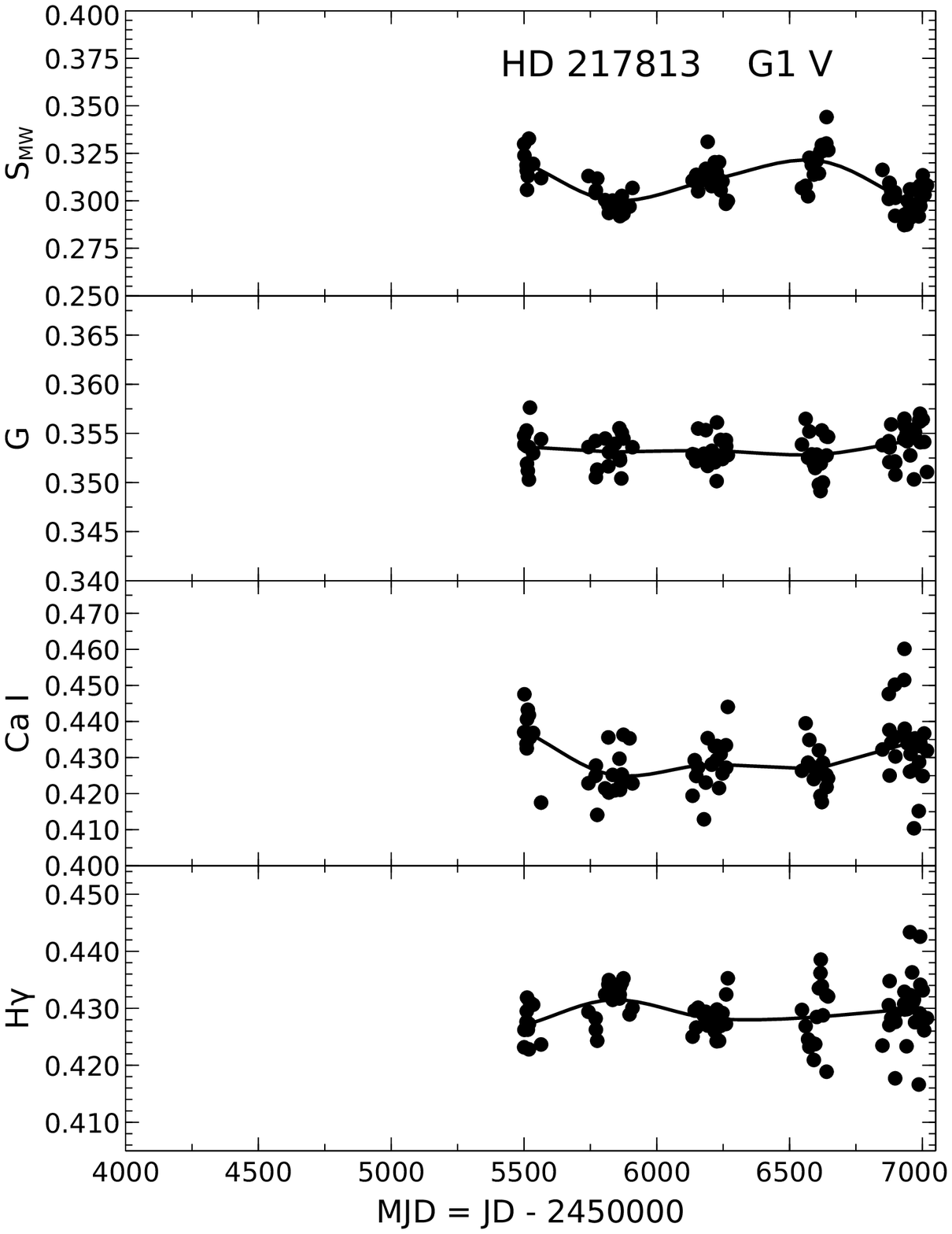} & \includegraphics[width=2.0in]{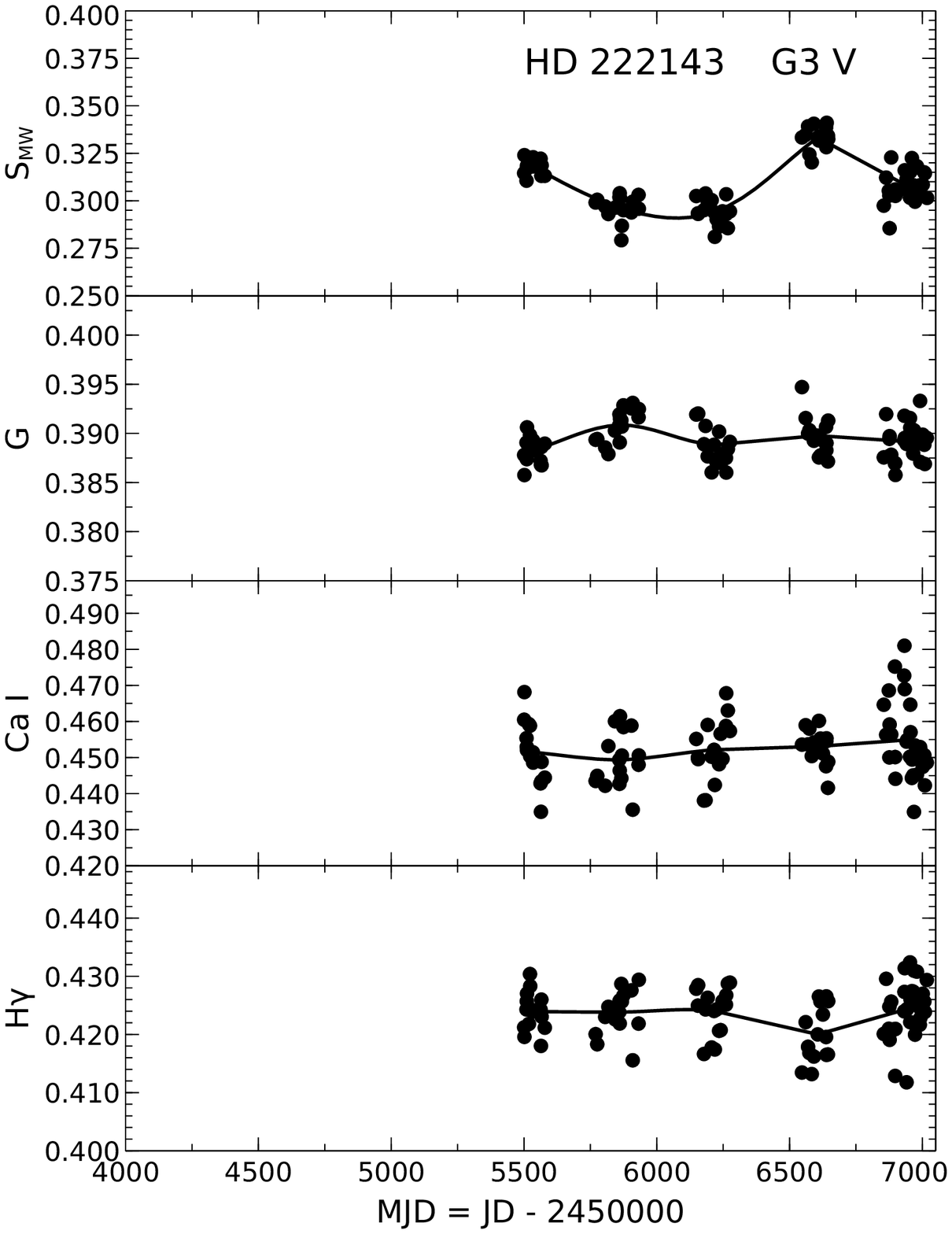} \\
\end{tabular}
\caption{Continuation of the montage in Figures \ref{fig:KHG1}, \ref{fig:KHG2},
and \ref{fig:KHG3}.}
\label{fig:KHG4}
\end{figure*}

\section{Statistical Analysis of the Spectroscopic Results}

\begin{deluxetable*}{llllllrllllll}
\tablecolumns{11}
\tablewidth{0pt}
\tablecaption{Young Solar Analog Stars\\Mean Activity Data, Predicted 
Rotational Periods in days, and Chromospheric Activity Ages\label{tbl:mad}}
\tablehead{
\colhead{Name} & 
\colhead{$\langle S_{\rm MW} \rangle$} & 
\colhead{$\sigma$} & 
\colhead{$\langle\log(R^\prime_{\rm HK})\rangle$} & 
\colhead{$P_{\rm rot}(R^\prime_{\rm HK})$} & 
\colhead{$P_{\rm max}(v\sin i)$} &
\colhead{Age} &
\colhead{$\langle$G$\rangle$} & \colhead{$\sigma$} &
\colhead{$\langle$\ion{Ca}{1}$\rangle$} & \colhead{$\sigma$} &
\colhead{$\langle$H$\gamma\rangle$} & \colhead{$\sigma$} \\
 & & & & \phn\phn(error) & \phn\phn(error) & (Myr) & & & & &}  
\startdata
HD 166      & 0.429  & 0.017  & -4.393 &  \phn7.52d (2.79) & 10.03d (0.45)                          & 375 &  0.459 & 0.002 &  0.539 & 0.006 &   0.384 & 0.005 \\
HD 5996     & 0.376  & 0.019  & -4.465 &  11.30\phantom{d} (2.80) & \phn\phn\phn\phn$\infty$  	    & 763 & 0.463 & 0.002 &  0.528 & 0.009 &   0.366 & 0.005 \\
HD 9472     & 0.322  & 0.011  & -4.495 &  10.06\phantom{d} (2.19) & 16.23\phantom{d} (1.05) 	    & 762 & 0.408 & 0.002 &  0.448 & 0.009 &   0.405 & 0.005 \\
HD 13531    & 0.369  & 0.013  & -4.431 &  \phn8.06\phantom{d} (2.37) & \phn7.38\phantom{d} (0.12)   & 486 & 0.436 & 0.002 &  0.488 & 0.009 &   0.391 & 0.005 \\
HD 27685    & 0.310  & 0.017  & -4.510 &  10.14\phantom{d} (2.08) & 32.71\phantom{d} (20.45) 	    & 810 & 0.425 & 0.002 &  0.456 & 0.010 &   0.410 & 0.006 \\
HD 27808    & 0.255  & 0.011  & -4.541 &  \phn4.36\phantom{d} (0.80) &  \phn5.15\phantom{d} (0.08)  & 548 & 0.275 & 0.002 &  0.330 & 0.010 &   0.472 & 0.004 \\
HD 27836    & 0.346  & 0.011  & -4.397 &  \phn4.03\phantom{d} (1.46) & \phn\phn\phn\nodata 	    & 210 & 0.343 & 0.003 &  0.387 & 0.011 &   0.431 & 0.008 \\
HD 27859    & 0.312  & 0.009  & -4.460 &  \phn5.80\phantom{d} (1.47) &  \phn8.14\phantom{d} (0.22)  & 390 & 0.360 & 0.002 &  0.391 & 0.011 &   0.436 & 0.005 \\
HD 28394    & 0.259  & 0.007  & -4.520 &  \phn3.46\phantom{d} (0.69) &  \phn3.01\phantom{d} (0.14)  & 876 & 0.241 & 0.002 &  0.307 & 0.011 &   0.479 & 0.006 \\
HD 42807    & 0.339  & 0.013  & -4.451 &  \phn7.58\phantom{d} (2.01) &  \phn8.91\phantom{d} (0.36)  & 494 & 0.416 & 0.002 &  0.450 & 0.008 &   0.402 & 0.005 \\ 
HD 76218    & 0.392  & 0.019  & -4.458 &  11.37\phantom{d} (2.91) & 12.54\phantom{d} (0.74) 	    & 753 & 0.470 & 0.003 &  0.570 & 0.010 &   0.362 & 0.006 \\
HD 82885    & 0.287  & 0.018  & -4.632 &  20.77\phantom{d} (2.91) & 16.00\phantom{d} (1.00) 	    & 2184 & 0.479 & 0.002 &  0.551 & 0.007 &   0.389 & 0.003 \\
HD 96064    & 0.476  & 0.028  & -4.355 &  \phn5.81\phantom{d} (1.82) & 15.61\phantom{d} (2.79)      & 230 & 0.453 & 0.003 &  0.544 & 0.012 &   0.368 & 0.008 \\
HD 101501   & 0.315  & 0.019  & -4.540 &  13.90\phantom{d} (2.56) & 15.55\phantom{d} (2.22) 	    & 1185 & 0.460 & 0.002 &  0.518 & 0.007 &   0.380 & 0.004 \\
HD 113319   & 0.303  & 0.020  & -4.511 &  \phn9.37\phantom{d} (1.92) & 12.77\phantom{d} (0.71)      & 743 & 0.408 & 0.002 &  0.439 & 0.008 &   0.406 & 0.007 \\
HD 117378   & 0.302  & 0.010  & -4.452 &  \phn4.20\phantom{d} (1.11) & \phn4.73\phantom{d} (0.09)   & 297 & 0.331 & 0.003 &  0.368 & 0.011 &   0.433 & 0.007 \\
HD 124694   & 0.282  & 0.010  & -4.473 &  \phn3.35\phantom{d} (0.80) & \phn3.08\phantom{d} (0.09)   & 344 & 0.281 & 0.002 &  0.331 & 0.008 &   0.453 & 0.006 \\
HD 130322   & 0.253  & 0.030  & -4.720 &  26.31\phantom{d} (3.01) & \phn\phn\phn\phn$\infty$  	    & 3202 & 0.484 & 0.004 &  0.564 & 0.011 &   0.363 & 0.004 \\
HD 131511   & 0.445  & 0.021  & -4.455 &  12.57\phantom{d} (3.27) &  \phn9.49\phantom{d} (0.40)     & 797 & 0.496 & 0.002 &  0.611 & 0.006 &   0.357 & 0.005 \\
HD 138763   & 0.316  & 0.012  & -4.436 &  \phn4.46\phantom{d} (1.28) & \phn\phn\phn\nodata          & 283 & 0.322 & 0.002 &  0.366 & 0.009 &   0.440 & 0.008 \\
HD 149661   & 0.303  & 0.019  & -4.651 &  24.23\phantom{d} (3.24) & 27.73\phantom{d} (3.70) 	    & 2581 & 0.506 & 0.002 &  0.623 & 0.007 &   0.352 & 0.004 \\
HD 152391   & 0.391  & 0.022  & -4.445 &  10.28\phantom{d} (2.81) & 10.35\phantom{d} (0.48) 	    & 651 & 0.467 & 0.002 &  0.538 & 0.007 &   0.371 & 0.005 \\
HD 154417   & 0.263  & 0.010  & -4.550 &  \phn6.92\phantom{d} (1.23) & \phn8.12\phantom{d} (0.24)   & 654 & 0.330 & 0.002 &  0.373 & 0.007 &   0.448 & 0.005 \\
HD 170778   & 0.315  & 0.014  & -4.444 &  \phn4.97\phantom{d} (1.37) & \phn6.35\phantom{d} (0.16)   & 322 & 0.344 & 0.003 &  0.392 & 0.008 &   0.430 & 0.007 \\
HD 189733   & 0.510  & 0.021  & -4.503 &  16.91\phantom{d} (3.57) & 13.58\phantom{d} (0.94) 	    & 1167 & 0.507 & 0.002 &  0.685 & 0.007 &   0.325 & 0.007 \\
HD 190771   & 0.326  & 0.011  & -4.462 &  \phn7.37\phantom{d} (1.85) & \phn9.61\phantom{d} (0.36)      & 492 & 0.401 & 0.002 &  0.453 & 0.007 &   0.421 & 0.005 \\
HD 206860   & 0.317  & 0.014  & -4.438 &  \phn4.73\phantom{d} (1.34) &  \phn5.01\phantom{d} (0.05)  & 305 & 0.339 & 0.002 &  0.384 & 0.007 &   0.433 & 0.004 \\
HD 209393   & 0.360  & 0.013  & -4.429 &  \phn7.38\phantom{d} (2.19) & 10.61\phantom{d} (0.53)      & 440 & 0.419 & 0.002 &  0.464 & 0.010 &   0.387 & 0.005 \\
HD 217813   & 0.310  & 0.012  & -4.462 &  \phn5.82\phantom{d} (1.47) & 11.89\phantom{d} (0.54)      & 397 & 0.353 & 0.002 &  0.430 & 0.008 &   0.429 & 0.004 \\
HD 222143   & 0.310  & 0.015  & -4.497 &  \phn8.87\phantom{d} (1.92) & 16.55\phantom{d} (1.03)      & 675 & 0.389 & 0.002 &  0.452 & 0.008 &   0.424 & 0.005
\enddata
\end{deluxetable*}

\begin{deluxetable*}{llllll}
\tabletypesize{\scriptsize}
\tablecolumns{6}
\tablewidth{0pc}
\tablecaption{Index-Variation Kolmogorov-Smirnov Significance Tests\\
Probability of the Null Hypothesis}
\tablehead{\colhead{Star ID} & \colhead{$S_{\rm MW}$} & \colhead{G-band} & 
\colhead{\ion{Ca}{1}} & \colhead{H$\gamma$} & \colhead{$N_{\rm seasons}$}}
\startdata
HD 166     & $< 10^{-5}$ & 0.020 & \nodata & \nodata & 4\\
HD 5996    & $< 10^{-5}$ & 0.028 & 0.0023  & \nodata & 5 \\
HD 9472    & 0.00075   & $< 10^{-5}$ & \nodata & \nodata & 5 \\
HD 13531   & $< 10^{-5}$ & 0.030 & \nodata & 0.024 & 5 \\
HD 27685   & $< 10^{-5}$ & 0.00002 & \nodata & \nodata & 6 \\
HD 27808   & \nodata  & $< 10^{-5}$ & \nodata & \nodata & 5 \\
HD 27836   & \nodata  & \nodata  & \nodata & \nodata & 5 \\
HD 27859   & 0.011    & \nodata  & \nodata & \nodata & 5 \\
HD 28394   & 0.014    & \nodata  & \nodata & \nodata & 5 \\
HD 42807   & $< 10^{-5}$ & 0.014   & $< 10^{-5}$ & 0.0153 & 6 \\
HD 76218   & $< 10^{-5}$ & \nodata & 0.00033 & \nodata & 7 \\
HD 82885   & $< 10^{-5}$ & 0.00059 & 0.0011  & 0.00026 & 7 \\
HD 96064   & $< 10^{-5}$ & 0.029 & 0.00019 & \nodata & 8 \\
HD 101501  & $< 10^{-5}$ & 0.00033 & 0.0045  & 0.013 & 7 \\
HD 113319  & $< 10^{-5}$ & 0.00046 & 0.0014 & 0.00045 & 6 \\
HD 117378  & \nodata  & \nodata & \nodata & \nodata & 4 \\
HD 124694  & $< 10^{-5}$ & \nodata & 0.00069 & 0.00016 & 5 \\
HD 130322  & $< 10^{-5}$ & \nodata & 0.033  & \nodata & 5 \\
HD 131511  & 0.025    & \nodata  & \nodata & 0.038 & 7 \\
HD 138763  & 0.00076  & \nodata  & \nodata & 0.023 & 6 \\
HD 149661  & 0.00043  & \nodata & \nodata & \nodata & 4 \\
HD 152391  & \nodata  & \nodata & \nodata & \nodata & 4 \\
HD 154417  & 0.00075  & \nodata & \nodata & \nodata & 5 \\
HD 170778  & 0.00022  & \nodata & \nodata & 0.0011  & 5 \\
HD 189733  & 0.00097  & 0.015   & \nodata & $< 10^{-5}$ & 8 \\
HD 190771  & 0.0005   & 0.024   & 0.049 & 0.010 & 6 \\
HD 206860  & $< 10^{-5}$ & \nodata & 0.0031 & \nodata & 6 \\
HD 209393  & 0.00036  & 0.048   & \nodata & \nodata & 5 \\
HD 217813  & $< 10^{-5}$ & \nodata & $< 10^{-5}$ & \nodata & 5 \\
HD 222143  & $< 10^{-5}$ & 0.0011  & \nodata & \nodata & 5 \\
\enddata
\end{deluxetable*}

\begin{deluxetable*}{lcccccc}
\tabletypesize{\scriptsize}
\tablecolumns{7}
\tablewidth{0pc}
\tablecaption{Pearson Statistical Tests of Index-to-Index Correlations}
\tablehead{\colhead{Star ID} & \colhead{$S_{\rm MW} - G$} & \colhead{$S_{\rm MW} - Ca~I$} & \colhead{$S_{\rm MW} - H\gamma$} & 
\colhead{$G - Ca~I$} & \colhead{$G - H\gamma$} & \colhead{$Ca~I - H\gamma$}}
\startdata
HD 166     & $+0.216,0.117$ & $-0.137,0.323$ & $-0.091,0.511$ & $-0.066,0.634$ & $+0.039,0.779$ & $-0.234,0.089$ \\
HD 5996    & $-0.190,0.081$ & $-0.102,0.351$ & $-0.236,0.030$ & $\mathbf{ +0.479,0.000}$ & $\mathbf{ +0.379,0.000}$ & $+0.226,0.038$\\
HD 9472    & $-0.025,0.843$ & $-0.170,0.175$ & $-0.066,0.603$ & $\mathbf{ +0.330,0.007}$ & $+0.214,0.086$ & $+0.222,0.075$ \\
HD 13531   & $-0.106,0.338$ & $+0.055,0.621$ & $-0.093,0.400$ & $+0.257,0.018$ & $\mathbf{ +0.489,0.000}$ & $+0.178,0.106$ \\
HD 27685   & $\mathbf{ -0.391,0.001}$ & $+0.267,0.033$ & $-0.205,0.104$ & $+0.133,0.295$ & $\mathbf{ +0.330,0.008}$ & $\mathbf{+0.360,0.004}$ \\
HD 27808   & $-0.215,0.102$ & $-0.146,0.268$ & $-0.188,0.153$ &  $+0.021,0.875$ &  $+0.091,0.494$ & $-0.104,0.435$ \\
HD 27836   & $+0.050,0.767$ & $-0.024,0.887$ & $-0.111,0.514$ &  $+0.363,0.027$ & $\mathbf{+0.558,0.000}$ & $\mathbf{+0.499,0.002}$ \\
HD 27859   & $+0.074,0.682$ & $+0.044,0.809$ & $-0.083,0.647$ &  $+0.062,0.731$ & $+0.158,0.381$ & $+0.258,0.147$ \\
HD 28394   & $-0.185,0.240$ & $-0.208,0.186$ & $\mathbf{-0.506,0.001}$ & $-0.057,0.722$ & $+0.363,0.018$ & $+0.221,0.160$ \\
HD 42807   & $\mathbf{-0.291,0.004}$ & $+0.044,0.675$ & $-0.245,0.017$ &  $+0.151,0.144$ & $\mathbf{+0.390,0.000}$ &  $+0.208,0.043$ \\
HD 76218   & $\mathbf{-0.323,0.000}$ & $+0.019,0.837$ & $-0.170,0.066$ &  $\mathbf{+0.496,0.000}$ & $\mathbf{+0.579,0.000}$ & $\mathbf{+0.570,0.000}$ \\
HD 82885   & $\mathbf{-0.464,0.000}$ & $\mathbf{-0.292,0.002}$ &  $+0.113,0.241$ & $+0.155,0.108$ &  $\mathbf{+0.249,0.009}$ & $+0.064,0.507$ \\
HD 96064   & $-0.164,0.142$ &  $-0.030,0.787$ &  $\mathbf{-0.296,0.007}$ &  $\mathbf{+0.425,0.000}$ & $\mathbf{+0.404,0.000}$ & $\mathbf{+0.279,0.011}$ \\
HD 101501  & $\mathbf{-0.351,0.000}$ &  $-0.135,0.168$ & $-0.021,0.830$ &  $-0.017,0.866$ &  $\mathbf{+0.388,0.000}$ & $-0.016,0.869$ \\
HD 113319  & $\mathbf{-0.508,0.000}$ &  $-0.038,0.731$ & $\mathbf{-0.493,0.000}$ & $+0.053,0.627$ & $\mathbf{+0.503,0.000}$ & $+0.135,0.212$ \\
HD 117378  & $+0.045,0.756$ &  $-0.187,0.193$ & $\mathbf{-0.501,0.000}$ & $-0.073,0.615$ & $-0.075,0.606$ & $\mathbf{+0.345,0.014}$ \\
HD 124694  & $-0.266,0.032$ &  $-0.053,0.677$ & $\mathbf{-0.457,0.000}$ & $+0.133,0.292$ & $+0.103,0.413$ & $+0.190,0.129$ \\
HD 130322  & $\mathbf{-0.429,0.003}$ &  $+0.046,0.762$ & $-0.010,0.943$ & $\mathbf{+0.390,0.007}$ & $+0.258,0.082$ & $\mathbf{+0.481,0.001}$ \\
HD 131511  & $\mathbf{-0.307,0.010}$ &  $-0.165,0.176$ & $-0.115,0.347$ & $+0.166,0.172$ & $\mathbf{+0.320,0.007}$ & $+0.181,0.138$ \\
HD 138763  & $+0.055,0.730$ & $-0.253,0.106$ & $-0.350,0.023$ & $\mathbf{+0.500,0.001}$ &  $\mathbf{+0.490,0.001}$ & $\mathbf{+0.520,0.000}$ \\
HD 149661  & $\mathbf{-0.442,0.005}$ &  $\mathbf-0.346,0.031$ &  $+0.102,0.536$ &  $+0.364,0.023$  & $+0.317,0.049$ & $-0.130,0.430$ \\ 
HD 152391  & $-0.414,0.026$ &  $-0.068,0.726$ &  $-0.281,0.140$ & $+0.194,0.314$ & $+0.129,0.505$ & $-0.085,0.662$ \\
HD 154417  & $-0.090,0.635$ &  $-0.041,0.831$ & $+0.072,0.706$ & $+0.187,0.323$ & $-0.005,0.980$ & $-0.106,0.578$ \\
HD 170778  & $-0.152,0.277$ & $\mathbf{-0.524,0.000}$ & $\mathbf{-0.385,0.004}$ & $+0.223,0.108$ & $+0.301,0.029$ & $\mathbf{+0.361,0.008}$ \\
HD 189733  & $-0.116,0.245$ & $-0.067,0.502$ & $-0.017,0.867$ & $\mathbf{+0.341,0.000}$ & $+0.210,0.033$ & $\mathbf{+0.262,0.007}$ \\
HD 190771  & $\mathbf{-0.307,0.003}$ & $\mathbf{-0.285,0.006}$ & $-0.240,0.022$ & $\mathbf{+0.289,0.006}$ & $+0.177,0.095$ & $\mathbf{+0.266,0.011}$ \\
HD 206860  & $-0.073,0.466$ &  $\mathbf{-0.470,0.000}$ & $\mathbf{-0.445,0.000}$ & $+0.164,0.102$ & $+0.066,0.513$ &  $+0.177,0.076$ \\
HD 209393  & $\mathbf{-0.270,0.011}$ & $+0.078,0.466$ &  $-0.208,0.051$ & $\mathbf{+0.473,0.000}$ & $+0.220,0.038$ & $+0.189,0.077$ \\
HD 217813  & $-0.228,0.038$ & $-0.192,0.081$ & $\mathbf{-0.284,0.009}$ & $+0.200,0.070$ & $+0.191,0.083$ & $-0.049,0.662$ \\
HD 222143  & $-0.089,0.411$ & $-0.006,0.959$ & $-0.211,0.050$ & $-0.053,0.623$ & $+0.163,0.131$ & $+0.097,0.372$ \\
\enddata
\end{deluxetable*}

Figures \ref{fig:KHG1}, \ref{fig:KHG2}, \ref{fig:KHG3}, and \ref{fig:KHG4} 
show montages (in order of HD number) of time series of the \ion{Ca}{2} H \& K
index (transformed to the Mount Wilson system index $S_{\rm MW}$), the G-band
index, the \ion{Ca}{1} index, and the H$\gamma$ index for our program stars.  

Table \ref{tbl:mad} lists mean values for the $S_{\rm MW}$ index and 
$\log(R^\prime_{\rm HK})$ from our observations of the program stars.  
The $S_{\rm MW}$ index measures
the flux in the cores of the \ion{Ca}{2} H \& K lines, but that flux includes
contributions from both the chromosphere and the photosphere.  A quantity
$R^\prime_{\rm HK}$, which is a useful measure of the chromospheric flux only, 
may be derived from $S_{\rm MW}$ using a method outlined in \citet{noyes84}.
The $\log(R^\prime_{\rm HK})$ index
may be calibrated against age and rotation period \citep{mamajek08}.  We
have included columns in Table \ref{tbl:mad} listing the expected rotation 
period (in days), $P_{\rm rot}(R^\prime_{\rm HK})$, based on the calibration of 
\citet{mamajek08}, as well as the upper limit to the rotation period derived 
from our $v\sin i$ 
values listed in Table \ref{tbl:bpp} ($P_{\rm max}(v\sin i)$), along with
associated errors.  Note that, with the possible exception of HD~82885, the
rotation periods derived from the activity levels are consistent,
within the errors, with the rotation period upper limits deduced from the
projected rotational velocities.  Chromospheric ``activity ages'' for our
program stars, based on the calibration of \citet{mamajek08}, are also included
in Table \ref{tbl:mad}.  As it turns out, all but three of our stars (HD~82885,
HD~130322, and HD~149661) lie within the target age limits for this project,
0.3 -- 1.5 Gyr.  However, the discrepancy between $P_{\rm rot}(R^\prime_{\rm HK})$ and
$P_{\rm max}(v\sin i)$ for HD~82885 suggests that the chromospheric activity
age for that star may not be accurate.  For instance, \citet{donahue96} quote
a rotation period for HD~82885 of 18.6 days.  This gives a gyrochronological
age, using the calibration of \citet{barnes07}, of 1.6 Gyr.

\subsection{The Sensitivity of the Photospheric Indices to 
Temperature Variations}

It was hypothesized in the Introduction that the three ``photospheric'' indices
defined in \S\ref{sec:Gband} -- \ref{sec:HG} will be primarily sensitive to 
temperature, and thus might be
useful in measuring integrated temperature changes on the stellar surface
arising from spots and/or photospheric faculae.  To determine the usefulness
of these indices for that purpose, we need to assess their sensitivity to
these changes.  Figure \ref{fig:BV} displays plots of these 
indices (using the Mount Wilson calibration stars from Table 5 of 
\citet{gray03}) versus 
$B-V$.  That Figure shows that in the realm of the late F-type
stars to the early K-type stars all three indices vary approximately linearly
with $B-V$.  The following equations are straight-line fits to the linear
portions of those curves:
\begin{align*}
B-V =\, & 0.259 + 0.966\, {\rm G}\\
B-V =\, & 0.212 + 1.010\,{\rm Ca\, I}\\
B-V =\, & 1.68 - 2.539\, {\rm H}\gamma
\end{align*}
where G, \ion{Ca}{1}, and H$\gamma$ refer to their respective indices, and
$B-V$ refers to the Johnson $B-V$ index.  Both the G-band index and the
\ion{Ca}{1} index have slopes of nearly unity with respect to $B-V$, and so
changes in those indices should translate directly into changes in $B-V$.
The H$\gamma$ index has a sensitivity that is smaller by about a factor of
2.5.  

We will report in Paper II that many of our stars vary $\le 0.03$ -- 0.07 magnitude 
in the Johnson $V$-band, and in the instances where we can measure color ($B-V$)
changes, those changes are generally $\le 0.01$ mag.  This is roughly what we
might expect if variations in brightness (due to sunspots and faculae) move 
the star parallel to the main sequence.  If the observed changes in the 
photospheric indices arise solely from temperature effects, we might therefore expect 
to observe variations in the G-band and \ion{Ca}{1} up to 0.01 in the index,
and by a factor of about 2.5 smaller in H$\gamma$.  Such changes should be
detectable in at least the G-band and \ion{Ca}{1} indices, as the measurement 
errors in those indices are on the order of 0.001 -- 0.003.  Indeed, because
of those measurement errors, these indices are potentially more useful in
measuring temperature changes than photometric colors where the errors are
larger.  Interestingly, the data
in Table 4 do indeed indicate variations in \ion{Ca}{1} of about the expected
magnitude ($\le 0.01$), but the observed variations in the G-band are smaller
by a factor of two or more ($\le 0.004$).  Hence, while it is plausible that
the observed variations in \ion{Ca}{1} are temperature related, it is clear
that the variations in the G-band may have a different or more complex origin.
The observed variations in H$\gamma$ are smaller than those observed in 
\ion{Ca}{1}, but not by the factor we would expect if those variations are 
governed by temperature alone.  We will examine these questions in more
detail in \S 5.4 below.
  
\subsection{Statistical Tests for Season-to-Season Variability}
\label{sec:var}

The $S_{\rm MW}$ plots are the traditional tool for detecting and characterizing
activity cycles in stars.   The detection and characterization of 
activity cycles in active stars requires time series observations that exceed,
preferably by a factor of two or more, the period or characteristic timescale
of the star in question.  That normally requires observations over decades, 
and so, except for stars that our program has in common with other long-term
surveys, such as the Mt. Wilson program, we are limited in what we can say
on that subject.  What can be done at the current stage of the project is to
1) evaluate the reality of the variations in the seasonal means and/or 
variances of the four ``activity'' indices -- the $S_{\rm MW}$, G-band, 
\ion{Ca}{1}, and H$\gamma$ indices -- that are suggested by the time series
montages and 2) to examine and try to understand the existence of correlations between 
those indices.   

To assess the significance of the variations in the four indices on a 
year-to-year basis (variations within a given observing season will be examined
in Paper II of this series where we will evaluate rotation periods for our
stars), we have employed the Kolmogorov-Smirnov (KS) 
statistical test.  KS tests are used in judging the significance of whether 
or not two experimental or observational distributions of a certain variable 
differ; the difference may arise from either a difference in the means or in 
the variances of the two distributions.  We may consider each set of seasonal 
data (the ``clumps'' in Figures \ref{fig:KHG1}, \ref{fig:KHG2}, 
\ref{fig:KHG3}, and \ref{fig:KHG4}) as independent samples of the index in 
question, and compare those samples for a given star on a pair-wise basis 
using the KS test to ascertain whether significant variation in the mean 
value and/or variance of the index has occured over the period we have 
observed the star.  The way we perform the tests is as follows.  Let us
suppose we have observed the star for four years (four observing seasons), 
seasons 1, 2, 3, and 4.  We then carry out KS tests on each of the following
six pair-wise comparisons: $1 \leftrightarrow 2$, $1 \leftrightarrow 3$, 
$1 \leftrightarrow 4$, $2 \leftrightarrow 3$, $2 \leftrightarrow 4$, and 
$3 \leftrightarrow 4$. The KS test yields a $p$-statistic for each comparison. 
Smaller values of $p$ indicate higher significance.  For instance, $p = 0.01$ 
indicates that the null hypothesis (no variation in the mean value or 
variance of the index) may be rejected with a confidence of 99\%.  But the fact
that we need to estimate the significance of variations in a time series rather
than simply between two seasons complicates the analysis.  For instance, let
us suppose we have five seasons of observations.  This results in a set
of 10 pair-wise comparisons.  If only one of those comparisons results in
$p = 0.01$, that does not rise to the level of significance because we would
expect, on the average, in a set of 10 comparisons, to encounter $p \le 0.01$ 
10\% of the time -- a significance of only 90\%.  However, if a given set
contains {\it multiple} comparisons with small $p$, we may then combine those
probabilities in assessing the significance of the observed variation.  

We use a Monte Carlo technique to evaluate these probabilities. A random number
generator was used to generate multiple gaussian distributions of an observational
variable, all with the same mean and variance (and thus for these artificial data 
the null hypothesis is true).  In total we generated 100,000 sets of 4-season data, 
each involving 6 pair-wise comparisons, for a total of 600,000 comparisons, and 
evaluated each comparison with KS statistics.  We did the same for sets of 5-season, 6-, 7-, and
8-season data, the latter involving 2.8 million pair-wise comparisons.  We were
able to verify, for instance, that comparisons with $p \le 0.01$ were encountered
with the expected frequency.  We then used these artificial data sets to evaluate 
the significance of variations in our observational data.  To take a real example, in
one of our 5-season data sets (HD~9472), we had, for the $S_{\rm MW}$ index, the following
values of $p$: 0.0116, 0.0132, 0.0188, 0.0281, and 0.0315.  The remaining five
comparisons had $p > 0.05$.  We then used the 5-season Monte Carlo data to
ask ``What proportion of 5-season sets have $p_{\rm min} \le 0.0116$ and four
other comparisons with $p \le 0.0315$?''  The result yields an overall $p = 0.00075$.
We have listed in Table 5 all the time series for which the overall $p \le 0.05$, 
indicating ``significant'' variability.

Spurious significant $p$ values can 
be created by outliers in the dataset.  We have reduced to a minimum the 
number of outliers in the dataset by rejecting all spectra with S/N $< 50$ and 
by examining each spectrum to eliminate those with obvious defects (such as 
cosmic rays) in the wavelength bands used for the calculation of the indices.
The remaining outliers cannot be rejected on a statistical basis (and may
indeed represent true excursions of the star) and so are included in the 
statistical tests.

It is clear from Table 5 that almost all of the program stars show significant 
season-to-season variations in $S_{\rm MW}$.  The ones that do not have only 4 or 5
seasons of data, so it is entirely possible that with a few more seasons of data
all will show significant variation.  50\% show significant variations in the G-band 
index, 40\% in the \ion{Ca}{1} index, and 37\% in the H$\gamma$ index.  We
expect that continuing the project for a few more years will increase those
proportions as well.  We emphasize that a lack of significant variation in 
the seasonal means and variances does not imply that the star is constant 
within a given season.  For instance, it  is well known, and we will further 
demonstrate in Paper II, that the ``scatter'' (at least for the $S_{\rm MW}$ index) 
within a given season can arise from rotational modulation in the index.

\subsection{Comments on the Nature of the Observed Variability}

A number of our stars that have significant season-to-season variations 
appear to be showing very short-term periodic or ``pseudo periodic''
behavior in the $S_{\rm MW}$ index.  Examples include HD~9472, HD~13531, 
HD~27685 (superimposed on a secular rise in activity), HD~217813,  
as well as some others.  Despite the shortness of the datasets, the 
above-mentioned stars show significant periods in the range of 2 -- 4 years 
with a
Lomb-Scargle analysis.  To judge the reality of such short periods, which are
considerably shorter than the periods found in \citet{baliunas95} we may refer
to other similar datasets.  For example, a number of stars in the
\citet{lockwood07} dataset appear to show very similar behavior (see the stars
HD~39587, HD~131156, HD~152391, HD~115404, HD~201092 for some possible 
examples).  This short-term variation appears to come 
and go and is often superimposed on longer timescale variations. Analysis of
the Lockwood et al. or similar datasets will be required to evaluate the reality of
these variations.

Stars in our dataset show a variety of behaviors associated with the dispersion
in the $S_{\rm MW}$ activity index within a given season.  For instance, the 
stars HD~27859 ($\langle\sigma_{S_{\rm MW}}\rangle = 0.007$), HD~124694 (0.008), 
HD~154417 (0.007), HD~217813 (0.008), and HD~222143 (0.006)
all show very tight activity dispersions within a given season.  On the other 
hand, HD~130322 (0.014), HD~131511 (0.018), and HD~189733 (0.018) show average 
seasonal dispersions greater by a factor of two or more within a given season.
This distinction appears to be intrinsic, as we are careful to achieve 
adequate S/N for all of our observations, and there are bright and 
``faint'' stars in both sets.  We note, however, that those stars
that have particularly low seasonal dispersions are F and early G-type stars,
while the three with the higher dispersions are all K-type stars.  If the seasonal
dispersions arise from rotational modulation, as active regions
rotate across the stellar disk, then this suggests that the late-type stars
mentioned above may be dominated by one or a small number of active regions, whereas 
for the F- and early G-type stars in the project sample, active regions are smaller and more 
dispersed across the stellar disk.  One caution should be noted: the activity
behavior of HD~189733 may not be typical of young active K-type stars, as it
has apparently been spun up by angular momentum transfer from its hot-jupiter
companion (see Introduction).  HD~131511 does, however, behave in quite a
similar way to HD~189733.  Even though HD~131511 is a spectroscopic binary, its 
stellar companion is in a much wider orbit 
\citep[$P_{\rm orb} = 125.4$d][]{nidever02,jancart05}, and 
probably has not yet had an important influence on the angular momentum of the 
primary.  We also note that in a recent Nordic Optical Telescope FIES spectrum
of HD~131511, the emission in the cores of the \ion{Ca}{2} H \& K lines appear
symmetrical, and so we see no evidence for emission from the companion.  This
will need to be verified by further spectra at different phases of the
companion's orbit.   

\begin{figure}
\includegraphics[width=3.25in]{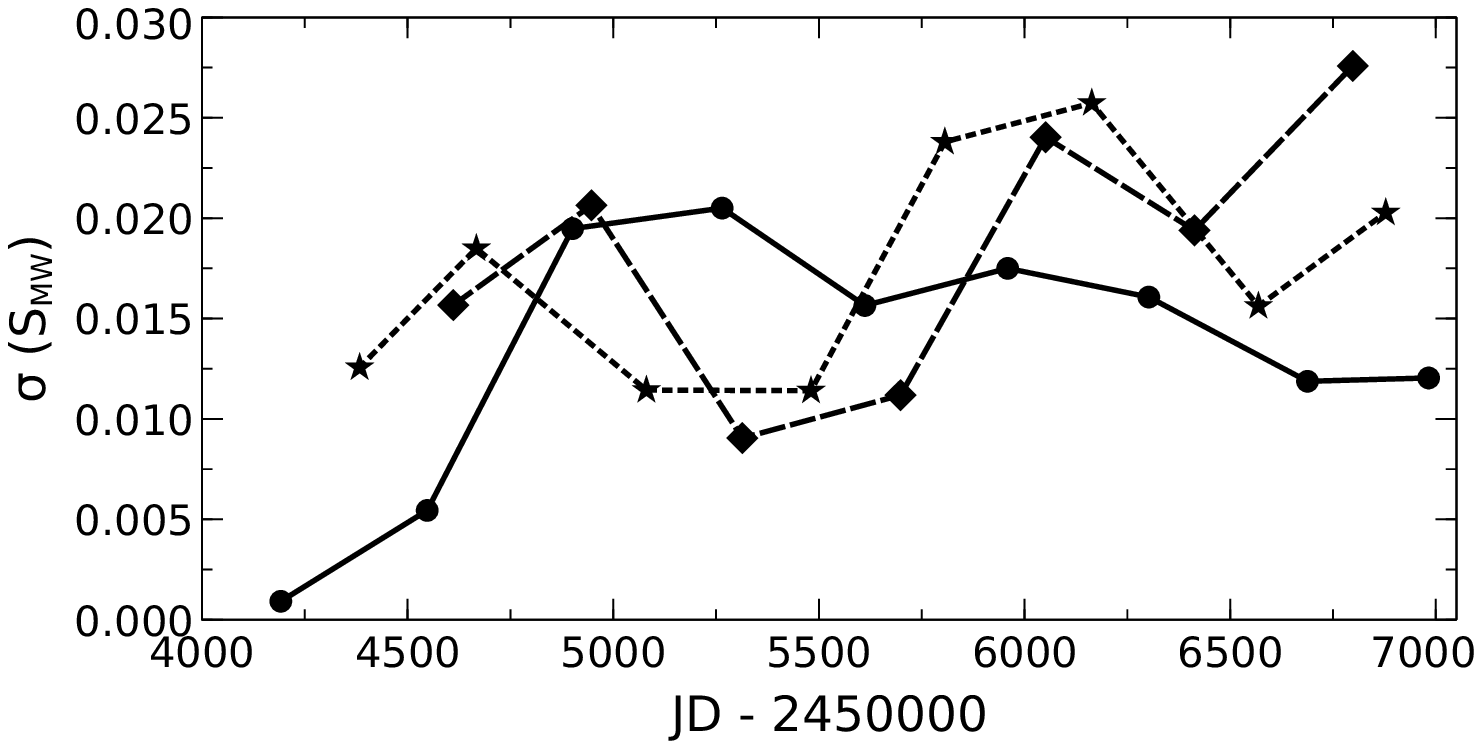}
\caption{The variation in the seasonal dispersion of $S_{\rm MW}$ with time for
three stars, HD~76218 (solid line), HD~131511 (dashed line), and HD~189733
(dotted line).}
\label{fig:vs}
\end{figure}

Interestingly, some of our stars appear to show variations in their seasonal 
dispersion behavior.  Whether that variation in dispersion is cyclical can
only be determined with longer time series.  The F-test is the appropriate
test for determining the statistical significance of season-to-season 
differences in the variance of an index.  HD~131511, for instance, appears
to alternate between seasons with high and moderate dispersions in the 
$S_{\rm MW}$ index.  Examination of the $S_{\rm MW}$ plot for HD~131511 (Figure
\ref{fig:KHG3}) shows four seasons with relatively high dispersions and 
three with moderate dispersions.  F-tests carried out on the
21 pair-wise comparisons between the seven seasons show highly significant
variations, with an overall $p = 0.0011$ (calculated using the same Monte
Carlo technique employed to evaluate the KS tests).
HD~189733 may be behaving in a similar way, although the statistics are
of lower significance ($p = 0.045$). HD~76218 apparently also varies in seasonal
dispersion ($p = 0.016$). The variation in HD~76218 is unusual in the sense 
that when the seasonal dispersion is the highest, the activity level is at 
or near a minimum.  This is opposite to the sun which shows the greatest dispersion 
in the \ion{Ca}{2} flux at activity maximum \citep{keil98}.  Figure \ref{fig:vs} 
shows the variation in seasonal dispersion with time for HD~76218, HD~131511, 
and HD~189733.  A possible interpretation of this
behavior is that these stars vary between a state in which the active regions
are relatively small, numerous, and dispersed (low-to-moderate seasonal 
dispersion) and a state which is dominated by one or a few large active 
regions (high seasonal dispersion).  This variation in seasonal dispersion 
may represent a novel type of activity cycle in stars, or it may be evidence
for a flipflop cycle \citep{jetsu93} and/or active longitudes.  More observations
will be required to fully characterize this behavior.  

\subsection{Correlations Between Indices}
    
A further question to address is whether or not significant correlations
exist between the four ``activity'' indices measured in this paper.  In the Introduction
we gave the rational for the three ``photospheric'' indices defined in this paper --
the G-band, \ion{Ca}{1}, and H$\gamma$ indices -- and suggested that these three
indices might vary in step with activity variations largely through related temperature
changes in the photosphere connected with changes in spots and photospheric faculae.  
How the photospheric indices would vary in relation to $S_{\rm MW}$ would then depend on 
whether cool spots or hot photospheric faculae dominate.  If the photospheric indices 
vary primarily on the basis of temperature, we would expect the G-band index to vary 
directly with the \ion{Ca}{1} index, and inversely with respect to the H$\gamma$ index.  
We will see below whether this is indeed the case.

Table 6 shows the results of Pearson's r-tests for linear correlations between the
four indices.  These comparisons are made with the original observations, and not with
the seasonal means.  Since all of these indices are measured in a single spectrum, we
do not have to worry about time differences between the observations of the different
indices.  The first column in Table 6 is the stellar ID, the second tabulates the results of the
Pearson r-test for correlations between $S_{\rm MW}$ and the G-band index, the third the
same for $S_{\rm MW}$ and \ion{Ca}{1}, the fourth for $S_{\rm MW}$ and H$\gamma$, the fifth
for the G-band and \ion{Ca}{1}, the sixth for the G-band and H$\gamma$, and the seventh
for \ion{Ca}{1} and H$\gamma$.  Each comparison consists of two numbers, Pearson's 
linear correlation coefficient $r$, and the $p$-statistic, from which the probability of
the null hypothesis (zero correlation) may be calculated.  Small $p$ indicates a
signficant correlation.  Correlations with $p \le 0.015$ are indicated with bold type
in Table 6.  We have adopted $p \le 0.015$ as a useful standard for judging the
significance of these correlations because, for a given index, and 30 tabulated stars,
we should expect at that significance level only 0.5 spurious correlations.

A glance at Table 6 shows the presence of multiple significant, in many cases highly significant, 
correlations, although none of those correlations are particularly strong ($r < 0.6$).  We have 
examined each of these correlations graphically to assure ourselves that none are caused by one 
or a few ``outliers''.  Could these correlations arise from instrumental effects?  We reject
that for a number of reasons: 1) We have not included in these tests data from the earlier
Photometrics CCD, and so that means that all of the observations involved in these tests 
have been carried out with the same CCD on the same spectrograph on the same telescope, and 
all have been reduced identically.  2) The passbands used in defining these indices do not overlap, 
and so a spectral defect (cosmic ray, etc.) that affects one index will not affect another. 3) While
some stars show highly significant correlations, others do not.  Instrumental effects would
lead to significant correlations (or not) in all stars, not just a limited number.

Let us now examine the nature of those correlations.  For the $S_{\rm MW}$ -- G-band comparison,
11 out of the 30 stars show significant ($p \le 0.015$) correlations, and all of those are negative correlations, 
meaning that in those stars $S_{\rm MW}$ and the G-band vary oppositely; when one increases, the other 
decreases.  Note that the correlations that do not rise to our level of significance are as well almost 
all negative.  For the 
$S_{\rm MW}$ -- \ion{Ca}{1} comparison, only 4 of the 30 stars show a significant correlation,
but again all of those are negative.  For $S_{\rm MW}$ -- H$\gamma$, 8 of the 30 stars show
significant correlations, and again all of those correlations are negative.  
So, a strong conclusion is that where significant correlations are present, the 
``photospheric'' indices are all negatively correlated with $S_{\rm MW}$.

What about correlations between the photospheric indices?  Examination of Table 6 shows the presence
of many highly significant correlations between these indices, all of which are {\it positive}.  So,
the tendency is, when the G-band weakens, so too do \ion{Ca}{1} and H$\gamma$.

This behavior is not consistent with the hypothesis that the photospheric indices are primarily 
affected by temperature changes in the photosphere arising from changes in spots
and photospheric faculae. What then are the possible physical causes behind the observed behaviors?

One possibility that we must consider is whether the \ion{Ca}{1} and H$\gamma$ indices are
affected by changes in CH opacity.  The G-band is a molecular feature, arising from the CH molecule, 
but CH absorption lines are ubiquitous in the region of the spectrum containing
the \ion{Ca}{1} 4227\AA\ resonance line and H$\gamma$.  To test this hypothesis we calculated a
number of synthetic spectra for late F, mid-G, and early K-type stars, all identical except for
differences in CH absorption strength (appropriate for the size of the variations we observe in 
the G-band index), and then measured the resulting \ion{Ca}{1} and H$\gamma$ indices.  Those indices showed 
very small changes compared to the resulting changes in the G-band index, and in the opposite sense, 
which would yield {\it negative} correlations instead of the positive ones observed.  

The negative correlation between $S_{\rm MW}$ and the H$\gamma$ index might be understood on the basis
of line emission.  In the spectrum of the solar chromosphere, both \ion{Ca}{2} H \& K and H$\gamma$ (as
well as, of course, H$\alpha$ and H$\beta$) are seen in emission, and so it is reasonable to
expect that $S_{\rm MW}$ and H$\gamma$ would be negatively correlated on this basis, as emission 
{\it fills in} the H$\gamma$ line, resulting in a index smaller than for a purely photospheric line, 
while chromospheric emission yields an increase in $S_{\rm MW}$ over what would be measured for 
pure absorption in \ion{Ca}{2} H \& K.  Neither the G-band nor \ion{Ca}{1} show 
up in any significant way in the chromospheric spectrum, so this mechanism does not help to explain 
the negative correlations of those indices with $S_{\rm MW}$ or their positive correlations with H$\gamma$. 

As noted above, the existence of direct correlations between all three photospheric indices
is difficult to understand on the basis of temperature differences.  This
suggests that the physical cause underlying those direct correlations does not 
depend on temperature.  Mechanisms that may be relevant here were noted 
by \citet{basri89} who observed that the equivalent widths of metallic lines (especially
low-excitation lines) in 
the blue-violet part of the spectrum were reduced in
certain active stars, apparently due either to continuum emission arising in the
chromosphere or upper photosphere leading to the phenomenon of ``veiling'' or to nonradiative heating in the
upper layers of the photosphere in plage regions resulting in weaker line cores 
\citep[see also][]{chapman68,giampapa79,labonte85,labonte86}.  
Indeed \citet{gray06} noted a similar phenomenon in the spectra of active K-type dwarfs, particularly 
in the vicinity of the \ion{Ca}{1} line.  Interestingly, they noted that some 
active K dwarfs show this phenomena, and other equally active dwarfs do not.  
Both of these mechanisms can help to explain not only the direct correlations between 
the G-band, the \ion{Ca}{1} line, and H$\gamma$, but also
are consistent with the negative correlations between those indices and $S_{\rm MW}$ 
because as stellar activity increases, both the veiling and/or core-weakening and the emission in
\ion{Ca}{2} H \& K  would presumably increase together.  Furthermore, a closer 
look at Table 6 reveals
that the most significant G-band anti-correlations with $S_{\rm MW}$ occur at
spectral types where the G-band is near its maximum strength, and most of the
significant H$\gamma$ anti-correlations appear in the late-F and early G-type stars
where H$\gamma$ is still a strong feature, exactly what one would expect if
the mechanisms suggested by \citet{basri89} were active.

We might then ask why temperature effects, hypothesized at the beginning of this paper
to be the primary drivers of changes in the ``photospheric'' indices do not appear to be important?
This question requires further investigation, but it may be that for the indices
considered in this paper, temperature effects arising from changes in both photospheric
faculae and spots -- which would tend to cancel -- sufficiently balance out so that 
temperature variations become only a secondary cause in driving changes in these indices.

While it may be disappointing that the purpose for which we 
designed these
indices has not been realized, it does appear that these indices can be used
to monitor the emission flux in the Paschen continuum arising from stellar
activity.  This suggests that these three indices may also prove to be useful 
proxies for monitoring emission in the ultraviolet {\it Balmer} continuum, 
which is largely inaccessible from the ground.  If that proves to be the case, 
these indices would be of direct utility in achieving the original goals of 
this project. 

\section{Conclusions}

This paper reports on initial results from the Young Solar Analogs project,
which began in 2007 and which is monitoring the stellar activity of 31 
late F-, G-, and early K-type stars with ages between 300 million and 1.5
billion years.  We have detailed the transformation between our 
instrumental \ion{Ca}{2} activity indices and the Mount Wilson $S$ activity
index.  In addition, we have defined three new photospheric indices based
on the G-band, the \ion{Ca}{1} resonance line in the blue-violet, and the
H$\gamma$ line, and have examined, on a detailed statistical basis, how those 
indices vary and how they are related.  All four indices show strong evidence
for variability on a multi-year timescale in our data.  The anti-correlations 
between $S_{\rm MW}$ and the photospheric indices and the positive correlations 
between the photospheric indices suggest the presence of varying continuum emission
and/or non-radiative heating of the upper layers of the photosphere 
in at least some of the program stars.  Further observations and modelling will
be required to better understand these physical mechanisms and
to evaluate the utility of the ``photospheric'' indices as proxies for
ultraviolet continuum emission.  Subsequent papers in this series 
will examine medium-term variations in these indices and the multi-band photometry, 
as well as short-term variations.  

\acknowledgments

The authors would like to thank an anonymous referee for careful and detailed
comments that resulted in a considerably improved paper.
The authors would also like to thank Lee Hawkins, Dark Sky Observatory engineer, 
for expert and enthusiastic technical assistance in the construction and 
maintenance of the Robotic dome. We are also pleased to acknowledge the 
assistance of Mike Hughes (Electronics technician), Dana Greene, Machinist, 
and David Sitar, all at Appalachian State University.  This project has been
supported by NSF grant AST-1109158.  We are also grateful for funding
provided by The Fund for Astrophysical Research during an early stage of
this project.  This research has made use of the Keck Observatory Archive 
(KOA), which is operated by the W. M. Keck Observatory and the NASA Exoplanet 
Science Institute (NExScI), under contract with the National Aeronautics and 
Space Administration.  This research has also made use of the Elodie Archive
(http://atlas.obs-hp.fr/elodie/).  We also acknowledge use of archival spectra
from the UVES Paranal Observatory Project (ESO DDT Program ID 266.D-5655).
It is also a pleasure to acknowledge the service observing program at the
Nordic Optical Telescope. This research has made use of the SIMBAD database,
operated at CDS, Strasbourg, France.  We acknowledge with pleasure the Veusz 
software package\footnote{http://home.gna.org/veusz/} which was used for the 
figures in this paper.

\newpage


\begin{thebibliography}{}

\bibitem[Andersen(1991)]{andersen91} Andersen, J. 1991, A \& A Rev., 3, 91

\bibitem[Bagnulo et al.\ (2003)]{bagnulo03} Bagnulo, S., Jehin, E., 
Ledoux, C., Cabanac, R., Melo, C., Gilmozzi, R., \& the ESO Paranal Science 
Operations Team 2003, The Messenger, 114, 10

\bibitem[Baliunas et al.\ (1995)]{baliunas95} Baliunas, S.L., Donahue, R.A., 
Soon, W.H., Horne, J.H., Frazer, J., Woodard-Eklund, L., Bradford, M., 
Rao, L.M., Wilson, O.C., Zhang, Q., Bennett, W., Briggs, J., Carroll, S.M., 
Duncan, D. K., Figueroa, D., Lanning, H.H., Misch, T., Mueller, J., 
Noyes, R.W., Poppe, D., Porter, A.C., Robinson, C.R., Russell, J., 
Shelton, J.C., Soyumer, T., Vaughan, A.H., \& Whitney, J.H. 1995,
\apj, 438, 269

\bibitem[Barnes(2007)]{barnes07} Barnes, S. 2007, \apj, 669, 1167.

\bibitem[Basri et al.\ (1989)]{basri89} Basri, G., Wilcots, E., \& 
Stout, N. 1989, \pasp, 101, 528

\bibitem[Berger \& Title(2001)]{berger01} Berger, T.E. \& Title, A.M. 2001,
\apj, 553, 449.

\bibitem[Cassagrande et al.\ (2010)]{cassagrande10} Cassagrande, L., 
Ram\'{i}rez,
I., Mel\'{e}ndez, J., Bessell, M. \& Asplund, M. 2010 \aap, 512, 54

\bibitem[Castelli \& Kurucz(2003)]{castelli03} Castelli, F., \& 
Kurucz, R. L. 2003, in Modelling of Stellar Atmospheres,
eds. N. E. Piskunov, W. W. Weiss, \& D. F. Gray (San Francisco: ASP), A20

\bibitem[Chapman \& Sheeley(1968)]{chapman68} Chapman, G.A. \& 
Sheeley, N.E. 1968, Solar Physics, 5, 442

\bibitem[Donahue, Saar, \& Baliunas(1996)]{donahue96} Donahue, R.A., 
Saar, S.H., \& Baliunas, S.L. 1996, \apj, 466, 384

\bibitem[Favata et al.\ (2004)]{favata04} Favata, F., Micela, G., 
Baliunas, S. L., Schmitt, J. H. M. M., G\"udel, M., 
Harnden, F. R., Jr., Sciortino, S. \& Stern, R. A. 2004, \aap, 418, 13

\bibitem[Flower(1996)]{flower96} Flower, P.J. 1996, \apj, 469, 355

\bibitem[Giampapa, Worden, \& Gilliam(1979)]{giampapa79} Giampapa, M.S.,
Worden, S.P., \& Gilliam, L.B. 1979, \apj, 229, 1143

\bibitem[{Gray \& Corbally(1994)}]{gray94} Gray, R.O. \& Corbally, C.J.
1994, \aj, 107, 742

\bibitem[{Gray \& Corbally(2009)}]{gray09} Gray, R.O. \& Corbally, C.J.
2009, Stellar Spectral Classification (Princeton: Princeton University 
Press)

\bibitem[Gray et al.\ (2003)]{gray03} Gray, R.O., Corbally, C.J.,
  Garrison, R.F., McFadden, M.T. \& Robinson, P.E. 2003, \aj, 126,
  2048.

\bibitem[{Gray et al.\ (2006)}]{gray06} Gray, R.O., Corbally, C.J., 
  Garrison, R.F., McFadden, M.T., Bubar, E.J., McGahee, C.E., 
  O'Donoghue, A.A., \& Knox, E.R. 2006, \aj, 132, 161 

\bibitem[Gray et al.\ (2011)]{gray11} Gray, R.O., McGahee, C.E., 
Griffin, R.E.M. \& Corbally, C.J. 2011, \aj, 141, 160.

\bibitem[Guinan \& Engle(2009)]{guinan09} Guinan, E.F. \& Engle, S.G. 
2009, in ``The Ages of Stars'', Proceedings of the International Astronomical 
Union, IAU Symposium, Vol. 258, p. 395

\bibitem[Hall(2008)]{hall08} Hall, J.C. 2008, Living Rev. Solar Phys. 5, 2
http://solarphysics.livingreviews.org/Articles/lrsp-2008-2/

\bibitem[Hall et al.\ (2009)]{hall09} Hall, J.C., Henry, G.W., Lockwood, G.W., 
Skiff, B.A. \& Saar, S.H. 2009, \aj, 138, 312.

\bibitem[Hempelmann et al.\ (2003)]{hempelmann03} Hempelmann, A., Schmitt, J, 
Baliunas, S.L. \& Donahue, R.A. 2003,  \aap, 406, 39.

\bibitem[Jancart et al.\ (2005)]{jancart05} Jancart, S., Jorissen, A., 
Babusiaux, C., \& Pourbaix, D. 2005, \aap, 442, 365

\bibitem[Jetsu, Pelt \& Tuominen(1993)]{jetsu93} Jetsu, L., Pelt, J., 
\& Tuominen, I. 1993, \aap, 278, 449

\bibitem[Keil, Henry, \& Fleck(1998)]{keil98} Keil, S.L., Henry, T.W., \&
Fleck, B. 1998, Synoptic Solar Physics -- 18th NSO/Sacramento Peak Summer 
Workshop held at Sunspot, New Mexico 8-12 September 1997. ASP Conference 
Series Vol. 140 ed. K. S. Balasubramaniam, J. Harvey \& D. Rabin, p.301

\bibitem[Kramida et al.\ (2014)]{nist} Kramida, A., Ralchenko, Yu., Reader, J. 
\& the NIST ASD Team 2014, NIST Atomic Spectra Database (version 5.2), 
http://physics.nist.gov/asd National Institute of Standards and Technology, 
Gaithersburg, MD.

\bibitem[Kurucz et al.\ (1984)]{kurucz84} Kurucz, R. L., Furenlid, I., 
Brault, J., \& Testerman, L. 1984, Solar Flux Atlas from 296 to 1300 nm 
(Sunspot, New Mexico: National Solar Observatory)

\bibitem[Labonte(1986)]{labonte86} Labonte, B.J. 1986, \apjs, 69, 229

\bibitem[Labonte \& Rose(1985)]{labonte85} Labonte, B.J. \& Rose, J.A.
1985, \pasp, 97, 790

\bibitem[Lockwood et al.\ (2007)]{lockwood07} Lockwood, G.W., Skiff, B.A., 
Henry, G.W., Henry, S., Radick, R.R., Baliunas, S.L., Donahue, R.A. 
\& Soon, W. 2007, \apjs, 171, 260

\bibitem[Luhman et al.\ (2007)]{luhman07} Luhman, K.L., Patten, B.M., 
Marengo, M., Schuster, M.T., Hora, J.L., Ellis, R.G., Stauffer, J.R., 
Sonnett, S.M., Winston, E., Gutermuth, R.A., Megeath, S.T., Backman, D.E., 
Henry, T.J., Werner, M.W., \& Fazio, G. G. 2007, \apj, 654, 570

\bibitem[Mamajek \& Hillenbrand(2008)]{mamajek08} Mamajek, E.E. \& 
Hillenbrand, L.A. 2008, \apj, 687, 1264

\bibitem[Melo et al.(2006)]{melo06} Melo, C., Santos, N.C., Pont, F. 
et al. 2006, \aap, 460, 251

\bibitem[Mermilliod, Mermilliod, \& Hauck(1997)]{mermilliod97} Mermilliod, 
J.-C., Mermilliod, M., \& Hauck, B. 1997, \aaps, 124, 349

\bibitem[Mishenina et al.\ (2012)]{mishenina12} Mishenina, T. V.,  
Soubiran, C.,  Kovtyukh, V. V.,  Katsova, M. M. \& Livshits, M. A. 2012,
\aap, 547, 106

\bibitem[Moultaka et al.\ (2004)]{moultaka04} Moultaka, J., Ilovaisky, S. A., 
Prugniel, P., \& Soubiran, C. 2004, \pasp, 116, 693

\bibitem[Nidever et al.\ (2002)]{nidever02} Nidever, D.L., Marcy, G.W., 
Butler, R.P., Fischer, D.A., \& Vogt, S. S. 2002, \apjs, 141, 503

\bibitem[Noyes et al.\ (1984)]{noyes84} Noyes, R. W., Hartmann, L. W., 
Baliunas, S. L., Duncan, D. K., \& Vaughan, A. H. 1984, \apj, 279, 763

\bibitem[Queloz et al.\ (1998)]{queloz98} Queloz, D., Allain, S., 
Mermilliod, J.-C., Bouvier, J., \& Mayor, M. 1998, \aap, 335, 183 

\bibitem[Phillips(1992)]{phillips92} Phillips, K.J.H. 1992, 
Guide to the Sun (Cambridge: Cambridge Univ. Press)

\bibitem[Pillitteri et al.\ (2011)]{pillitteri11} Pillitteri, I., 
G\"{u}nther, H.M., Wolk, S.J., Kashyap, V.L., \& Cohen, O. 2011,
\apjl, 741, L18

\bibitem[Santapaga et al.\ (2011)]{santapaga11} Santapaga, T., Guinan, E.F., 
Ballouz, R., Engle, S.G., \& Dewarf, L. 2011, Bulletin of the American 
Astronomical Society 217, 343.12

\bibitem[Sch\"{u}ssler et al.\ (2003)]{schussler03} Sch\"{u}ssler, M., 
Shelyag, S.,  Berdyugina, S.,  Vögler, A. \& Solanki, S. K. 2003, \apj, 
597, 173.

\bibitem[Skrutskie et al.\ (2006)]{skrutskie06} Skrutskie, M. F., 
Cutri, R. M., Stiening, R., et al. 2006, \aj, 131, 1163

\bibitem[Smith \& Redenbaugh(2010)]{smith10} Smith, G.H. \& Redenbaugh, A.K.
2010, \pasp, 122, 1303

\bibitem[Soderblom, Duncan \& Johnson(1991)]{soderblom91} Soderblom, D.R.,
Duncan, D.K., \& Johnson, D.R.H. 1991, \apj, 372, 722.

\bibitem[Telting et al.\ (2014)]{telting14} Telting, J. H., Avila, G., 
Buchhave, L., Frandsen, S., Gandolfi, D., Lindberg, B., Stempels, H. C., 
Prins, S., \& the Nordic Optical Telescope staff 2014, AN, 335, 41

\bibitem[van Leeuwen(2007)]{vanleeuwen07} van Leeuwen, F. 2007, \aap, 474, 
653

\bibitem[Vaughan, Preston, \& Wilson(1978)]{vaughan78} Vaughan, A.H., 
Preston, G.W., \& Wilson, O.C. 1978, \pasp, 90, 267

\bibitem[Wilson(1968)]{wilson68} Wilson, O.C. 1968, \apj, 153, 221.

\bibitem[Wilson(1978)]{wilson78} Wilson, O.C. 1978, \apj, 226, 379.

\end{thebibliography}
\end{document}